\documentclass[review]{elsarticle}

\usepackage{lineno}
\usepackage[colorlinks,citecolor=blue,linktoc=all,linkcolor=cyan]{hyperref}
\usepackage{graphicx}

\usepackage[T1]{fontenc}
\usepackage{dsfont}               % use mathds instead of mathbb for outline fonts
\usepackage{mathrsfs}             % provides mathscr without overwriting mathcal
\usepackage{slashed}              % For Dirac slash notation.
\usepackage{amsmath}
\usepackage{amssymb}
\usepackage{amsbsy}
\usepackage{amsfonts}

\numberwithin{equation}{section}
\numberwithin{table}{section}
\numberwithin{figure}{section}

\journal{Progress in Particle and Nuclear Physics}

\usepackage{titlesec}
\usepackage{sectsty}
\titleformat{\section}{\normalfont\Large\bfseries}{\thesection}{1em}{}
\titleformat{\subsection}{\normalfont\large\bfseries}{\thesubsection}{1em}{}
\titleformat{\subsubsection}{\normalfont\normalsize\bfseries}{\thesubsubsection}{1em}{}

\begin{document}

\begin{frontmatter}

\title{Lattice perspectives on doubly heavy tetraquarks}

%authors, affiliations, corresponding author mention 
\author[mymainaddress]{Anthony Francis}

%\author[mysecondaryaddress]{Corresponding author\corref{mycorrespondingauthor}}
\ead{afrancis@nycu.edu.tw}

\address[mymainaddress]{Institute of Physics, National Yang Ming Chiao Tung University, 1001 Daxue Road, 30010 Hsinchu City, Taiwan}
%\address[mysecondaryaddress]{Second institution (if applicable)}

\begin{abstract}

Doubly heavy tetraquarks have emerged as new probes to study the heavy hadron spectrum. With the experimental observation of the $J^P=1^+$ $T_{cc}^+$, they pose a unique opportunity to bring together efforts in experiment, phenomenology, and lattice QCD. In lattice calculations they are accessible as ground states, unlike  hidden flavor tetraquarks, and this enables accurate determinations of the scattering parameters alongside the binding energies of these tetraquarks. 
Today, lattice calculations firmly predict $J^P=1^+$ $T_{bb}^{ud}$ and $T_{bb}^{us}$ as QCD bound states, while recent studies approaching the $J^P=1^+$ $T_{cc}^{ud}$ find it to be a virtual bound state at slightly non-physical input quark masses. Studies of the $J^P=1^+$ $T_{bc}^{ud}$ are ongoing and  a new focus area. In light of these developments the evolution of this field until this point is reviewed. Emphasis is put on the methods in lattice spectroscopy that enable a robust evaluation of the lattice studies gathered. They are further reviewed towards their limitations and achievements. Current challenges and opportunities are discussed, including possibilities to approach the left-hand cut in the scattering analysis of the charm candidates and towards understanding the structure of those including two bottom quarks. 

\end{abstract}


\end{frontmatter}

\newpage

\thispagestyle{empty}
\tableofcontents

%to begin the line numbers: 
%\linenumbers

%beginning of the core of the manuscript
%newpage
\section{Motivation}

In recent years, the study of doubly heavy tetraquarks on the lattice has been gaining momentum, with more and more sophisticated studies being published at a high rate. Arguably, the discovery of the $I(J^P)=0(1^+)$ $T_{cc}^{ud}$ doubly heavy tetraquark by LHCb \cite{LHCb:2021vvq,LHCb:2021auc} in 2021 in the middle of the COVID-19 pandemic provided a boost for this research. However, at the same time, it added to an already ongoing discussion that had been growing in intensity in the community since especially the robust predictions of deeply bound $I(J^P)=0(1^+)$ $T_{bb}$ doubly heavy tetraquarks in the lattice QCD calculations \cite{Bicudo:2012qt} and \cite{Francis:2016hui}. Here, the attempt is made to gather the materials and provide an overview of the ongoing discussion, starting with the first studies of inter-meson potentials in systems of two static-light mesons in 1990 \cite{Richards:1990xf}. 
A simultaneous focus is on practical aspects of the calculations and analyses done, highlighting the development of the research program in sync with the evolution of lattice QCD as a field.

This review is aimed at QCD experts with a basic knowledge of lattice QCD and practitioners just getting into this research. Emphasis is put on describing the possible systematics inherent to the methods used and providing a practical guide to understanding the newest lattice QCD studies and publications on this topic. For example, the attempt is made to motivate an understanding of the typically presented finite volume spectrum plots that are becoming common and how to interpret them. 
One reason for this is the subjective motivation that while lattice QCD results are entering increasingly important particle physics applications, often their details are not that well known, and a type of black box effect appears. This is dangerous, especially regarding lattice QCD results, due to the inherent aspect of controlling all systematics. Indeed, even in the most current and advanced calculations, there are many ways in which they can be improved, and studies are just scratching the surface. The field is dynamic and reactive to new approaches, better ways of calculating an observable, or adapting a combination of tools other groups have seen to lead to superior control over systematics and statistics. Many calculations being done today were not on the horizon of possibilities just a few years ago, and keeping track of things can be challenging; this review hopes to make this easier. Nevertheless, it cannot claim to be exhaustive or complete, and any opinions expressed are the author's own. 
In this regard, there are more reviews and whitepapers on the topic of tetraquarks with different focus points in phenomenology, experiment and lattice that are highly recommended as further literature \cite{Richard:2016eis,Esposito:2016noz,Ali:2017jda,Karliner:2017qhf,Belle-II:2018jsg,Cerri:2018ypt,Chen:2022asf,Mai:2022eur,Yasui:2022brf,ParticleDataGroup:2022pth,Gross:2022hyw,Brambilla:2022ura,Bicudo:2022cqi,Bulava:2022ovd}.
For those interested in more background on lattice QCD as an approach, one could consider \cite{Montvay:1994cy,Luscher:1998pe,DeGrand:2006zz,Gattringer:2010zz} as starting points.

%\newpage
\section{Situational overview}\label{intro}

\begin{figure}
\centering
\includegraphics[width=0.6\textwidth]{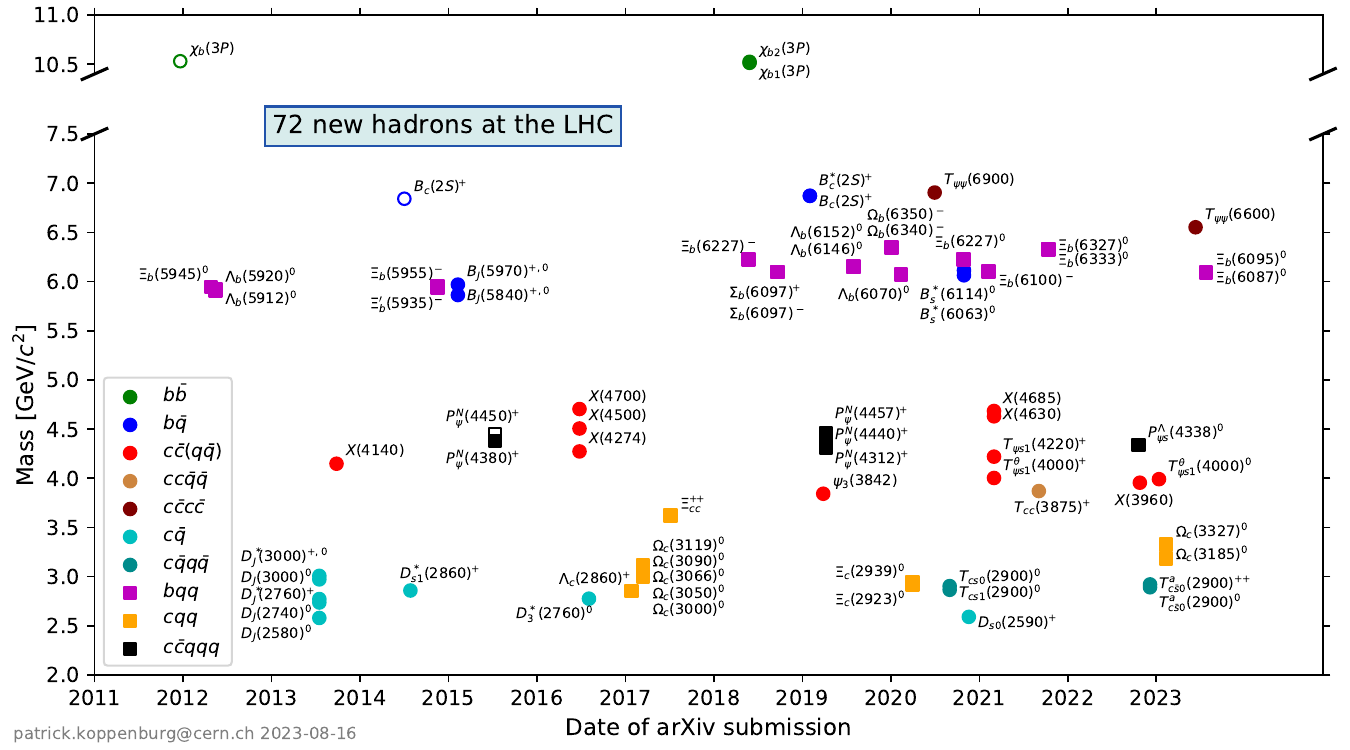}
\caption{\textit{Collected masses of new hadrons observed at LHC \cite{LHCbzoo} at the time of writing.
}}
\label{fig:lhcbzoo}
 \end{figure}

Few would doubt the impact that the study of the heavy hadron spectrum has had in developing and understanding the strong force.
Concepts such as the heavy quark potential, color blocking, and diquarks predate QCD as the complete theory description of the strong force and helped formulate it.
It is all the more puzzling that today we find our perception of the heavy hadron spectrum questioned or incomplete. The need for new input became evident with the discovery of the $X(3872)$ at BELLE in 2003 \cite{Belle:2003nnu}. Prior to this event the predicted and measured hadron masses agreed, the potential model as a description of the hadrons worked well, and the OZI rule applied without exceptions. Now there was an unexpected hadron, that was consistent with a tetraquark and the flavor content $\bar{c}c\bar{u}d$.
Furthermore, this discovery was not an isolated event: 
At the time of writing, LHCb reports the observation of 74 states in the heavy hadron spectrum \cite{LHCbzoo}. This includes the 2023 update, and the masses of the observed new states are shown in Fig.~\ref{fig:lhcbzoo}. Although it could well be that some of these states will ultimately go away as the experimental search is refined, the pace of discovery is staggering, with ten new states from 2022 to 2023 alone.
Color-coded by their likely quark flavor compositions, the figure gathers a large number of candidates for ordinary hadrons alongside the not-so-ordinary tetraquarks and pentaquarks. 
Today there are about a dozen of these states and
both types are unexpected in traditional quark models. 

Many explanation attempts exist for them; however, in most cases, QCD is approximated by quark model extensions aimed at phenomenologically capturing missing dynamics effects. 
Such extensions usually proceed by considering and adding new substructures as effective degrees of freedom or constraints and modifying model interaction terms for them.
For example, 
one could consider including substructures such as diquarks - which will play a role later in this manuscript - or special molecular structures in the model Hamiltonians.
Given the many possible extensions and connected interpretations, it is difficult to make clear-cut statements. This is partly due to the inherent flexibility of such models.
The way forward is to aim for fully non-perturbative insights into the hadron spectrum in an approach that does not need to approximate the QCD dynamics. Lattice QCD calculations provide such an approach and enable the study of tetra- and pentaquarks without having to model any part of the full dynamics.

\subsection{A new family of tetraquarks}

\begin{figure}
\centering
\includegraphics[width=0.5\textwidth]{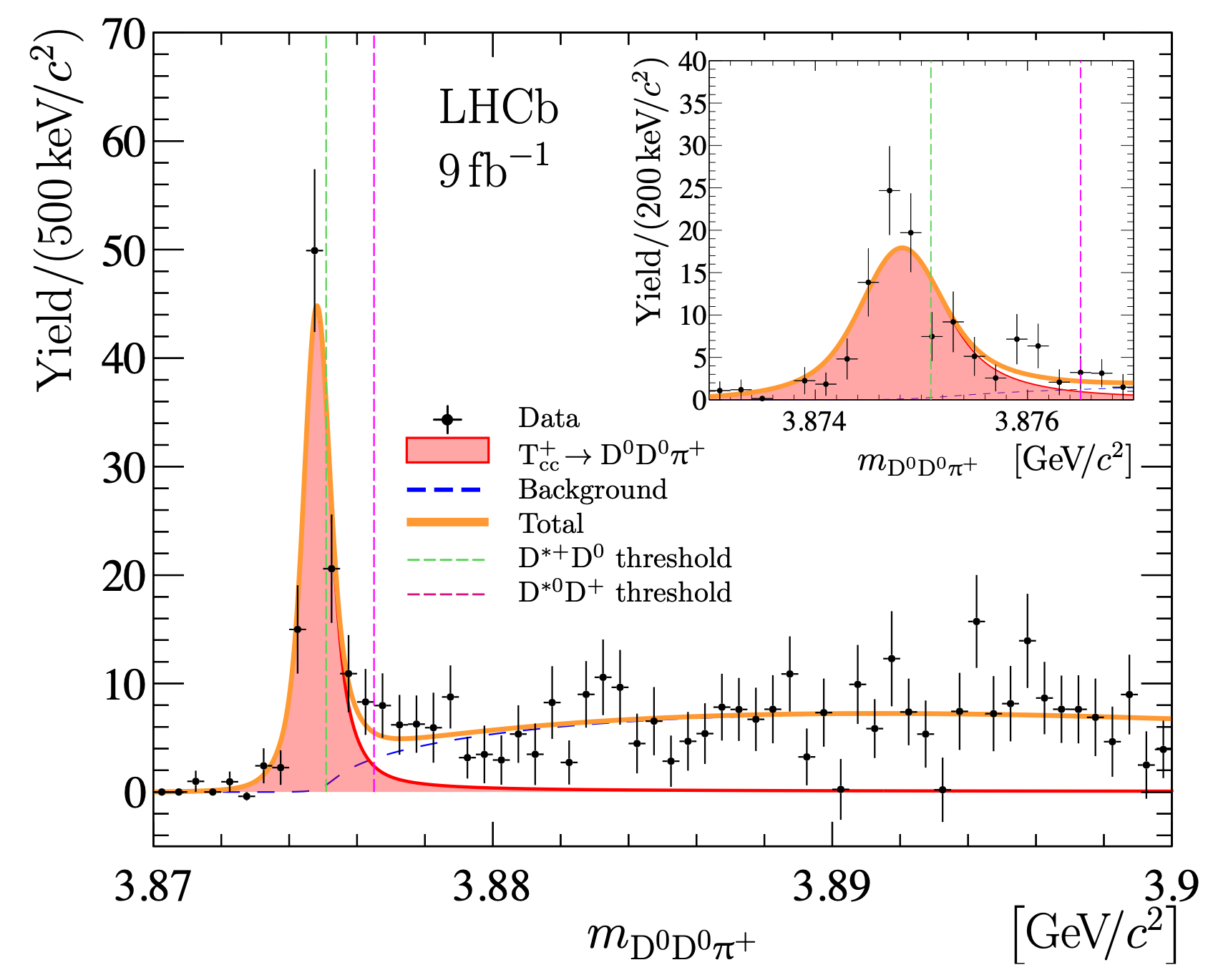}
\caption{\textit{ Subtracted distribution of the $D^0D^0\pi^+$ mass as reported in \cite{LHCb:2021vvq}.
The  $D^{*+}D^0$ and $D^{*0}D^+$ thresholds are shown as vertical dashed lines and the Breit-Wigner peaked line-shape is given in red. The insert shows a zoom of the peak area.
}}
\label{fig:lhcb-tcc}
 \end{figure}

This review focuses on doubly heavy tetraquarks as a new family of heavy hadrons. They are exciting because they are accessible by experiment, QCD phenomenology, and lattice QCD in controlled setups with relatively good statistical signal and without having to perform many difficult extrapolations to control systematics.
One member of this family is the $T_{cc}^{+}$, which was experimentally discovered by LHCb in 2021 \cite{LHCb:2021vvq,LHCb:2021auc}. 
In this study, the authors reported observing a narrow peak structure below the $DD^*$ threshold in the decay channel of $X$ going to $D^0D^0\pi^+$. Fitting the peak to a Breit-Wigner line shape could establish both the binding energy and the width of this state to be
\begin{align}
E_{B}  =-273 \pm 61 \pm 5_{-14}^{+11} ~\mathrm{keV}  ~,~~~~
\Gamma  =410 \pm 165 \pm 43_{-38}^{+18} ~\mathrm{keV}~~.
\end{align}
The corresponding invariant mass plot is shown in Fig.~\ref{fig:lhcb-tcc}. In the follow-up study \cite{LHCb:2021auc}, the state was found to be consistent with $cc\bar{u}\bar{d}$ quark flavor content and quantum numbers $J^P=1^+$. According to the naming conventions of \cite{Gershon:2022xnn}, the observed state is a $J^P=1^+$ $T_{cc}^+$ tetraquark. In the following, we will adopt a similar naming convention, whereby we additionally cite the light flavor content and isospin but drop the charge, for example the $J^P=1^+$ $T_{cc}^+$ will be denoted as $I(J^P)=0(1^+)$ $T_{cc}^{ud}$ in the following.
This discovery catalyzed and added to an ongoing discussion on the theory side in the QCD phenomenology and lattice QCD communities. Indeed, as we will see, today the theory side predicts an entire family of doubly heavy tetraquark states with currently four, perhaps five members in nature. Of these, the experimentally observed $I(J^P)=0(1^+)$ $T_{cc}^{ud}$ is the lightest and most shallow bound one so far. More deeply bound candidates are expected as the heavy quarks are exchanged for heavier partners and especially to $I(J^P)=0(1^+)$ $T_{bb}^{ud}$. 

\paragraph{Diquarks as building blocks.}
On the phenomenological side, one way to understand, and perhaps expect, doubly heavy tetraquarks is as a consequence of the appearance of effective QCD degrees of freedom called diquarks.  
They were initially introduced as a concept to describe the masses of low-lying baryons \cite{Lichtenberg:1967zz}, and their success in correctly reproducing them has been a compelling argument to consider diquarks and their properties as key features of QCD \cite{Jaffe:2004ph}. Arguably, their most interesting property is that two quarks can take on several different diquark configurations, whereby those that are of concern here are:
\begin{align}
& \left|\{q_1 q_2\}=ud;~\overline{\mathbf{3}}_{\mathrm{c}};~\overline{\mathbf{3}}_{\mathrm{f}};~ J^P=0^{+}\right\rangle \nonumber\\
& \left|\{q_1 q_2\}=us,~ds;~\overline{\mathbf{3}}_{\mathrm{c}};~\overline{\mathbf{3}}_{\mathrm{f}};~ J^P=0^{+}\right\rangle \\
& \left|\{q q\}; q\in{u,d,s,c,b};~\overline{\mathbf{3}}_{\mathrm{c}};~{\mathbf{6}}_{\mathrm{f}};~ J^P=1^{+}\right\rangle~~. \nonumber
\end{align}
The first two are in the scalar, so-called "good" diquark configuration and have the potential for a significant attractive effect that is absent in the other vector or "bad" diquark configuration.
 This can be argued from one gluon exchange and instanton interactions \cite{Jaffe:2004ph}.
Indeed, the mass splitting between the good and the bad diquark channels can be calculated in lattice QCD and is found to be $198(4)~$MeV in the $ud$-channel at the physical pion mass, as is shown in Sec.~\ref{sec:diquarks}. 
Further details presented there broadly confirm this picture and the attractive effect in the good diquark. In particular, a splitting of $145(5)~$MeV was found for the $us$ diquark, whereby isospin symmetric quarks $u=d$ were assumed. This implies a more substantial attractive effect for lighter diquark components.

\paragraph{The case for doubly heavy tetraquarks.} A good diquark of this type could then act as a kind of substructure within a doubly heavy tetraquark and, through its attraction, lead to the energy of the tetraquark being reduced below that of two non-interacting mesons, thereby making it a bound state. 
Note that a diquark is not a gauge-invariant object by itself in QCD. So, to introduce a good diquark into a tetraquark-type operator, its open color indices need to be contracted with an antidiquark. This could be of any quark content, however, the doubly heavy option is most appealing for this purpose. 
This seems counter-intuitive because one might expect two good light diquarks to benefit most from the attraction. However, the non-interacting mesons contain $\pi$'s or $K$'s in this scenario. They in turn have low masses through other dynamics and it is difficult for the diquark attraction to compete. Furthermore, the interactions between the constituent quarks in the tetraquark reduce the effectiveness of the attraction.

However, if the quarks are instead heavy and, in particular, close enough to the heavy quark limit for heavy quark spin symmetry (HQSS) to be a good approximation, then one may expect the diquark attraction to become effective in the sense described, see e.g. \cite{Jaffe:2004ph}. Despite only the vector diquark configuration being available when choosing two identical heavy quarks, this is an advantageous combination since the heavy quark limit also suppresses further interactions of the constituents of the would-be tetraquark proportional to $\sim 1/m_Q$, where $Q$ is the heavy quark. This picture of a binding mechanism then implies three properties for doubly heavy tetraquarks: (a) the most likely doubly heavy tetraquarks would have $I(J^P)=0(1^+)$, (b) they would have deeper binding energies the lighter the quarks in the scalar antidiquark component are $E_B(T_{QQ}^{ud})<E_B(T_{QQ}^{us})$, (c) they would have deeper binding energies the heavier the heavy quarks in the vector diquark component are. 
Note at this point that it is convention to consider the tetraquark operator of this type as a combination of a scalar light antidiquark and a vector heavy diquark.
See also \cite{Karliner:2017qjm,Eichten:2017ffp} for predictions for the doubly heavy tetraquarks picture from the phenomenological side.

\paragraph{Limitations of this picture away from the heavy quark limit.}
Within this picture of a binding mechanism, an immediate question is how heavy the quarks need to be. It is a priori not clear whether charm quarks or bottom quarks, for that matter, are close enough to the heavy quark limit.
However, some intuition can be derived from the Particle Data Booklet \cite{ParticleDataGroup:2022pth} by comparing splittings of heavy-heavy-light baryons with splittings of heavy-light mesons. 
Following \cite{Eichten:2017ffp} HQSS seems a good approximation for the bottom quark case but not so much for the charm quark case\footnote{As a side remark, another option is to calculate HQSS-related properties more directly in lattice QCD by embedding the heavy quarks in hadrons with static quarks, which cancel out in mass differences.}.
As such the argument for the good diquark heavy quark spin symmetry binding mechanism is the strongest for $I(J^P)=0(1^+)$ $T_{bb}^{ud}$ doubly heavy tetraquarks. 
This poses the question if the picture holds, or has anything to say, in flavor combinations away from this case, especially for the $I(J^P)=0(1^+)$ $T_{cc}^{ud}$. That heavy quark spin symmetry is not a good approximation here, implies that the interaction between both the heavy quarks and those between the heavy and the light quarks could not be heavily enough suppressed to invoke the suggested binding mechanism. As further word of caution, other areas in hadron physics containing charm quarks have proven that heavy quark arguments can fail to capture the essential physics \cite{Cheng:2021vca}.
Finally, note that in the case of the $I(J^P)=0(1^+)$ $T_{bc}^{ud}$, in principle, the good diquark configuration is accessible to the heavy diquark component. This could imply binding opportunities for other quantum number combinations, particularly $J^P=0^+$.
Although the binding mechanism could be different in both cases just mentioned. An important point is that in all cases, the hope is to see a ground state tetraquark candidate. %This significantly impacts the prospects of determining the tetraquark energies directly in lattice QCD calculations. 

\subsection{Doubly heavy tetraquarks on the lattice}

To illustrate why doubly heavy tetraquarks lend themselves to lattice QCD calculations, consider the comparison with other types of tetraquarks, such as
the $X(3872)$. Recall that the flavor structure of this state is $\bar{c}c\bar{u}d$. On the lattice, this requires a different handling using an all-to-all propagator, as explained further below.
In lattice studies of this state, see e.g. \cite{Prelovsek:2013cra,Li:2024pfg}, the next fundamental difficulty is that it is very close to threshold and couples to the excitation spectrum of $c\bar{c}$. This makes their study extremely challenging as a coupled channel analysis is required, and many finite volume energies need to be determined at high precision.
For doubly heavy tetraquarks, the situation is different. The good diquark heavy quark spin symmetry picture implies a deeply bound state below the corresponding meson-meson threshold. 
If they exist, this means the doubly heavy tetraquarks are the ground states and can be read off from the asymptotic-$t$ behavior of their lattice correlation functions. 

\paragraph{Lattice motivation beyond tetraquarks.} 
Although doubly heavy tetraquarks are interesting in themselves from a physics point of view, from a lattice point, they offer another perspective on multi-hadron or -quark systems and can serve as test-beds to develop new techniques. For example, it is a long-term goal of the lattice community to perform an ab initio study of multi-nucleon systems and especially the deuteron \cite{Tews:2022yfb}. However, apart from the computational challenges, there are also challenges in performing the necessary analyses. 
As will be discussed in Sec.~\ref{sec:lefthandcut} the deuteron suffers from the so-called "left-hand cut problem". In short this means that there is another decay channel open that is close to the deuteron and threshold of two non-interacting nucleons that we want to study. This limits the validity of the effective range expansion, which is usually used to determine the scattering parameters. The situation in the $T_{cc}^{ud}$ is very related and a good example: It is a state below the $DD^*$ threshold, however the $D^*$ itself decays into $D\pi$. This needs to be consistently handled. As such developing a formalism in the doubly heavy tetraquarks first, where data is available and new practical insights gained, can be part of a broader strategy to access multi-hadron systems in lattice QCD.

\paragraph{Lattice potentials and spectra.}
One can follow one of two paths to access doubly heavy tetraquarks on the lattice: The first is to aim to extract a lattice potential, and the second is to extract the lattice spectra. 
Below is a short overview, details and extended references are given in the more extensive discussions throughout the remainder of this review.

In the potential-based approach, one typically works with interpolating operators where at least some of the quarks are static, which means their mass is infinite. As explained in Sec.~\ref{sec:potentials}, the extracted lattice data may be fitted using an appropriate ansatz for the potential. The fitted functional form can then be input into a Schr\"odinger equation and further processed to yield scattering information and energy levels, see e.g. \cite{Bicudo:2012qt}. 
An extension of this approach away from static quarks is the so-called HALQCD method \cite{Aoki:2009ji}. To determine lattice potentials, this method identifies a certain lattice correlation function with the Nambu-Bethe-Salpeter (NBS) wave function. This can be used to, in turn, determine the potential. However, this lattice potential is nonlocal, and the latter step 
requires a derivative expansion of which often only the first, local, term is determined. The extracted local potential can be further analyzed for binding energies and scattering parameters. 
The potential approach has contributed significantly to the success of the study of doubly heavy tetraquarks.
However, the potential data needs to be fitted to an ansatz and controlling the systematics can be difficult. This is discussed in more detail in Sec.~\ref{sec:potentials}.

The spectrum-based approach follows a different strategy. It aims to determine the finite volume energy levels of the target system. 
In the case of a deeply bound doubly heavy tetraquark the first step could be to equate the observed lowest-lying energy level with the ground state energy. 
Comparing this result with the corresponding threshold, defined by two non-interacting mesons, one can unambiguously determine the binding energy, assuming that the systematic uncertainties of the finite volume, chiral, and discretisation effects are under control.
The extension of this approach is to expand towards a scattering analysis \cite{Luscher:1985dn,Luscher:1986pf}. For this multiple finite volume energies, the ground state and excited energy levels, have to be extracted. They can then be converted into scattering phase shifts using finite volume quantization conditions. The scattering phase shifts, in turn, allow for the determination of the effective range and scattering length. In more complicated systems, a parametrization of the scattering matrix is also an option \cite{Guo:2012hv}. Note that this introduces a model in principle and can lead to a systematic that is difficult to estimate, see also Sec.~\ref{sec:scatt-phase}. 

In all cases, the first step is to identify a set of interpolating operators to determine a large number of lattice correlation functions. Especially for the spectrum approach the aim is to realize all combinations of these interpolating operators at the source and at the sink to generate a large correlation matrix \cite{Michael:1985ne,Blossier:2009kd}. Diagonalizing this correlation matrix gives access to a number of finite volume energy levels that is, in principle, as large as the rank of the matrix, see Sec.~\ref{sec:gevp} for details. 
The spectrum approach directly evaluates these energy levels as key outputs and at the first instance some information on the binding energy can be extracted. 
In the potential approach instead these measurements yield information on attraction or repulsion in a given channel. Later on we will see how this can be further used to address questions in the structure of the tetraquark states.

\paragraph{Outline of this work and brief synopsis of the community status.}
Exploiting both approaches and using the freedom that the lattice calculations have to choose and exchange quark flavors, quantum numbers, and masses at will opens the unique opportunity to map out the binding properties of doubly heavy tetraquarks; thereby acting as constraints for phenomenological models designed to describe them.
Today, the lattice community has mapped a large part of this parameter space to gain qualitative insights. It is beginning to perform precision calculations for the most prominent doubly heavy tetraquark candidates. 

Below in Sec.~\ref{three}, the methods briefly sketched are explained in detail. In Sec.~\ref{four}, the development of doubly heavy tetraquark lattice calculations from the early days until the first few months of 2024 is given. Then, in Sec.~\ref{five}, challenges and opportunities are highlighted, and finally, a short conclusion is given.

%\newpage
\section{Tetraquark correlators on the lattice: Definitions and methods }
\label{three}

%Invariant mass plots are of central importance in experiments as a key tool for understanding the hadrons spectrum. Through their study, new hadrons are observed, such as in the case of the $T_{CC}$ shown in Fig.~\ref{FIG}, and their properties.
%On the theory side, the goal is to have or develop a tool that connects as closely to them as possible from a cleanly defined observable in quantum field theory. Lattice correlation functions encode this information in terms of a spectral function $\rho(\omega)$. However, they do so in an integrated way, as lattice correlation functions are the result of the integral over the convolution of the spectral function with an application-dependent kernel. In the following section, we will briefly introduce lattice correlation functions and their relation to spectral functions. We will further examine how the finite volume impacts their study, and what opportunities and challenges this leads to. 

From the most general starting point, lattice QCD is the formulation of quantum field theory in Euclidean space-time and placed onto a finite 4-dimensional hypercubic lattice.
This regulates the theory and cleanly defines the expectation values of observables in terms of functionals of the type $\langle O \rangle = \frac{1}{Z}\int dA d\bar{\psi} d\psi \,O\,e^{-S_{QCD}(A,\bar{\psi},\psi)}$ with the QCD action $S_{QCD}$ and the partition function $Z=\int dA d\bar{\psi} d\psi \,e^{-S_{QCD}(A,\bar{\psi},\psi)}$.
The finiteness of the lattice volume introduces a discreteness for the lattice momenta, and with periodic boundary conditions in space, the dispersion relation is limited to the Brillouin zone. As such, all lattice momenta are in the range $-2\pi/L<p\leq 2\pi/L$ in lattice units. This is also true in the limit $a\rightarrow 0$, and the discreteness of the spectrum remains intact as long as the volume stays finite.
As such, all lattice-derived observables must be connected to the continuum, $a\rightarrow 0$, and infinite volume, $L\rightarrow \infty$, limits. In particular in the case of the volume this can be done by either extrapolating the results from a sample of different lattice setups or by including analytic knowledge that guides the finite lattice results to their infinite volume counterparts.
   
As mentioned, one benefit of lattice QCD is that practitioners can freely choose the physics parameters that are the input into the calculation, i.e. the number of dynamical quark flavors, the quark masses, volumes and lattice spacings. They all have an impact on the spectrum and as we will see in Sec.~\ref{sec:latlimit} there are also some numerical limitations.
Of course, in principle, we are interested only in setups as close to QCD as possible and with all limits taken, so that only the quark masses are open parameters. Such lattice calculations are not limited to the physical quark flavors; they can study hypothetical flavor combinations and mass dependencies at will.
However, we will see in the review of historical results that how close to QCD we get changed significantly over time, making the inclusion of the other parameters necessary for the discussion. 
Precision QCD calculations are extremely demanding because the gap between the values for all parameters in the lattice calculation and nature has to be understood and quantified to a very high level.
In the case where a qualitative understanding is desired, for example, wanting to decide upon a molecular or elementary, compact structure of a state, the freedom to also move away from the QCD values is a feature: The extra parameter dimensions pose constraints on a suggested binding mechanism or structure and can yield valuable insights in themselves.

\subsection{Correlation and spectral functions}

In experimental particle physics, invariant mass plots are of central importance to understanding the hadron spectrum, as it is through their study that new hadrons are observed like for example in the case of the $J^P=1^+$ $T_{cc}^{ud}$ shown in Fig.~\ref{fig:lhcb-tcc}.
In principle, lattice QCD correlation functions encode this information in terms of a spectral function $\rho(\omega)$. Indeed in general terms the spectral function can be linked to the experimentally observed cross-sections. 
However, lattice calculations have only access to them in an integrated way: Measurable lattice correlation functions are the result of the integral over the convolution of the spectral function with an application-dependent kernel. In the following section, we will briefly introduce lattice correlation functions and their relation to spectral functions.
If not stated otherwise, $\vec p=0$ is considered and a generalization to $\vec p\neq 0$ is usually straightforward.

Most lattice QCD calculations relevant for this review are performed on Euclidean signature $L^3\times T$ lattices with periodic boundary conditions for the spatial directions $L$, anti-periodic boundary conditions\footnote{Some calculations in \cite{Hudspith:2023loy} also use open boundary conditions.} in $T$ and $T\gtrsim 2L$. This choice of boundary formally defines the lattice calculation to correspond to QCD at finite temperature. Staying in the continuum infinite volume situation for now, the 2-point correlation function with hadron interpolating operators $\mathcal{O}_i$ can then be written as:
 \begin{align}
C_{\mathcal{O}_1 \mathcal{O}_2}(t,\mathcal{T})=\int d^3 x \left\langle\mathcal{O}_1(t,\vec x) \mathcal{O}_2^{\dagger}(0,\vec 0)\right\rangle_\mathcal{T} ~~,
\end{align}
where the $\mathcal{T}=1/T$ is the temperature of the system and the expectation values are taken with respect to $e^{-H/T}/Z(T)$. The Euclidean time evolution of the operators is $\mathcal{O}(t)=e^{H t} \mathcal{O} e^{-H t}$ and the spectral representation of the correlator is, see e.g., \cite{Meyer:2011gj}:
\begin{align}
G_{\mathcal{O}_1 \mathcal{O}_2}(t, \mathcal{T}) \stackrel{\mathcal{O}_1 = \mathcal{O}_2}{=} \int_0^{\infty} \frac{d \omega}{2 \pi} ~\rho(\omega, \mathcal{T}) \frac{\cosh [\omega(1 / 2\mathcal{T}-t)]}{\sinh (\omega/ 2\mathcal{T})}  \nonumber
\end{align}
where the spectral function is defined as
\begin{equation}
\rho(\omega, \mathcal{T})=\int d t d^3 x \,e^{i \omega  t}\,\Big\langle\left[\mathcal{O}_1(t, \vec{x}), \mathcal{O}_2(0)^{\dagger}\right]\Big\rangle_\mathcal{T}~.
\end{equation}
Note that here, the operators evolve as $\mathcal{O}(t) = e^{i H t} \mathcal{O} e^{-i H t}$. In the limit of $\mathcal{T}=0$, the correlator reduces to:
\begin{align}
G_{\mathcal{O}_1 \mathcal{O}_2}(t) = \int_0^{\infty} \frac{d \omega}{2 \pi} ~\rho(\omega)~e^{-\omega t}~~.
\end{align}
Because $T$ is large in the following, i.e. $\mathcal{T}\sim 0$, this form is commonly used in lattice calculations of doubly heavy tetraquarks with the limitation that $t\lesssim T/2$. 
In these cases, the $\cosh$-behavior is only important when wanting to benefit from the backward propagating component to boost statistics. The spectral function $\rho$ has the properties $\rho(-\omega)=\rho(\omega)$ and Im$(\rho)=0$. To reconstruct the spectral function from the measurable lattice correlator, one needs to perform an inverse Laplace transform from a finite set of correlator data with errors. This is very challenging and often called a numerically ill-posed problem\footnote{See \cite{Cuniberti:2001hm} for a detailed discussion and the possibility that it is actually ill-conditioned.}.

Focusing on the spectral function, given the continuum infinite volume theory expressions above the situation is modified by the discrete space-time in the lattice calculation. In particular the finite lattice spacing forces the spectral function to be zero at the cut-off energy scale, irrespective of the volume, $\omega\in[0:\omega_{max}]$. At the same time, the finite volume reduces the support of allowed energies $\omega\rightarrow\omega_i$ to a subset given by the available lattice momenta. Effectively, this turns the spectral function into a collection of $\delta$-functions that becomes more dense as the volume is increased, the gaps between the allowed energies thereby being $\Delta \omega_i \sim (2\pi/L)^2$.
This implies that it can be a benefit for a lattice calculation to work in a small volume as the collection of $\delta$-functions becomes more separated, with a relatively sparser area especially in the low-$\omega$ regime. As we will see in Sec.~\ref{sec:gevp} with a sparse spectrum it becomes feasible to robustly determine the lowest few finite volume energies that can be further analyzed and thereby circumventing the inverse problem to some extent.
Given these general features of lattice spectral functions, it becomes clear that integrating over each $\delta$-function in them contributes one term to the correlator:
\begin{align}
G_{\mathcal{O}_1 \mathcal{O}_2}(t) = \int_0^{\infty} \frac{d \omega}{2 \pi} ~\rho(\omega)~e^{-\omega t}~\rightarrow ~~Z_n \sum_n e^{-E_n t}~~,
\end{align}
whereby the positivity condition of the spectral function implies $Z_n\geq 0$. In the more standard derivation \cite{Gattringer:2010zz} the $Z_n$ are identified with the matrix elements $\langle 0| \mathcal{O}_1  |n \rangle\langle n| \mathcal{O}_2  |0 \rangle$, which for $\mathcal{O}_1=\mathcal{O}_2$ becomes $|\langle 0| \mathcal{O}_1  |n \rangle|^2$.
In the asymptotic limit $t\rightarrow \infty$, we see that the lowest energy level above the vacuum will dominate the correlator. This is the ground state of the system we are studying. In the interacting theory, there are sharp peaks above the continuous background for the hadronic states as their matrix elements are enhanced. As a consequence, the ground state energy can be extracted from the long $t$ behavior of the correlation function without having to perform a spectral reconstruction.

\paragraph{Infinite volume limit, scattering and a no-go theorem.}

Because the lattice spectral function is always a collection of delta functions, arriving at the infinite volume limit is subtle and in principle not well defined. 
However, in \cite{Hansen:2017mnd} the authors could show that by introducing a smearing function of width $\Delta$, a clean infinite volume limit can be defined via:
\begin{equation}
\rho(\omega)=\lim _{\Delta \rightarrow 0} \lim _{L \rightarrow \infty} \widehat{\rho}(\omega, L, \Delta)~~,
\end{equation}
where $\widehat{\rho}(\omega, L, \Delta)$ is the finite volume spectral function smoothed by the smearing function. Taking the double limit in the correct order is essential. 
Although they have not been applied to doubly heavy tetraquarks yet, 
these new insights have led to many new applications, see e.g., \cite{Bulava:2019kbi} for an idea to extract scattering amplitudes or the determination of properties from transition spectral functions \cite{Hansen:2017mnd,Bulava:2019kbi,Gambino:2020crt,Evangelista:2023fmt,Alexandrou:2024gpl}  

Apart from pointing at new directions to research, these developments are mentioned here because they address a fundamental issue with studying scattering, and by extension doubly heavy tetraquarks, on the lattice: They seemingly circumvent the no-go theorem by Maiani and Testa \cite{Maiani:1990ca} in 1990, which stated that no useful scattering information can be obtained from Euclidean correlation functions.
On the contrary, as clarified on general grounds in \cite{Bruno:2020kyl} in 2020, certain scattering information can be extracted from certain types of smeared spectral functions. 
An alternative, much more widely used approach, to study scattering on the lattice that circumvents the no-go theorem was put forward by L\"uscher in \cite{Luscher:1985dn,Luscher:1986pf} in 1985. In these studies, the author showed that the scattering information for a 2-to-2 process can be extracted indirectly from the finite volume energy levels. The connection to the above discussion on spectral functions in finite volumes then becomes clear.
This is the finite volume scattering formalism, which we will review in the next section. Note that in the finite volume scattering formalism, we usually assume that the continuum limit has already been taken.

\subsubsection{Finite volume energy levels, bound states and scattering phase shifts}
\label{sec:scatt-phase}

Determining scattering phase shifts in lattice QCD is only briefly introduced here with a focus on the practical recipe following \cite{Morningstar:2015sna,Francis:2018qch}.
Derivations of the full formalism are available in dedicated resources, see e.g. \cite{Briceno:2014oea,Briceno:2017max} for a review or \cite{BerkoScatt} for a hands-on introduction in $D=1$ quantum mechanics. 
The practical recipe in mind is the exploitation of finite volume quantization conditions that relate the finite volume energies determined in lattice QCD calculations to the infinite volume scattering phase shift. Even though many new theoretical developments are happening now, see Sec.~\ref{sec:lefthandcut}, the basic idea dates back as early as 1955 \cite{Reifman:1955ca,DeWitt:1956be}. There, it was observed that the finite volume energies can, in principle, be linked to scattering phase shifts, even though this early work did consider the infinite volume limit of the system.
At this point, we identify two interesting questions: The first is how an individual finite volume energy behaves with respect to the size of the volume, and the second is how finite volume energies can be used to extract infinite volume information in the form of scattering phase shifts. These questions were first clarified in \cite{Luscher:1985dn,Luscher:1986pf} and \cite{Luscher:1990ux,Luscher:1990ck}. The details were further worked out and expanded in \cite{Rummukainen:1995vs,Lellouch:2000pv,Gockeler:2012yj} both at rest and in moving frames. 
A milestone was reached with the reformulation and rederivation within a field theory-based approach \cite{Kim:2005gf}, which has since sparked multiple extensions and generalizations, see e.g., \cite{Briceno:2017max,Hansen:2012tf,Briceno:2014uqa} and the material in Sec.~\ref{sec:lefthandcut}.

\paragraph{Determining a bound state without scattering analysis.} The first question is essential for understanding how a ground state energy determined from a lattice calculation may be related to the binding energy $E_B= E_{\textrm{candidate}} - E_{\textrm{threshold}}$.
The two scenarios to consider are the volume dependence of a scattering and of a true bound state. \\
\textit{Scenario 1: Scattering state.} In \cite{Luscher:1986pf}, it was found on rather general grounds that if a finite volume energy belongs to a scattering state in $S$-wave, then the volume dependence behaves as 
\begin{equation} 
E_{B,L}\sim E_{B,\infty} \cdot \Big[~ 1 + c_1\frac{a_0}{L^3} + \mathcal{O}(\frac{1}{L^4}) + ...~\Big]~~,
\label{eq:fvol-plaw-formula}
\end{equation}
where $a_0$ denotes the $S$-wave scattering length. The coefficient $c_1$ introduced here depends on the details of the scattering process and energy level, it will be discussed more below.
An important aspect of this coefficient is that it may be negative.
In the case of the ground state determined from a given lattice correlation function, this means that a negative binding energy could be observed that does not correspond to a non-zero binding energy in the infinite volume. However, without further knowledge of $c_1$ and higher-order coefficients, this so-called level-repulsion would disappear with a power law dependence as the volume of the calculation is increased.\\
\textit{Scenario 2: Stable state.} In the bound state scenario, it was shown \cite{Luscher:1985dn,Luscher:1986pf,Luscher:1985dn} that the volume dependence is governed by the binding momentum $\kappa$ of the state. In the case of the doubly heavy tetraquarks with the two mesons energies $m_1$ and $m_2$ that set the threshold, this is defined by
$E= 2\sqrt{ m_1\cdot m_2 - \kappa^2}$.
Furthermore, the volume dependence of the corresponding finite volume energy level was shown to go as
\begin{equation}
 E_{B,L}\sim E_{B,\infty}  \cdot \Big[ 1 + A e^{-\kappa L}  \Big]  ~~.
 \label{eq:fvol-exp-formula}
\end{equation}
This means observing a ground state energy with an exponential instead of a power law volume dependence indicates a stable bound state.
As we will see below, the threshold repulsion effect cannot be very large, which implies that if a ground state far below the threshold is observed in a finite volume, there is a good chance that it is, in fact, a bound state. This argument was used a few times in Sec.~\ref{four} when multiple volumes were not available. In \cite{Hudspith:2023loy}, it was also used to extrapolate the binding energy to the infinite volume limit.

\paragraph{Scattering analysis.} The recipe outlined above aims at interpreting just the ground state finite volume energy level. In a lattice calculation, this is the level with the highest chance of being resolved with high accuracy and control. However, for a complete picture, this is not sufficient. 
For this, one needs to perform broad-range parameter scans in volumes $L$, external momenta $\vec p$ and the number of states resolved as well as track the dependence of the binding energy to finally determine the scattering parameters robustly. At lowest order these are the scattering length $a_0$[fm] and the effective range $r_0$[fm]. Whereby this assumes that the wave function of the state can be well described by its $\ell=0$ partial wave component.

The key development of \cite{Luscher:1990ux} is to relate the infinite volume scattering phase shift of a given scattering process to finite volume quantization conditions:
\begin{equation}
\operatorname{det}[M(\vec p,L)-\cot \delta]=0~~,
\end{equation}
where $\vec p$ is the spatial momentum of the scattering particles in the center-of-mass frame satisfying $E_{cm}=\sqrt{E^2-\vec p ^2}=2\sqrt{m_1\cdot m_2 +\vec p^2}$. The matrix $M$ depends on the details of the scattering process and is a combination of generalized Riemann $\zeta-$functions\footnote{To determine the $\zeta-$function often the numerical implementation of \cite{Gockeler:2012yj} is used.}. In 2-to-2 scattering, the relation to the scattering phase shift becomes:
\begin{equation}
p \cot \delta(p)=\frac{2}{\sqrt{\pi} L \gamma} Z_{00}^{\mathrm{D}}\left(1,\left(\frac{p L}{2 \pi}\right)^2\right)~~,
\end{equation}
where $p=|\vec p|$ and $\gamma=E/E_{cm}$ is the moving frame boost factor \cite{Rummukainen:1995vs}. With this relation a finite volume energy determined on the lattice can be converted into a scattering phase shift. All states in the finite volume spectrum enter $p \cot \delta(p)$ and lie on branches determined by the $\zeta$-function. These branches are separated by singularities and become denser as the volume increases; see the next section.
A bound state in this formalism corresponds to a pole in the scattering amplitude on the real $p^2$-axis below zero. Its location can be extracted from the point where the so-called pole condition is met: 
\begin{equation}
p \cot \delta(p)=-\sqrt{-p^2}~~.
\end{equation}
Going further, the phase shift can be described in an effective range expansion 
\begin{equation}
p \cot \delta=-\frac{1}{a_0}+\frac{1}{2}  r_0 p^2+\cdots
\end{equation}
with the scattering length $a_0$ and effective range $r_0$ as parameters. Determining a number of scattering phase shifts on the lattice, i.e. converting multiple finite volume lattice energy levels, the scattering parameters can be determined by a fit. In this form, the fit would be appropriate for describing $S$-wave scattering, and we see it being used a few times in Sec.~\ref{four}. %However, there are also extensions of this to higher-order partial waves. 
Note that an effective range expansion of this type is not always appropriate. For example, it is primarily a valid approach at small values $p^2$. An alternative approach becomes necessary when non-analyticities appear, e.g., a left-hand cut see Sec.~\ref{sec:lefthandcut}.
Finally, when coupled channels must be considered, one may need to parametrize the scattering matrix to determine the scattering phase shifts, see e.g., \cite{Briceno:2017qmb}.

\subsubsection{A toy model illustration of the procedure}

Sidelining the concrete lattice implementation for the moment, in the following, we go through the procedure for a simple model spectral function to illustrate the individual steps and what problems have to be tackled. To recap, assuming we have somehow determined a number of finite volume energies, the first step is to identify whether they belong to a free, bound, or resonance state spectrum. The second is to convert this information into scattering parameters. 
The model we consider is a simple Breit-Wigner type spectral function that is characterized by a mass and a width:
\begin{equation}
\rho(\omega) \sim \frac{\Gamma/2}{\omega - m - i\Gamma/2}~~,
\end{equation}
where $m$ is the energy of the resonance or bound state and $\Gamma$ is its width. For this toy model we assume $\Gamma\ll m$.
Once more, such a spectral function can be related to experimental cross sections, $\rho(\omega)\sim\sigma(\omega)$ and is broadly applicable.
For example, apart from the doubly heavy tetraquark scenario it could also be a model for the vector-vector spectral function where it is directly related to the tree-level cross-section $\sigma_{e^+e^-\rightarrow \mu^+\mu^-}(\omega)$. 

Next, the cross-section can be connected with the scattering amplitude $f_\ell$ in the partial wave expansion. Assuming only elastic scattering \cite{ParticleDataGroup:2022pth,Friman:2000} the amplitude may be written as:
\begin{equation}
f_\ell = \frac{1}{2ip}\Big( \eta e^{2i\delta} -1\Big) ~\longrightarrow~ \frac{1}{p} e^{2i\delta} \sin\delta = \frac{1}{p}\frac{1}{\cot \delta -i}
\end{equation}
The phase shift $\delta$ is close to $\pi/2$, and therefore $\cot\delta\simeq 0$, once $\omega\approx m$. In this case, we may expand and identify:
\begin{align}
\cot\delta(\omega) &=  \cot\delta(m) + (\omega-m) \frac{d\,\cot\delta(\omega)}{d\omega}\Big{|}_{\omega=m} + ...\\ \nonumber
&= 0 + (\omega -m)\frac{2}{\Gamma} + ...\\
\Rightarrow ~&\delta(\omega) \simeq \rm{arctan}\Big(\frac{\Gamma/2}{\omega-m}\Big)
\end{align}
The scattering amplitude one finds by using this result reads:
\begin{equation}
f_\ell = \frac{1}{p}\frac{\Gamma/2}{\omega - m - i\Gamma/2} 
\end{equation}
which in turn is directly proportional to a cross-section $f_\ell\sim \sigma_\ell$.
At the same time, for $S$-wave scattering, one may apply an effective range expansion to obtain an alternative denominator description.
%\begin{equation}
%p \cot \delta=-\frac{1}{a_0}+\frac{1}{2}  r_0 p^2+\cdots
%\end{equation}
%where the scattering parameters are the effective range $r_0$ and scattering length $a_0$, as usual. 
The two expressions for $\cot\delta$ show the relation between the Breit-Wigner spectral function and the scattering parameters.
The upshot is that by setting an effective range and scattering length we are choosing different scenarios for the Breit-Wigner model, modifying its width and mass peak location.
Using this model in the next step, we will work backward by choosing first a set of parameters for $a_0,~r_0$ that correspond to the three scenarios of a free, bound and resonant state. In particular we set (a) $a_0=0,~r_0=0$ for the free; (b) $a_0=-10,~ r_0=0.2$ for the bound and (c) $a_0=+2.5, r_0=-9.0$ for the resonant case.
In principle, these parameters have units. However, to keep the discussion general, we drop them here and all shown results should be seen as relative to each other. 
The parameters are chosen purely for illustrative purposes based on considerations that in the resonant case the scattering length should be positive while the effective range should be negative, in the bound case they should be negative and positive as well as small, respectively, while in the free case they should both be zero, see also the discussion on the Weinberg criterion in Sec.~\ref{sec:weinberg}. Their values have been chosen such that their respective spectra are visibly distinguishable. 
Assuming once more exclusively two-particle scattering and using the $\zeta$-function for a number of volumes, we can determine the corresponding values of the finite volume energies by reading off the intersections of the $\zeta$-function branches with the scattering phase shifts of the three scenarios.
Remember that this procedure will have to be done in reverse in the actual lattice study, and we present the results of this model in that order. 

\begin{figure}
\centering
\includegraphics[width=0.25\columnwidth]{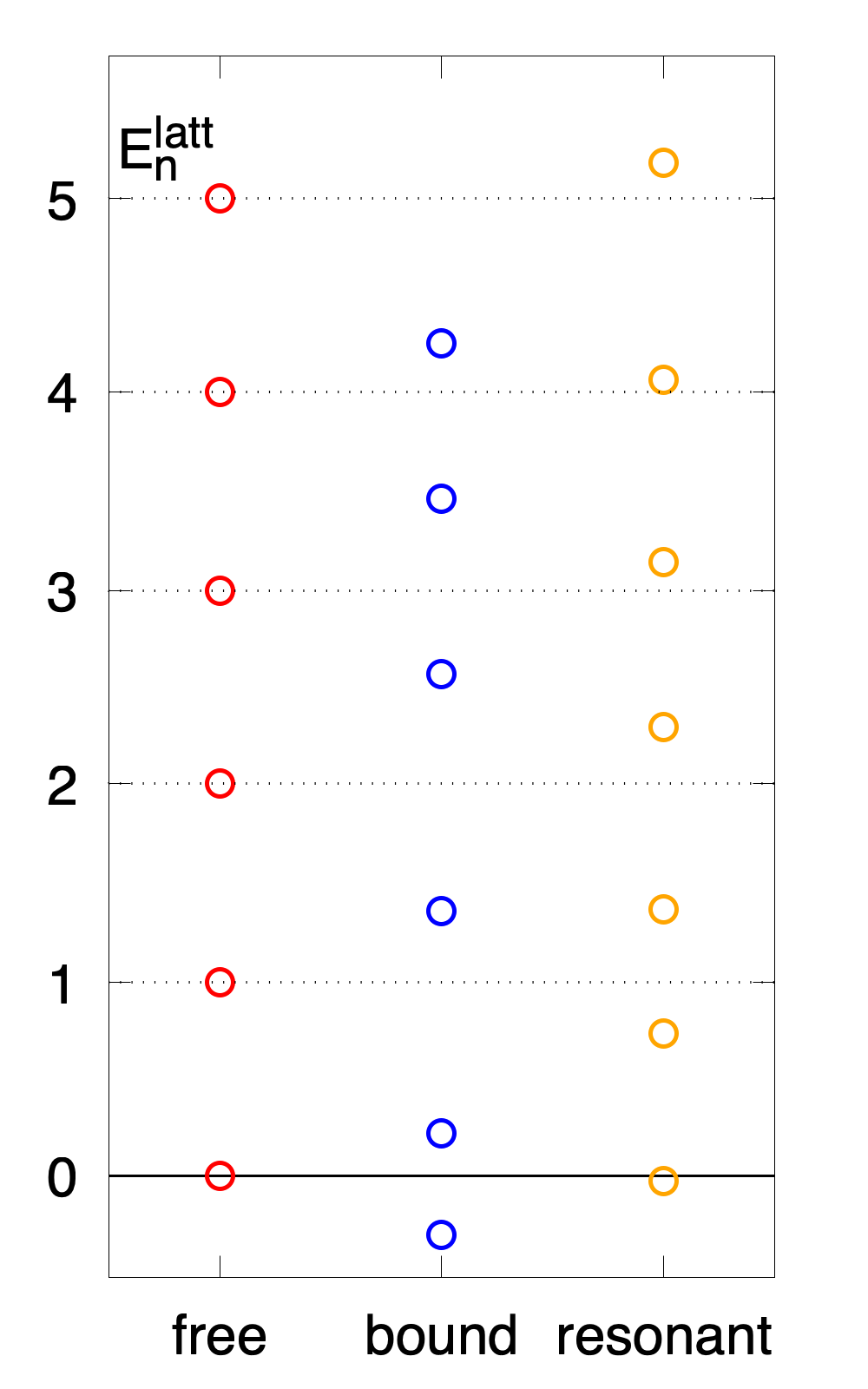}
\includegraphics[width=0.22\columnwidth]{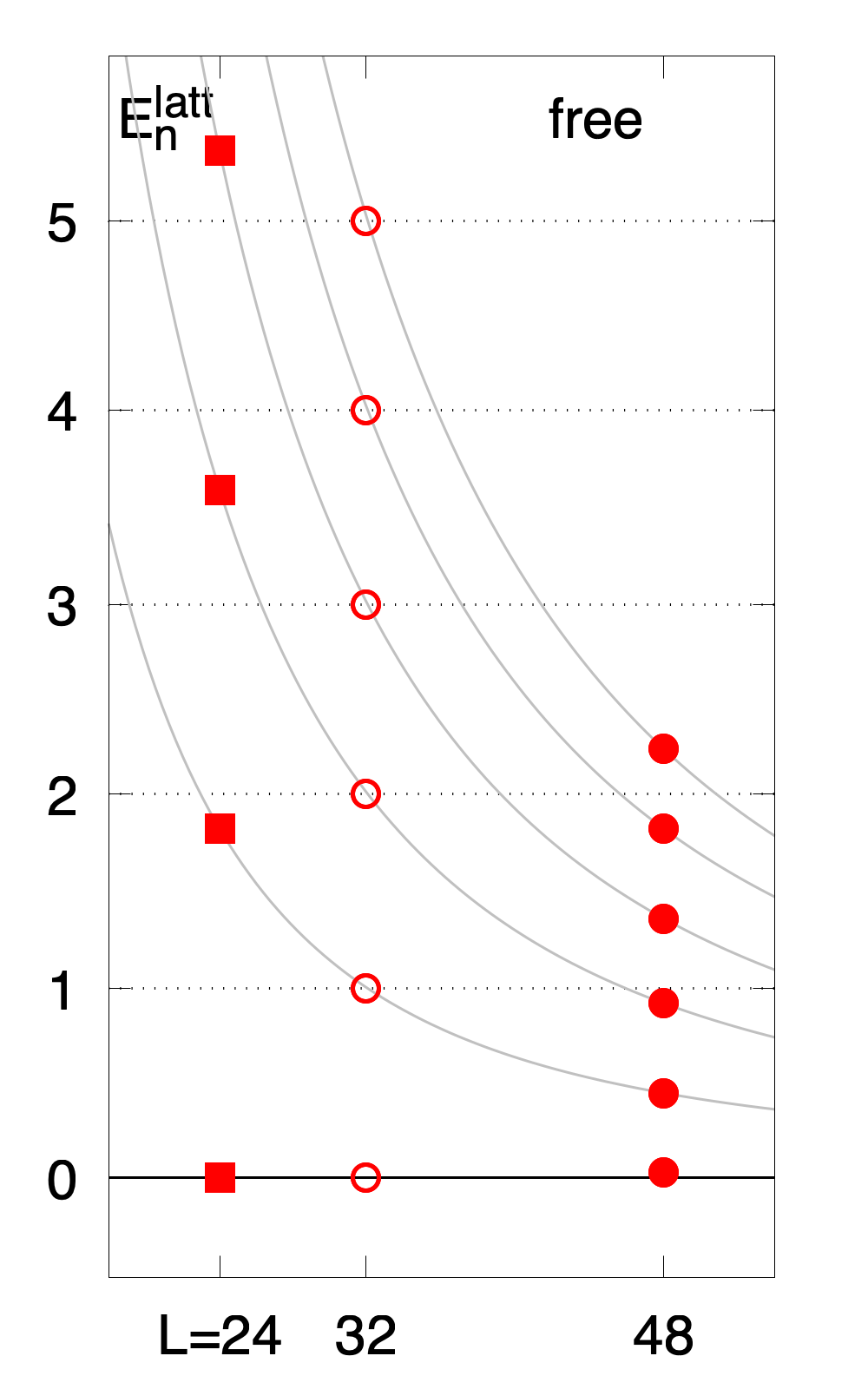}
\hspace{-3ex}
\includegraphics[width=0.22\columnwidth]{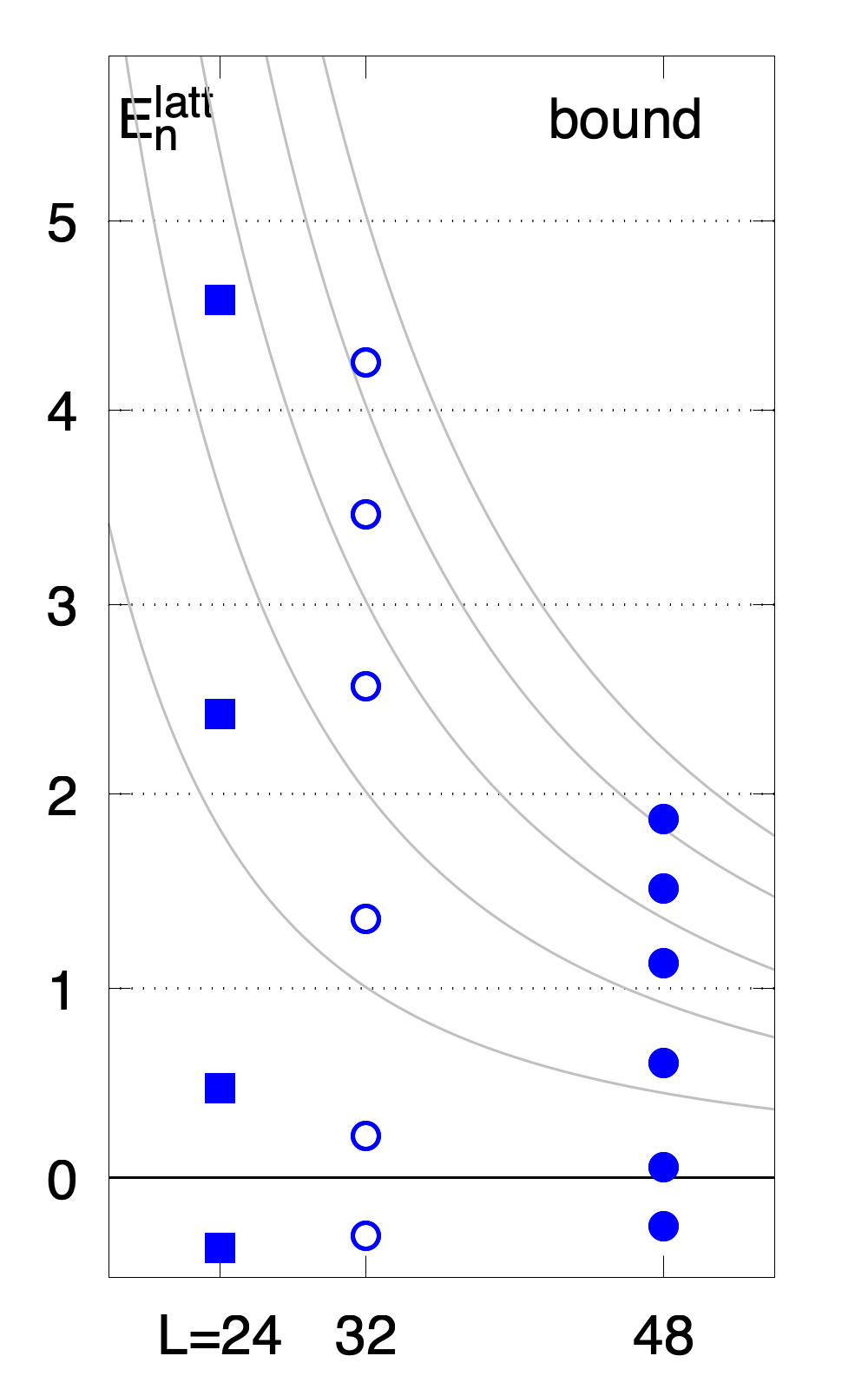}
\hspace{-3ex}
\includegraphics[width=0.22\columnwidth]{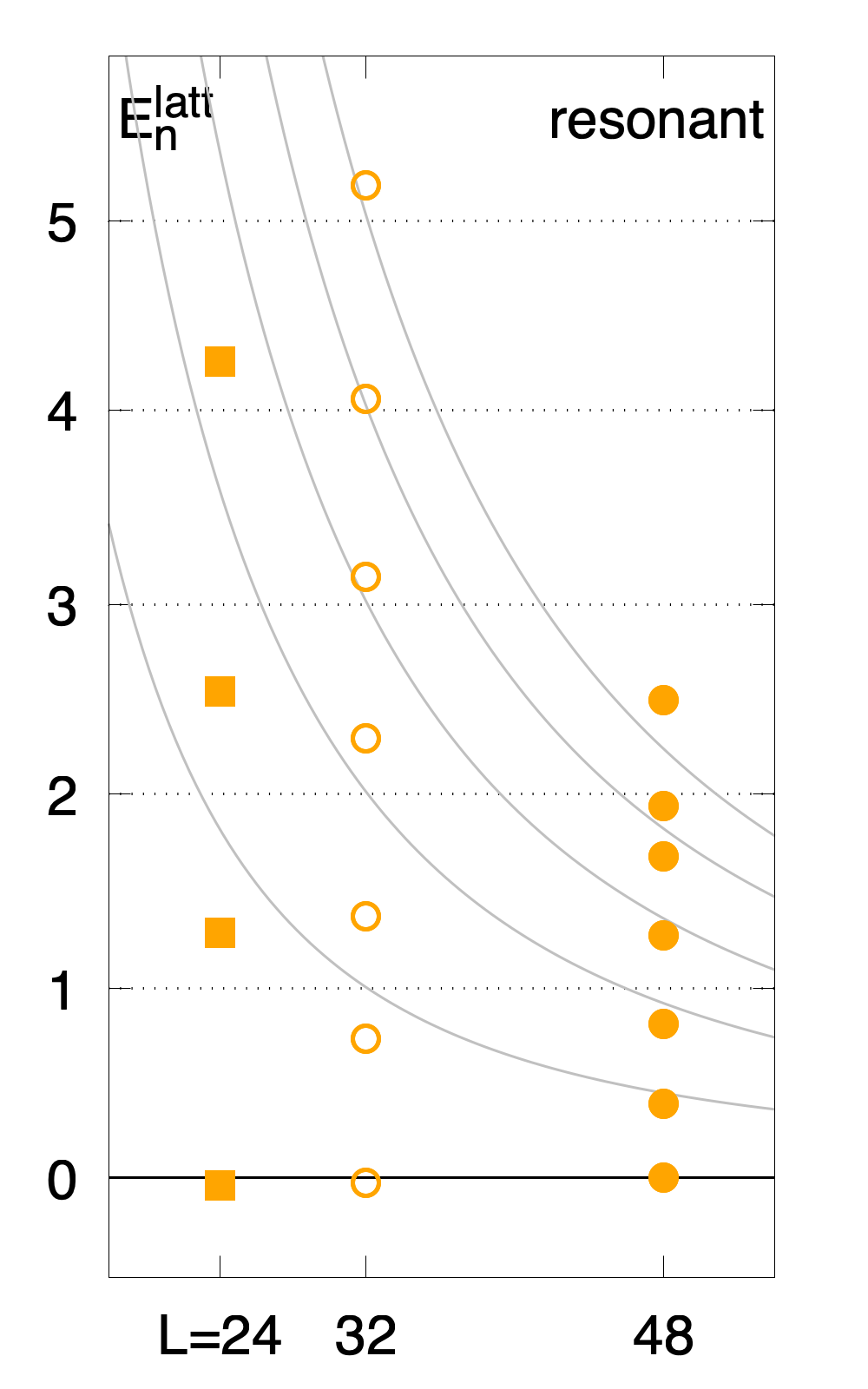}
\caption{\textit{Left: Finite volume energies at the same reference volume for the free, bound and resonant scenarios in the toy model. The dotted lines denote the free energy levels in this setup. Three figures from center to right: In sequence, the free, bound and resonant state scenarios for different reference volumes. The volume dependence in the free case is given as the grey lines in all three figures. For more details, see the text and compare with Fig.~\ref{fig:sasa-1} for an example in a real lattice calculation.
}}
\label{fig:toy-1}
 \end{figure}

\paragraph{Finite volume energies.}

The results for the first six, respective seven in the resonant case, finite volume energies are shown in Fig.~\ref{fig:toy-1} in the leftmost panel for the free, bound and resonant scenarios for a reference lattice volume ($L=32$) that has been chosen to make the spectrum relatively sparse. The $y$-axis is labeled with respect to the $n$-th level in the free case, which is also given using a dotted line to make the comparison with the other two cases easier. 
Recall that in the actual lattice study, the energies would be the outcomes of the numerical calculation and analysis; unlike the situation here, they would have uncertainties and error bars associated with them.  A good example of this can be seen in Fig.~\ref{fig:historic-tantalo} where the uncertainties progressively increase from the ground to the higher finite volume energies.

When deciding which of the three scenarios the state in question is, the free case seems the most straightforward. Here, the observed spectrum of the more complicated hadron, e.g. the tetraquark, is the same as that of the underlying decomposition of hadrons, two mesons that do not interact. In particular, the ground state energy should be the same as the threshold.
Typically, this threshold is computed on the lattice by separately determining the ground state energies of the individual hadrons involved and adding their masses. The result can then be shown as a reference line or subtracted from the combined state energy. Again, this comes with statistical and systematic uncertainties that must be controlled. 
Next, identifying a bound state entails observing a clearly separated downward-shifted state below the corresponding threshold. We see that the entire finite volume spectrum is shifted downward. In the model, we have chosen a set of parameters where the excited spectrum looks clearly dissimilar from the free case. However, it should be noted that this is not generally the case. Especially when there is a shallow bound state, with uncertainties included, it can be difficult to establish the key distinguishing features of this scenario.
Finally, in the case of a resonance an additional finite volume energy level appears, compared to the free scenario. All other states have a slight downward shift, but with uncertainties included, the extra energy level is the surest sign. Note that because this extra level appears as an excited state energy, determining it in a lattice calculation while having enough control for precision results can be challenging.
Both this and the previous paragraphs indicate why a prediction of the $T_{cc}^{ud}$ was challenging before the observation in experiment. As will be detailed further below, studies of this candidate observed a spectrum that, within errors, overlapped with the free scenario and did not observe an extra energy level. As such, it seemed unlikely to be a bound state, although it was not ruled out entirely, while a resonance could not yet be entirely ruled out either due to not having enough energy levels resolved with certainty. Note that in nature, the state is bound by less than $<1$~MeV, which implies the corresponding lattice ground state energy would have to be determined to better than that precision. This also has to include the possibility of at least a one-sigma shift to account for fluctuations from the mean in the finite sample of the lattice calculation. 

\begin{figure}
\centering
\includegraphics[width=0.39\columnwidth]{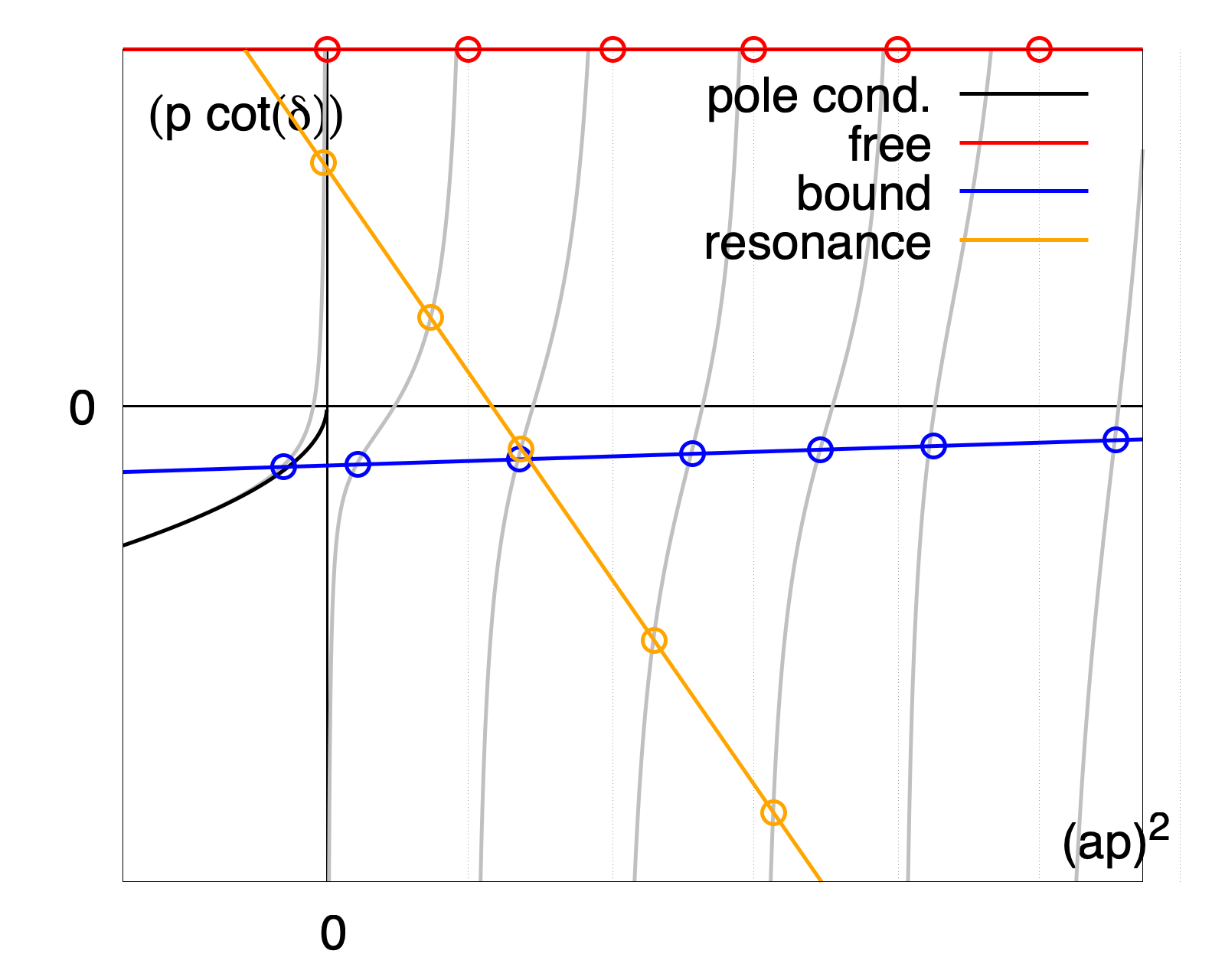}
\includegraphics[width=0.39\columnwidth]{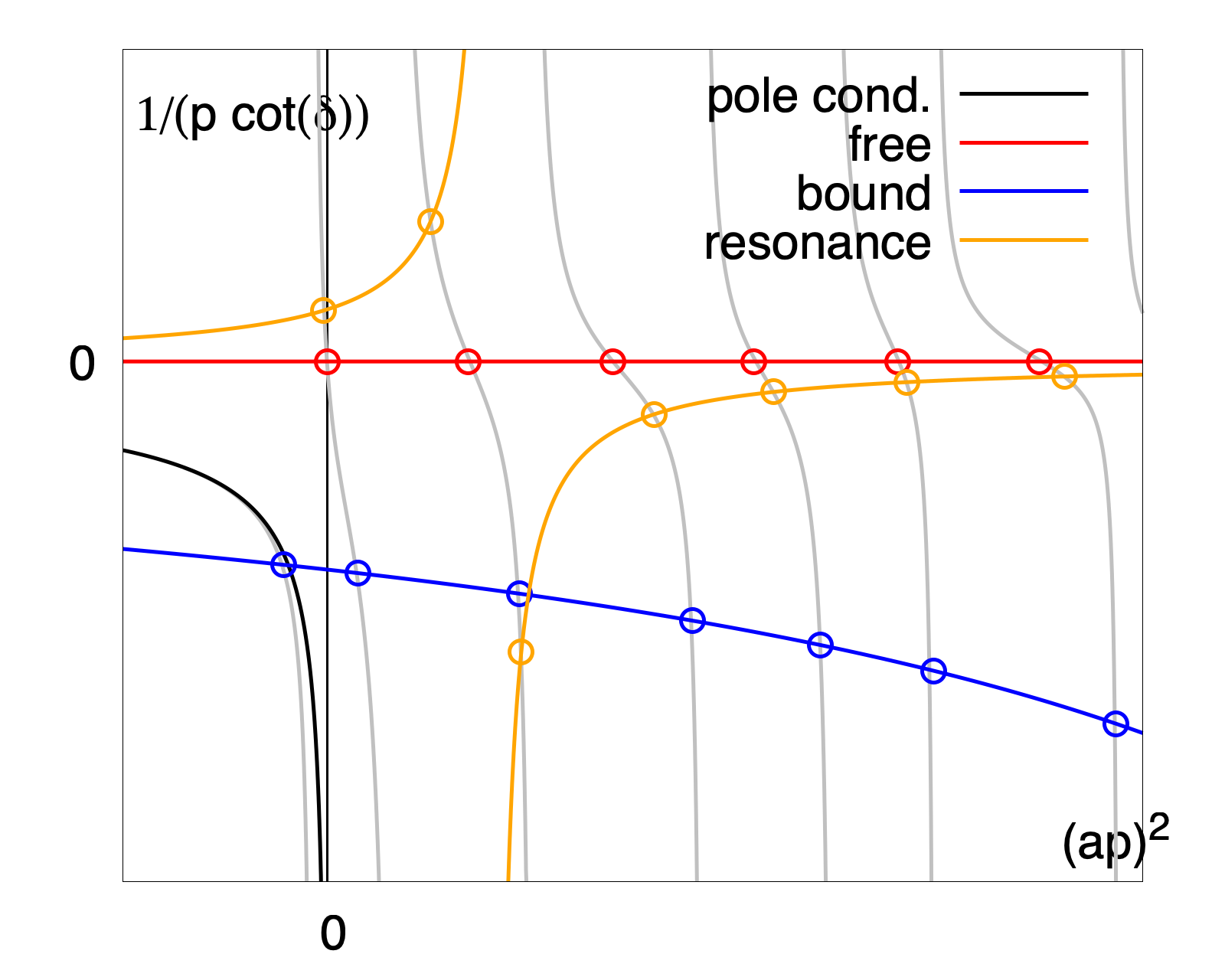}
\caption{\textit{Left: Scattering phase shifts $p\cot{\delta}$ converted from the finite volume energies in Fig.~\ref{fig:toy-1}. The colored lines represent would-be fits to these results in an effective range expansion to determine $a_0$ and $r_0$. Right: The same for $1/(p\cot{\delta})$, we see how the extra level for the resonance case sits on the same branch. 
}}
\label{fig:toy-2}
 \end{figure}

\paragraph{Volume dependence of finite volume energies.}

Reading off the first six or more finite volume energies in multiple volumes is straightforward in the toy model. This is shown in the three slightly smaller figures on the right of Fig.~\ref{fig:toy-1} for reference volumes $\tilde{L}=24,~32$ and 48, which are typical lattice geometries. In the first left central panel, the free scenario is shown for the first few states, with the finite volume dependence of a given state being highlighted by the grey lines. Since this dependence is known from the quantization condition, it is often given as a reference, as in the other two panels for the resonant and bound state cases here. Alternatively, a fitted volume dependence from a scattering matrix parametrization may be given.
The figures clearly show how the spectrum becomes more dense with increasing volume. 
Looking ahead, note that in the lattice calculation, the finite volume energies are typically determined via the GEVP method, see Sec.~\ref{sec:gevp}, and for fixed-size interpolating operator bases, the denser spectrum can represent a difficulty, especially when considering uncertainties. With multiple states close together, resolving them individually when the uncertainties are not small enough can lead to results that are uncontrolled and wrongly identify energies.
As an example of this type of plot in a full calculation consider Fig.~\ref{fig:sasa-1} which presents the finite volume spectrum determined in the study \cite{Padmanath:2022cvl} on the $I(J^P)=0(1^+)$ $T_{cc}^{ud}$ tetraquark in this way.

\paragraph{Scattering phase shifts and parameters.}

Finally, assuming the finite volume quantization conditions are known for the scattering process being studied in the lattice calculation, the finite volume lattice energies can be converted to scattering phase shifts.
While we focus on the L\"uscher-type formalism to do this here, alternatives exist, see Sec.~\ref{sec:lefthandcut}. 
The results from this conversion to $p\cot{\delta}$ in the toy model are  shown in Fig.~\ref{fig:toy-2} in the left and the inverse $1/(p\cot{\delta})$ in the right panel. The circles denote the results from converting the finite volume energies in Fig.~\ref{fig:toy-1}(left) in the same color coding. They lie on the branches of the $\zeta$-function, represented by grey lines.
The colored lines correspond to would-be fits that one would do to determine the scattering parameters $a_0$ and $r_0$. Finally, the black lines denote the pole condition $-\sqrt{-p^2}$, the intercept of the fitted effective range expansion would determine the binding energy of the state.
The inverse on the right of the figure representation has the benefit that the free scenario can be plotted more easily since, in the left-hand case, their values would sit at $\infty$. 
We see how in the $p\cot{\delta}$ presentation the resonance shows a steep linear dependence. In the $1/(p\cot{\delta})$ presentation, it becomes very visible how the extra level sits on the same branch of the $\zeta$-function.
In the actual calculation this analysis step would include deciding the model to fit the determined $p\cot{\delta}$. In the case of the model corresponding to a Breit-Wigner type peak like the one used here, the scattering phase shifts can be fitted to a straight line with the scattering parameters extracted determining the intercept and slope, as they represent the first two terms in the effective range expansion.
Depending on the studied system, this can be much more complicated, especially when more than the $\ell=0$ partial wave is being considered or a coupled channel analysis performed.
%Indeed, one often has to parametrize the scattering matrix in a way that takes the finite volume energies as input and produces the final results.
%This is, for example, necessary when one can expect a non-analyticity in the effective range expansion. This is the case for the $T_{cc}$ because a nearby left-hand cut needs to be considered, and progress in this direction is briefly discussed in Sec.~\ref{sec:lefthandcut}.
%
In the actual calculation, the uncertainties of the energies involved add a complication to the simple scenario presented here. Typically, the ground state energy is the most precisely resolved one, with all others having successively larger errors. Converting these into scattering phase shifts then spreads them along the branches determined by the finite volume quantization condition. Note that these branches become more steep and dense for higher states. As such, the errors are further increased through the conversion. 
Even a simple linear fit, as required here, can become difficult or effectively unconstrained. For an example see Fig.~\ref{fig:leskovec-2019} (bottom). Here the lattice determined $p\cot \delta$ for $I(J^P)=0(1^+)$ $T_{bb}^{ud}$ from the study \cite{Leskovec:2019ioa} are shown and we can see how the uncertainties in the first excited state are amplified when comparing to the top panel in the same figure. As a result, the fit cannot determine the effective range parameter $r_0$ with sufficient precision to make robust statements.
In a sense, this harks back to the underlying connection to the inverse problem: The effective range connects with the width of the underlying Breit-Wigner-type spectral function, and widths are very difficult to extract robustly.

% ----------------------------------------------------
\subsection{Techniques in lattice spectroscopy}

\paragraph{Sources and propagators.}

Calculating the lattice correlation function means to evaluate numerically 
\begin{equation}
G_{\mathcal{O}_1 \mathcal{O}_2}(t, \vec p)=\sum_{\vec x} e^{i \vec p \cdot \vec x}\left\langle\mathcal{O}_1(t,\vec x) \mathcal{O}_2^{\dagger}(0,\vec 0)\right\rangle\sim \sum_n Z_n e^{-E_n(|\vec p|) t},
\end{equation}
where $\mathcal{O}_i$ are interpolating operators that project onto the quantum numbers of the target hadronic system.
For example, in the simple case of a meson $M$ with $\vec p=0$ the correlator can be computed via:
\begin{align}
G_{PP}(t)=\sum_{\vec x} \left\langle M(t,\vec x) M^\dagger(0,\vec 0)\right\rangle&= \sum_{\vec x} \left\langle
\Big[ \bar{d}_a^\alpha(t,\vec x) \Gamma^{\alpha \beta} u_b^\beta(t,\vec x) \Big]\Big[\bar{d}_c^\gamma(0,\vec 0) \Gamma^{\gamma \delta} u_d^\delta(0,\vec 0)\Big]^\dagger\right\rangle~~\nonumber\\
& =\sum_{\vec x}\Gamma^{\alpha \beta} (\Gamma^{\gamma \delta})^\dagger\left\langle u(t,\vec x)_{b}^\beta \bar{u}(0,\vec 0)_{d}^\delta\right\rangle\left\langle d(0,\vec 0)_{c}^\gamma \bar{d}(t,\vec x)_{a}^\alpha\right\rangle~~\nonumber\\
& =\sum_{\vec x}\textrm{Tr} \,\tilde{\Gamma}\left[D_u^{-1}\Big((0,\vec 0)|(t,\vec x)\Big) D_d^{-1}\Big((t,\vec x)|(0,\vec 0)\Big)\right] ~~,
\end{align}
where Greek indices denote Dirac components and Roman indices colors. The choice of meson was kept open and by fixing $\Gamma=\gamma_5,\gamma_i,...$ the above equation can be used to determine the spectrum of particles with $\pi$, $\rho$, ... quantum numbers. 
We performed the necessary Wick contractions and introduced the quark propagator:
\begin{equation}
D_l^{-1}(x|y)=\langle \,l(0,\vec 0\,)_{a}^\alpha \,\,\bar{l}(t,\vec x\,)_{b}^\beta\,\rangle~~~(=:S_{l})~~,
\end{equation}
where $D$ is the lattice Dirac operator. The propagator is $\gamma_5$ Hermitian enabling the determination of the above correlator from just one propagator when working in the isospin symmetric limit $u=d$.
Omitting details on how to compute it, considering the equation for the quark propagator, note that it is given by the inverse of the lattice Dirac operator. Numerically, the inversion requires setting a source, which, once the inversion is complete, gives as output the propagator connecting the input source point(s) with all other points in the lattice.

The simplest source choice is a point source. As its name suggests, it corresponds to setting the source vector at a single point on the lattice to one. Picking a single location for the source localizes it in configuration space but leaves its momentum properties open.
This source type is versatile as it often overlaps with much of the spectrum strongly. As a downside, this also means that to reach the asymptotic value and, therefore, the ground state energy, can require a long distance in $t$.
In this light we might want to choose different sources that have advantageous properties in this regard. One example is the Coulomb gauge fixed wall source. Here, one first fixes the gauge configuration, or at least the source timeslice, to Coulomb gauge\footnote{Algorithms to do so are commonly available, examples are \cite{Boyle:2015tjk,GPT,GLUcode}. }.
Once that is done, one sets every entry in the source timeslice to one (for $\vec p=0$). Performing an inversion with this source leads to a maximally nonlocal source with fixed momentum properties. \\
Another option is to fill the source timeslice with random noise drawn from some distribution. For mesons noise is usually drawn as random entries of $U(1)$ or $Z(2)$ elements, while for baryons $Z(3)$ is used. In the case of doubly heavy tetraquarks the authors of \cite{Alexandrou:2023cqg} used $Z(2)\times Z(2)$.
Introducing these so-called noise sources underscores a particular problem of calculating lattice correlators: When one needs to perform momentum projections on the source side, one needs to sum over the entire source spatial three-volume. However, in the point source, there are no contributions, and in the gauge fixed wall source, they are already projected.
This is particularly important for tetraquarks because many interpolating operators are combinations of two individual mesons, as we will discuss further below. 
%Unfortunately, using noise sources often leads to unfavorable signal-to-%noise ratios, and high statistics are required.

An alternative solution to the all-to-all, or in most cases, the timeslice-to-all problem is the distillation framework \cite{HadronSpectrum:2009krc}.
In distillation, the standard quark fields are replaced with Laplacian Heaviside (LapH)-smeared quark fields. This smearing procedure is based on calculating the $n$ lowest eigenmodes of the spatially gauge-covariant Laplacian defined on each timeslice. The spatially gauge-covariant Laplacian is constructed using spatially stout smeared gauge links \cite{Morningstar:2003gk}.
This defines the distillation operator:
\begin{equation}
\square_{i j}(t)=\sum_{n} \nu_{i,\vec x}^{(n,t)} \nu_{j,\vec y}^{(n,t) \dagger}~~,
\end{equation}
which in the limit of having determined all eigenmodes is the unit matrix. The smeared quark fields are obtained by projecting the standard quark fields into the space of the Laplacian eigenmodes:
\begin{equation}
\tilde{q}_i (t, \vec{x}) = \sum_{n,j,\vec{y}}  \nu_{i,\vec x}^{(n, t)} \nu_{j,\vec y}^{(n, t) \dagger } q_j(t, \vec{y})
\end{equation}
The smeared quark fields living in this so-called distillation space can also be inverted to give distillation space propagators. These are usually called perambulators and are given by:
\begin{equation}
\mathcal{S}_{\alpha \beta}^{n m}\left(t, t_0\right) \equiv \sum_{i, j, \vec{y}, \vec{x}} \nu_{i,\vec y}^{\left(n, t\right) \dagger} \, D_{\alpha i, \beta j}^{-1}\left(t, \vec{y} \,| t_0, \vec{x}\right) \,\nu_{j,\vec x}^{\left(m, t_0\right)} .
\end{equation}
The required Wick contractions to form doubly heavy tetraquarks are then performed using the perambulators instead of the standard propagators.
Because LapH smearing projects onto a much smaller subspace, it becomes possible to calculate the full timeslice-to-all propagator without having to resort to noise source methods. This significantly improves the signal-to-noise properties, and the much smaller size of the object to be stored also increases its re-usability. The downside is that algorithms aimed at evaluating eigenmodes scale poorly with the volume, limiting the number of eigenmodes that can be obtained in practice. At the same time keeping the smearing width constant requires scaling the number of eigenmodes determined with the spatial volume. To circumvent this issue, a stochastic variant of distillation has been put forward \cite{Morningstar:2011ka}.
Furthermore, to include an unsmeared or only little smeared quark requires the determination of many eigenmodes, or the combination with other methods. As such distillation in principle introduces a bias towards nonlocal operator structures and interpreting overlaps in terms of trial states can be difficult.

\subsubsection{Operator construction}

An essential first step in the lattice calculation of doubly heavy tetraquarks is to choose the interpolating operators for the correlation functions.
To start, consider the desired quantum numbers and flavor content of the target state, for example, $I(J^P)=0(1^+)$ and $c,~c,~\bar{u},~\bar{d}$ and for the $T_{cc}$, observed in experiment, or $b,~b,~\bar{u},~\bar{d}$ for the $T_{bb}$, predicted to have a deep binding energy.
From the point of view of scattering, the simplest process that could create such a state would be $DD^*$ or $BB^*$ scattering, respectively, and the most directly related interpolating operator would be a di-meson type made up of a pseudoscalar and a vector heavy-light meson:
\begin{equation}
P(x)=\bar{Q}_a^\alpha(x) \gamma_5^{\alpha \beta} u_a^\beta(x)~,~~ V(x)=\bar{Q}_a^\alpha(x) \gamma_i^{\alpha \beta} d_a^\beta(x),
\end{equation}
where, again, Greek indices denote Dirac components, Roman indices colors, and $Q$ denotes the heavy quark involved, here either charm or bottom $Q=c,~b$.
The di-meson operator then reads:
\begin{align}
O_{PV}(x)=  P(x) V(x) 
= \Big[ \bar{b}_a^\alpha(x) \gamma_5^{\alpha \beta} u_a^\beta(x) \Big]\Big[\bar{b}_b^\kappa(x) \gamma_i^{\kappa \rho} d_b^\rho(x) \Big] -\Big[ \bar{b}_a^\alpha(x) \gamma_5^{\alpha \beta} d_a^\beta(x) \Big]\Big[\bar{b}_b^\kappa(x) \gamma_i^{\kappa \rho} u_b^\rho(x) \Big]~,
\end{align}
where the second term originates from exchanging $u\leftrightarrow d$, implying isospin symmetric light quarks, which is the case in all lattice calculations of doubly heavy tetraquarks. Following the previous discussion, the correlation function then is
\begin{equation}
G_{\mathcal{O}_1 \mathcal{O}_2}(t, \vec p)=\sum_{\vec x} e^{i\vec p \vec x} \left\langle\mathcal{O}_1(t,\vec x) \mathcal{O}_2^{\dagger}(0,\vec 0)\right\rangle =\sum_n\left\langle 0\left|\mathcal{O}_1\right| n\right\rangle\left\langle n\left|\mathcal{O}_2\right| 0\right\rangle e^{-E_n(|\vec p|)\, t}~~.
\end{equation}
This is the most common operator structure used, especially at the source and in earlier studies. One reason is that it can be evaluated using a standard point-to-all propagator determined on the lattice. 
One extra channel with these quantum numbers is available from considering the discussion below, and that is the $O_{VV}$ channel. Further, when considering $J^P=0^+$, one chooses the $O_{PP}$ channel.

\begin{table}
\centering
\begin{tabular}{cclll}
\hline\hline $\vec{P}$ &  $\Lambda^P$ & $J^P$ & $\ell$ & interpolating operators: $P\left(\vec{p}_1^2\right),~V\left(\vec{p}_2^2\right)$ \\
\hline $(0,0,0)$ &  $T_1^{+}$ & $1^{+}$ & 0,2 & $P(0) V(0),~ P(1) V(1),~ V(0)V(0)$ \\
 $(0,0,0)$ &  $A_1^{-}$ & $0^{-}$ & 1 & $P(1)V(1)$ \\
 $(0,0,1) \frac{2 \pi}{L}$ & $A_2$ & $0^{-}, 1^{+}, 2^{-}$ & $0,1,2$ & $P(0) V(1),~ P(1)V(0)$ \\
 $(1,1,0) \frac{2 \pi}{L}$ &  $A_2$ & $0^{-}, 1^{+}, 2^{-}, 2^{+}$ & $0,1,2$ & $P(0)V(2),~ P(1)V(1),~ P(2)V(0)$ \\
 $(0,0,2) \frac{2 \pi}{L}$ &  $A_2$ & $0^{-}, 1^{+}, 2^{-}$ & $0,1,2$ & $P(1) V(1)$ \\
\hline \hline
\end{tabular}
\caption{\textit{Total momenta $\vec{P}$, irreducible representations $\left(\Lambda^P\right)$ and di-meson type interpolating operators for the doubly heavy tetraquarks, modified from \cite{Padmanath:2022cvl}. 
Total $J^P$ and partial wave $\ell\leq 2$ are also listed. See also \cite{Cheung:2017tnt} for a discussion.}}
\label{tab:sasa-ops1}
\end{table}

Once techniques to handle all-to-all propagators become available, the discussion on the di-meson operator can also be expanded to consider the mesons to be individually projected onto a definite momentum $\vec p$:
\begin{equation}
P(t,\vec p_1)=\sum_{\vec x} e^{i\vec p_1 \vec x} P(x)~,~~ V(t,\vec p_2)=\sum_{\vec x} e^{i\vec p_2 \vec x} V(x)~.
\end{equation}
This is important because, while the tetraquark operator with this di-meson structure has definite continuum spin $J$, this is no longer true on the lattice because it breaks rotational symmetry. To identify continuum spin, the lattice operator needs to connect to the correct irreducible representation $\lambda$ of the octahedral group, which makes up the lattice symmetries.
Following \cite{Padmanath:2022cvl} and \cite{Cheung:2017tnt} at zero total spatial momentum, the relevant irreps are $J=0\rightarrow A_1$, $J=1\rightarrow T_1$. Considering also non-zero total spatial momenta, one arrives at the relations of Tab.~\ref{tab:sasa-ops1}.
In keeping with the pedagogical aspect of this review, concretely the operators constructed for the $T_1^+$ case with zero momentum $\vec p=(0,0,0)$, omitting the $t$ index, and considering the $z$-direction component, read \cite{Padmanath:2022cvl}:
\begin{align}
O^{l=0} & =P(0,0,0) V_z(0,0,0) \\
O^{l=0} & =P(1,0,0) V_z(-1,0,0)+P(-1,0,0) V_z(1,0,0) \nonumber\\
& +P(0,1,0) V_z(0,-1,0)+P(0,-1,0) V_z(0,1,0) \nonumber\\
& \left.+P(0,0,1) V_z(0,0,-1)+P(0,0,-1) V_z(0,0,1)\right] \\
O^{l=2} & =P(1,0,0) V_z(-1,0,0)+P(-1,0,0) V_z(1,0,0) \nonumber\\
& +P(0,1,0) V_z(0,-1,0)+P(0,-1,0) V_z(0,1,0) \nonumber\\
& -2\left[P(0,0,1) V_z(0,0,-1)+P(0,0,-1) V_z(0,0,1)\right] \\
O^{l=0} & =V_{1 x}[0,0,0] V_{2_y}[0,0,0]-V_{1 y}[0,0,0] V_{2 x}[0,0,0]~~,
\end{align}
while for the $A_1^-$ one has:
\begin{align}
O& =P(1,0,0) V_x(-1,0,0)-P(-1,0,0) V_x(1,0,0) \nonumber\\
& +P(0,1,0) V_y(0,-1,0)-P(0,-1,0) V_y(0,1,0) \nonumber\\
& +P(0,0,1) V_z(0,0,-1)-P(0,0,-1) V_z(0,0,1)~~.
\end{align}
Once more, building these operators at the source of a correlator requires a method that can provide the individual momentum projected mesons on that side. If one were to consider point-to-all propagators or a grid of them, the momentum projection is incomplete since the required sum over the spatial volume is (heavily) truncated, as such an all-to-all or timeslice-to-all propagator is necessary. 

An alternative operator structure can be motivated from the binding mechanism for the doubly heavy tetraquarks possibly relying on the quarks forming a diquark and an antidiquark as a substructure \cite{Francis:2016hui,Eichten:2017ffp,Karliner:2017qhf}. In this picture, formally, the diquark operators for the light as well as heavy quark pairs may be written as
\begin{equation}
D_{ud}(x)=\left(u_a^\alpha(x)\right)^T\left(C \gamma_5\right)^{\alpha \beta} q_b^\beta(x)~,~~ D_{QQ}(x)= \left({Q}_a^\kappa(x)\right)^T\left(C \gamma_i\right)^{\kappa \rho}{Q}_b^\rho(x)~~.
\end{equation}
Recall that $\gamma_5$, corresponding to $J^P=0^+$, is only available when the two quarks are not the same. The good diquark attraction is only active when it is in this channel. For doubly heavy tetraquarks with degenerate heavy quarks only $\gamma_i$, $J^P=1^+$, is possible. 
Combining these two structures leads to the diquark-antidiquark interpolating operators for the doubly heavy tetraquarks:
\begin{align}
D_1(x)&=\Big[\left(u_a^\alpha(x)\right)^T  \left(C \gamma_5\right)^{\alpha \beta} q_b^\beta(x) \Big] \, \Big[\bar{Q}_a^\kappa(x)\left(C \gamma_i\right)^{\kappa \rho}\left(\bar{Q}_b^\rho(x)\right)^T \Big]~ \\
D_2(x)&= \Big[\left(u_a^\alpha(x)\right)^T  \left(C \gamma_5\right)^{\alpha \beta} q_b^\beta(x) \Big] \, \left(\Big[\bar{Q}_a^\kappa(x)\left(C \gamma_i\right)^{\kappa \rho}\left(\bar{Q}_b^\rho(x)\right)^T \Big]- \Big[\bar{Q}_b^\kappa(x)\left(C \gamma_i\right)^{\kappa \rho}\left(\bar{Q}_a^\rho(x)\right)^T \Big]\right)~~.
\end{align}
Further variations on both the di-meson and diquark-antidiquark operators are available through exchanging $\gamma_5\leftrightarrow \gamma_t\gamma_5$ and $\gamma_i\leftrightarrow \gamma_t\gamma_i$ \cite{Hudspith:2020tdf}.

The spatial structure of the source and sink operators are an important aspect to consider. When using point-to-all propagators, i.e. where only local sources are available, the quarks in the source operator are all located at the same location $x$. On the sink side, however, we may choose to split them. If this is done in a gauge invariant way, the operator attains a nonlocal component. One way to achieve this is to smear the propagators along the lines presented in \cite{Hudspith:2020tdf} and discussed in Sec.~\ref{sec:wall-smear-1}. Another is to project the individual color singlets to zero momentum individually, as explained above, but only on the sink side\footnote{Another option is to spatially dislocate the quarks and connect them explicitly with links, see \cite{Cook:2002am}.}. An example where the latter approach was used can be found in \cite{Leskovec:2019ioa,Meinel:2022lzo} and is discussed in Sec.~\ref{sec:wall-smear-1}.
Finally, note that many of the different interpolating operators in the local case are Fierz-related, implying very little if any, extra information on the state under study from such operators. This is also a warning for considering overlap factors as indicators of which operator structures are preferred, for example, diquark-antidiquark over di-meson. In practice, there is often insufficient control over the operator basis to make a robust statement about this. The problem here are the many structures one could consider, not just molecular versus compact, but also their radii and other properties. Furthermore their relative normalization can be difficult to disentangle but have large effects on the overlap factors.

\subsubsection{Ground state overlap, excited states and noise}

Once the operators are constructed, the general discussion on spectral functions and also the equation for the correlation functions $G(t,\vec p=0) = \sum_n ~A_n~e^{-m_n t}$ shows that the lattice QCD calculated correlator data has contributions from all states that carry the target quantum numbers. 
In the asymptotic limit where $t\rightarrow\infty$ the ground state dominates with all others exponentially suppressed. However, in practice the lattice is often either too short for this to be achieved, or the signal-to-noise ratio is not controlled enough to reach the long time regime without the signal being washed out.

In this regard, note that a rough estimate of the time dependence of the noise can be obtained from the Parisi-Lepage argument \cite{Parisi:1983ae,Lepage:1989hd}. Here we follow a recent discussion in \cite{Bruno:2023vhs} (Appendix. A): The essential aspect of the Parisi-Lepage argument is that while the correlation function contains $n_q$ quark propagators its variance $\sigma(t)^2 = \Big\langle G(t)^2 - \langle G(t)^2 \rangle\Big\rangle$ contains contributions from $n_q$ quark and $n_q$ antiquark propagators:
\begin{equation}
G(t) \sim \textrm{Re} \int d^3 \vec{x}\left\langle S(t,\vec{x}|0,\vec{0})^{n_q}\right\rangle ~~,~~~\sigma\left(G(t)\right) \supset \int d^3 \vec x \int d^3 \vec{y}\left\langle S(t,\vec{x}|0,\vec{0})^{n_q} S^*(t,\vec{y}|0,\vec{0})^{n_q}\right\rangle~~.
\end{equation}
These contributions do not cancel and can form valid diagrams in themselves. The lightest states that can appear in this way are pion-like pseudoscalar mesons.
To the author's knowledge this was researched most thoroughly in the case of the nucleon where, see e.g. \cite{Wagman:2016bam,Wagman:2017xfh}:
\begin{equation}
G(t) \sim e^{-m_N t} ~~, ~~~\delta G(t):=\sqrt{\sigma\left(G(t)\right)} \sim e^{-\frac{3}{2} m_\pi t}~~, ~~~\frac{G(t)}{\delta G(t)}\sim e^{-\big(m_N - \frac{3}{2}m_\pi\big)\,t}
\end{equation}
that is with $n_q=3$ the signal-to-noise ratio is exponential and dominated by a contribution proportional to the much lighter mass of the pion. As a result, the signal of the much heavier nucleon is quickly washed out.
This affects every hadron, with the only exception being the pion itself, where the signal-to-noise ratio is constant instead.
Furthermore, there are contributions of this type for all lightest available combinations of quarks involved, for example, in combination with strange or charm quarks. 
However, they are suppressed compared to the leading pion mass-dependent term.
The situation is particularly severe in the nucleon and multi-nucleon case due to the large number of light quarks involved and the subsequent large prefactor in the exponent. The situation is less problematic once not all quarks are light, for example, in the $uuddss$ $H$-dibaryon.
Following this argumentation and extending it to the case of doubly heavy tetraquarks, the noise is governed by a leading term that is $\sim e^{-m_\pi t}$ for the $T_{QQ}^{ud}$ channels with further contributions for admixtures of pseudoscalar type mesons made up of all quarks involved.

\paragraph{Ground state and binding energy determination.}

Focusing on the situation where only the ground state needs to be determined, the first thought is to compute the so-called log-effective mass:
\begin{equation}
\tilde{m}_{\rm{eff}}(t) := \frac{1}{\delta t}\log\left(\frac{G(t)}{G(t+\delta t)}\right)~,
\end{equation}
where most commonly $\delta t=1$. Formally in a temperature $\mathcal{T}=0$ calculation of a Euclidean correlation function this is asymptotically an estimator of the mass of the ground state energy $\tilde{m}_{\rm{eff}}(t)\stackrel{t\rightarrow \infty}{=}E_0:=m_{\rm eff}$ where it appears as a constant. At shorter distances, however, there are exponentially decaying distortions which go as $\sim \sum_n e^{- (E_n-E_0)t}$. 
In certain cases, it can be beneficial to choose larger values of $\delta t$, as typically the short distance behavior is made more flat and when the data is precise enough for a long enough range in $t$ the resulting effective mass is more smooth and less prone to local fluctuations. This can be useful when cross-checking results to ensure the asymptotic plateau is correctly extracted or to study a local fluctuation or autocorrelations in the data from one timestep to the next. \\
When wanting to exploit the region near $T/2$, the effective mass formula should be adapted to take into account the boundary conditions and correct kernel for a temperature $\mathcal{T}\neq 0$ system, that is $e^{-\omega t}\rightarrow \cosh(\omega(1/2\mathcal{T}-t))/\sinh(\omega/2\mathcal{T})$.
In the case where $\delta t=1$ the effective mass requires finding the solution of:
\begin{equation}
\frac{G\left(t\right)}{G\left(t+1\right)}=\frac{\cosh \left(m_{\mathrm{eff}}\left(t - 1/2\mathcal{T}\right)\right)}{\cosh \left(m_{\mathrm{eff}}\left( t+1 - 1/2\mathcal{T} \right)\right)}~~.
\end{equation}
Solving this equation numerically requires finding the roots, which can be more difficult when the noise is not under control. Finding sensitivity to the two definitions of the effective masses can be used as a simple cross-check for control over errors and how far to trust the temporal data.

With this in hand, determining whether or not a doubly heavy tetraquark is bound in its simplest form then proceeds by (a) determining the effective masses of the two mesons ($M_{1,2}$) that define the non-interacting threshold and (b) determining the effective masses of the tetraquark correlator $T_{QQ}$. Taking all three results together, the binding energy simply is 
\begin{equation}
E_B= m_{\rm eff}^{T_{QQ}} - m_{\rm eff}^{M_1} - m_{\rm eff}^{M_2}~~.
\end{equation}
In most lattice calculations, these results will have come from the same propagators and configurations, implying they are correlated. 
This requires a more sophisticated data analysis, whereby the Bootstrap and Jackknife methods are popular in the community. 
One may also be tempted to think that the correlations can be exploited. Especially when taking ratios, correlated errors can be canceled. In some studies, the authors then first define a ratio of the tetraquark and the two meson correlators to define a "binding correlator"\footnote{For an example where this was used, see \cite{Francis:2016hui}.}:
\begin{equation}
G_{T_{QQ}}^B(t)=\frac{G_{T_{QQ}}(t)}{G_{M_1}(t) C_{M_2}(t)}~~.
\end{equation}
 Then, the binding energy is the effective mass of this correlator:
 \begin{equation}
E_B \stackrel{t\rightarrow \infty}{=} \log\left(\frac{G_{T_{QQ}}^B(t)}{G_{T_{QQ}}^B(t+\delta t)}\right)~.
\end{equation}
In principle this works, however, in practice, taking the ratio of the tetraquark and meson correlators means subtracting their respective spectra. 
In the language of spectral functions, if there is a peak in the tetraquark channel there are now also peaks with negative signs for the threshold mesons. This makes the spectrum more dense and makes it more difficult to distinguish features in the correlation functions given the noise. Observing dominance of the ground state plateau, which determines the binding energy, can appear delayed or obscured.

An alternative to determining the effective mass in this recipe is to instead directly fit the correlators to a number of exponentials:
\begin{equation}
G_{\rm fit}(t) \sim \sum_n Z_n e^{-E_n\,t}~~.
\end{equation}
Hereby, often $n=1$ or 2, depending on the number of points available in the correlator and how precise they are. When considering multiple correlators and combined fits, $n\simeq 4$ is not uncommon. Furthermore, the fit parameters are often ordered in some way for a better success rate for the fit to converge and to achieve a good $\chi^2$. For example, one could enforce that $E_0$ has the smallest value and $E_0<E_1<E_2<...<E_n$. Some times also minimal gaps are included, like $E_1\gtrsim E_0 +m_\pi$. As mentioned, ultimately, fits to exponentials are often unstable, and success can depend on parameter input values and this type of ordering. 

A systematic for both effective masses and correlator fits is the fit-window dependence. It is important to correctly identify the most extended stable plateau to control excited state contamination and dominance of the noise. At times, relying on $\chi^2$ can be deceiving, especially when the long-distance points are dominated by noise and do not contribute anything meaningful to it. One way to approach this problem is to perform fits with two different forms, for example, a single exponential and a two-exponential fit, and to impose some criterion on the fit window depending on whether they lead to the same ground state energy, see e.g. \cite{Bruno:2014jqa}.

\paragraph{Ground state enhancements.}

Given these difficulties, it is desirable to improve the lattice correlators somehow so that they (a) have longer ground state plateaus, (b) reach the plateau earlier, and (c) exhibit smaller errors. In most cases, these are anti-correlated, e.g., improving (a) leads to problems in (c), and one has to choose a balance.
Generally, a method that improves (b) is to introduce smearing. There are different types of smearing, but smearing out point source quark propagators, in particular, often leads to an earlier onset of the ground state plateau. One says that such a setup overlaps with the ground state better, or the overlap is increased, but it is clear that the smearing affects the entire spectrum and modifies the overlaps with the other states. 
Imprinting a smearing function that mimics a specific wave function is, in principle, also possible using so-called "free-form smearing", see \cite{vonHippel:2013yfa,Wurtz:2015mqa}. A downside of smearing is that it often increases noise (c), and the plateau signal is lost at earlier times (a).
An example of a method that improves (c) is using Coulomb gauge fixed wall sources, which were mentioned before. When a wall source propagator of this type is contracted with a local sink operator, the signal-to-noise ratio changes significantly, and correlator signals can be followed much further. Signal enhancements of an order of magnitude or more over a local source propagator are the norm. The downside of such correlators is that the spectral decomposition is changed, and the ground state is not reached from above as usual but from below. This behavior returns to normal when wall-wall correlators are considered; however, then the noise grows significantly.
The approach from below introduces an important systematic to control, especially when determining binding energies since the method will likely overestimate them. This is discussed in more depth in Sec.~\ref{sec:wall-smear-1}. Overall, while this method improves (c) significantly, it does not improve (b), and finding the ground state plateau can be more demanding (a).
Naturally, all of these statements depend on the system being studied and there is no one go-to answer or best setup. As will be seen in the review section, part of the evolution of the field has been to converge on which tools to combine for the best, i.e., most controlled, results.

\subsubsection{Methods for state extraction: GEVP}
\label{sec:gevp}

%The preceding discussion was based on analyzing just one or a small number of lattice correlation functions connected to the same quantum numbers. 
A significant methodology evolution can be achieved by systematically combining different source and sink operators into a correlation matrix that can be evaluated using a variational analysis \cite{Luscher:1990ck,Michael:1985ne,Blossier:2009kd}. The key is to use that each interpolating operator $O_i$, or $O_j$, that projects onto the same quantum numbers connects the same energy levels $E_n$. At the same time the matrix elements are generally different $Z_i\neq Z_j$. Cross-correlating all combinations of source and sink operators, we may define a correlation matrix whose entries read:
\begin{equation}
G_{ij}(t) = \sum_n Z_i^n Z_j^n e^{-E_n\,t}~~\rightarrow~~
C(t)=\left(\begin{array}{ccc}
G_{00}(t) & \ldots & G_{0j}(t) \\
\vdots & \ddots & \vdots\\
G_{i0}(t) & \ldots & G_{ij}(t)
\end{array}\right)
\end{equation}
Assuming that this correlation matrix $C(t)$ is square and Hermitian, i.e., symmetric, it can be shown that diagonalizing it can yield information on the individual energy levels. As such, the determined eigenvalues read
\begin{equation}
\lambda_{n}\left(t\right) \sim e^{-E_n t}\left(1+\mathcal{O}\left(e^{- (E_{n+1}  - E_n) t}\right)\right)~~,
\end{equation}
which means that the time dependence of the $n$-th eigenvalue $\lambda_n(t)$ shows an exponential decay with $E_n$ as exponent.
In principle, evaluating the effective mass of $\lambda_n(t)$ would yield $E_n$ in the asymptotic-$t$ limit.
The disentangling of the energy levels can be further improved by considering the generalized eigenvalue equation:
\begin{equation}
C\left(t\right) \nu=\lambda\left(t\right) C\left(t_0\right) \nu~~,
\end{equation}
where $\nu$ are the eigenvectors and $t_0$ defines a reference value.
The eigenvalues determined this way have the same result as above, but the prefactor for the contamination from neighboring energy levels is typically smaller.
In short, by solving the generalized eigenvalue problem (GEVP), one can in principle identify the first $N$ energy levels separately. This is assuming the analysis has been done carefully with precise data and a large value for $t_0$ so that the separation of states is robust and effective. 
As mentioned, to some extent, this can be controlled by judiciously choosing $t_0$ or the separation of $t-t_0$, and one can show \cite{Blossier:2016vgh} that with $t_0\geq t/2$ the full $N$ levels of the GEVP can be extracted and interpreted while the corrections to each level $n$ should have a minimal prefactor (in $\sim e^{-(E_{n+1}-E_{n})t}$). However, in practice, the data is usually not precise enough, and one needs to fix $t_0$ at a smaller distance. This is a problem because at short distances the higher excited states contribute more to the correlators and the solution of the GEVP may become inaccurate. When this happens and the eigenvectors are not sufficiently orthogonal the separation of the states at longer $t$ can suffer and introduce contamination to the extracted states. Controlling these effects can be difficult and they have to be systematically checked.
Note that choosing $t_0$ also normalizes the data at this point.
Once the GEVP has been solved, the discussion of the previous section transfers over, and one is faced with analyzing robustly the log effective masses of the eigenvalue time dependencies $\lambda_n(t)$ or fitting their exponential decays. 
It is then also possible to form ratios of the different $\lambda_n(t)$ to form differences of energy levels that should coincide with the energy gaps, see, e.g., \cite{Francis:2017bjr}. Although this can lead to some cancellations, in the end, one needs to base these energy gaps on a well-determined ground state.

As a side remark, from the point of the signal-to-noise ratio the GEVP improves upon the issue because with a good choice of $t_0$ it relies on the information in the intermediate distance regime to construct the estimates of the energy levels. 

\paragraph{Non-symmetric GEVPs.}
In the case of non-symmetric GEVPs, $i\neq j$, there is currently no standard method for handling them. Prony's method has been used in the context of multi-nucleon lattice correlation functions, e.g., \cite{Beane:2010em}. Also multi-exponential fits
\begin{equation}
G_{ij}(t) = \sum_n Z_i^n Z_j^n e^{-E_n\,t}~~.
\end{equation}
with different parameters for each $Z_i^n$ and $Z_j^n$ are an option \cite{Leskovec:2019ioa}. Often, this leads to an underconstrained fit, and extra information has to be included, for example, by enforcing conditions on the parameters as in single correlator fits or by including Bayesian priors and interpreting the results as part of a Bayesian analysis.

\paragraph{Further exploiting the eigenvectors.} The eigenvectors are usually not further analyzed. However, in principle, they do contain information on which entries in the basis that defined the GEVP contributed the most to a given eigenvalue.
This can be used to define what is sometimes called "optimized operators", where one combines the correlation functions $G_{ij}(t)$ according to the eigenvectors in such a way that the overlap with a desired eigenvalue is maximized. 
It is tempting to interpret these weights as statements about which operator structure might be dominant in the given state. If, for example, a diquark-antidiquark operator contributes a dominant fraction of the weights required to maximize the overlap with the ground state one would like to claim that this is the dominant structure of the ground state.
Here, one should proceed with caution though. As mentioned above, the operator structures are often Fierz-related, especially when they are defined locally. In addition their definition in terms of lattice quantum numbers, or (non)local structures when smearing, and also normalization are often not as clean as one might like. This means that the operator couples to the desired quantum numbers, but which structure it is actually coupling to is not that clear. The relation becomes even more obscured when the GEVPs are not symmetric, and the correlation functions have local sources and nonlocal sinks or variations thereof. This is discussed in situ for the tetraquark in Sec.~\ref{sec:structure}.

\subsubsection{Methods for potential extraction: Static quarks and HALQCD}
\label{sec:potentials}

\paragraph{Method in the static limit.} From the very beginnings of lattice QCD research, the static limit has played a crucial role in advancements.
This limit represents the situation of infinitely heavy quarks, and the quark propagator reduces to just a string of gauge links; see \cite{Gattringer:2010zz} for an introduction. 
Combining these quark propagators into gauge-invariant observables gives the Polyakov loop, which is used to understand the phase structure of QCD, and the Wilson loop. The latter connects to the static quark potential and is constructed by including the transverse displacements of one forward and one backward static quark propagator and connecting them through transverse link strings. 

\begin{figure}[t!]
\centering
\includegraphics[width=0.41\textwidth]{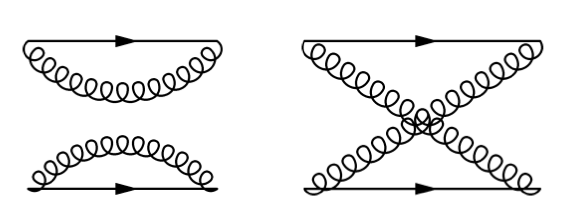}
%\hspace{5ex}
%\includegraphics[width=0.41\textwidth]{./figures/static-diagram-2}
\caption{\textit{Illustration of the diagrams in a static-static-light-light system, the two diagrams correspond to the $\mathcal{BB}$ channel relevant to doubly heavy tetraquarks and have been taken from \cite{Pennanen:1999xi}.}}
\label{fig:static-diagrams}
\end{figure}

Another way to see the form of the static quark propagators is to consider also the discussion in Sec.~\ref{sec:heavyquarks} and Eq.~\ref{eq:heavyprop-evo}, where the static limit can be reached through setting $H_0=0$ and $\delta H=0$. In this language, the static quark propagator can simply be slotted into the already formulated correlator instead of the heavy quark, whereby one usually does not sum over the spatial volume and leaves this variable open. Furthermore the identification of $J^P$ needs to be adjusted \cite{Pennanen:1999xi,Wagner:2010ad,Bali:2010xa}.
Specifically, the tetraquark correlation function reads:
\begin{align}
W_{\mathcal{BB}}(t, \vec r)&= \sum_{\vec x} 
\Big\langle \Big(\Big[ \bar{Q}(t,\vec x+\vec r) \,\Gamma_1\, u(t,\vec x+\vec r) \Big]\Big[\bar{Q}(t,\vec x) \,\Gamma_2\, d(t,\vec x) \Big] \Big)  
 \cdot \Big( \Big[ \bar{Q}(0,\vec r) \,\Gamma_1\, u(0,\vec r) \Big]\Big[\bar{Q}(0,\vec 0) \,\Gamma_2\, d(0,\vec 0) \Big]\Big)^\dagger  \Big\rangle~~,
\end{align}
whereby we used $\mathcal{B}$ to distinguish the static-light meson from a physical one and left $\Gamma_i$ open instead of explicitly choosing the $P$ and $V$ cases. The correlation function contains two light and two static quark propagators. To illustrate, the diagrams that need to be evaluated for the $\mathcal{BB}$ channel are given in Fig.~\ref{fig:static-diagrams}. It is clear that this calculation requires a timeslice-to-all propagator in order to have access to the light quark propagators originating at the values of $\vec r$ one wants to include in a study. Typically, noise wall sources are used for this task. To improve the signal-to-noise ratio, they can be diluted to spatial grids on the source timeslice with support only on the target $\vec r$. 

Evaluating this correlator and projecting $\vec r$ leads to a data set with $t$ and $r=|\vec r|$ entries. It is connected to the lattice potential $V_{lat}(r)$ via
\begin{align}
W(t,r) \sim e^{- V_{lat}(r) t} ~\rightarrow ~~
V_{lat}(r) \stackrel{t\rightarrow \infty}{:=} \log \frac{W(t,r) }{W(t+1,r) }  ~~.
\end{align}
As such, for every value of $r$, it is given by the long $t$-plateau of the effective mass of the static-static-light-light correlator. For further analysis, the lattice potential is often fitted to a phenomenologically motivated fit form for the continuum potential. The fitted potential can then be input into a Schr\"odinger equation to determine energy levels or scattering phase shifts. 
From the data point of view, the difficulties to address are: the signal-to-noise ratio and ground state saturation in the temporal direction, as well as the range of available values in the spatial separation. 

The first issue can be approached using the same methods as in the non-static case. For example some form of spatial smearing may be introduced, however, this also renders points in $r<r_{\rm smear}$ unusable. Since static quarks are involved, the signal can also be improved through temporal smearing \cite{Hasenfratz:2001hp}, inducing an analogous caveat for $t<t_{\rm smear}$.
Alternatively GEVP based methods can be used, either by reformulating the entire procedure in a GEVP, or by determining first the optimized correlators before determining the potential.
The latter approach is more attractive because no valuable information at short distances is lost. This is connected to the second issue, the available range in $r$. The shortest range is $r=1$ in lattice units and directly gives a bound driven by the lattice spacing on the shortest distance interactions that can be probed. However, this is a crucial region for determining the spectra of the heavy tetraquark states, and without guidance from the lattice data, the potential relies almost entirely on the information put into the fit ansatz in this region. The only constraint is indirectly given through the $V_{lat}(r=1)$ value. This is a difficult systematic to control. 
Results in this approach are extensively discussed below; a good example of extracted lattice potentials and fits to them is shown in \ref{fig:historic-bicudo}.

\paragraph{HALQCD method.}

The HALQCD method aims to extend the method to the non-static case. It was first developed and reported on in \cite{Ishii:2006ec,Aoki:2009ji} by members of the HALQCD collaboration. 
From the lattice point of view its starting point is the same tetraquark correlation function as in the static case, but now with heavy quarks formally being handled by standard propagators. For example in the $I(J^P)=0(1^+)$ $T_{bb}^{ud}$ channel:
\begin{align}
W_{PV}(t, \vec r)&= \sum_{\vec x} 
\Big\langle \Big(\Big[ \bar{b}(t,\vec x+\vec r) \,\gamma_5\, u(t,\vec x+\vec r) \Big]_{\rm local}\Big[\bar{b}(t,\vec x) \,\gamma_i\, d(t,\vec x) \Big]_{\rm local} \Big) \nonumber \\
 &\qquad\cdot \Big( \Big[ \bar{b}(0) \,\gamma_5\, u(0) \Big]_{\rm wall}\Big[\bar{b}(0) \,\gamma_i\, d(0) \Big]_{\rm wall}\Big)^\dagger  \Big\rangle~~,
\end{align}
where the labels wall and local have been introduced to indicate that Coulomb gauge fixed wall sources are combined with local sink operators.
As before, this is an exponentially decaying correlator; however, the key insight is to identify the prefactor, or amplitude, of it with the Nambu-Bethe-Salpeter (NBS) wave function:
\begin{align}
\Phi(\vec{r}) :=\frac{1}{\sqrt{Z_{P}}} \frac{1}{\sqrt{Z_{V}}} ~W_{PV}(t, \vec r)~ e^{-(E_{P}+E_{V})\,t}~,
\end{align}
whereby one assumes that $t$ is large. The NBS wave function can be formally related to a potential $U(\vec r,\vec r')$ via
\begin{align}
\left(\frac{\nabla^2}{2 \mu}+\frac{p^2}{2\mu}\right) \Phi(\vec{r})=\int d^3 \vec{r}^{\prime} \,U\left(\vec{r}, \vec{r}^{\prime}\right) \Phi\left(\vec{r}^{\prime}\right)~~,
\end{align}
where $\mu$ is the reduced mass of the di-meson system and $p$ its total momentum. The potential defined in this way is generally nonlocal and needs to be expanded. Identifying $E=\frac{p^2}{2\mu}$ as the asymptotic state energies, the leading order local potential in this expansion reads:
\begin{equation}
V_0(r)=\Phi^{-1}(\vec{r}, t)\left[-E_{tot} + \frac{\vec\nabla^2}{2\mu}\right] \Phi(\vec{r}, t)~~,
\end{equation}
here we recognize the Schr\"odinger equation and notice that one difficulty is the estimation of $E_{tot}$. The "time-dependent HALQCD" method \cite{Ishii:2012ssm} was designed to deal with this, and the potential then becomes:
\begin{equation}
V_0(r)=\Phi^{-1}(\vec{r}, t)\left[\frac{d}{8\mu}\partial_t^2-\partial_t - + \frac{\vec\nabla^2}{2\mu}\right] \Phi(\vec{r}, t)~~,
\end{equation}
where the prefactor $d$ is often so small that the $\partial_t^2$-term can be neglected.

Once the local potential is determined on the lattice, the procedure follows that of the static case. The potential is fitted to an ansatz, which is subsequently analyzed towards binding energies and scattering parameters. As such, the approaches to improve signals and their limitations are also similar. An example of the extracted and fitted potential for reference is shown in Fig.~\ref{fig:hal-tcc-1}.
However, the key systematic in this approach is the derivative expansion of the potential and the limitation of extracting only the leading local term. To the author's knowledge, there is no standardized way to quantify this systematic and its possible effects on the finalized results.

\subsubsection{Limitations in lattice parameter space}
\label{sec:latlimit}

Although the quark masses, volume and lattice spacing are inputs into lattice studies, and we are free to vary them alongside other parameters of the calculational setup, there are limitations. They are often imposed by numerical aspects and are often common to all lattice setups and actions. Consequently most lattice calculations are limited to a "common" parameter window. Here we briefly introduce some of these limitations.

\paragraph{Quenched versus full QCD.}
Since we are also reviewing historical studies as early as 1990, we will also encounter so-called "quenched" studies. When considering the partition function the fermions are usually integrated out and it becomes $Z=\int dA \,\textrm{det}[M(A)]\, d\psi \,e^{-S_{gauge}(A)}$, where $M$ denotes the fermion matrix and the required determinant is related to the inverse of the lattice Dirac operator. 
The sole driving cost factor in generating lattice data is the evaluation of this determinant. Since it is connected to the lattice Dirac operator, it can be shown that omitting all sea quark loop effects amounts to setting $\textrm{det}[M]=1$. Calculations in this approximation are called quenched for this reason, and they are orders of magnitude cheaper computationally speaking. Given the low resource availability at the time they were the dominant form of calculation in the early days of lattice QCD. 
The evolution of this was to instead handle two light quark flavors ($N_f=2$) without approximation and to address their determinant fully. Today the standard are $N_f=2+1$ quark flavor calculations with two isospin symmetric light quarks and a strange quark. For some lattice actions also $N_f=2+1+1$ are available.

\paragraph{Quark mass window.}
Once the general number of flavors is decided on, the first limitation is the sensible range of quark masses, both in the sea and valence sectors. The upper edge is driven by the discretization effects, where when using a lattice action that has discretization effects starting at $\mathcal{O}(a)$, the leading discretization effects on particles with heavy masses are $am_Q$, for example. Therefore, to keep these effects under control, the condition $am_Q\lesssim \mathcal{O}(1)$ needs to be satisfied. An action like this is the Wilson-type fermion action, and the rule of thumb is to set $\mathcal{O}(1)\simeq 0.6$ at most. In the case of an action with leading order $\mathcal{O}(a^2)$ discretization effects, such as Domain Wall or staggered-type fermion actions, one sees $\mathcal{O}(1)\simeq 0.8$. Further below we will see that currently this puts a rough upper bound of $m_Q\lesssim 3.4$ or $4.5$~GeV for Wilson-type and Domain Wall fermion actions, respectively.
At the same time, these quarks typically do not need to have sea quark contributions included explicitly since they are typically small and go as $\mathcal{O}(1/m_Q)$.
Nevertheless, performing calculations with (relativistic) heavy quarks on (too) coarse lattice spacings is difficult. 

The lower edge is driven by numerical considerations. 
As such, the cost of generating a gauge configuration is mainly determined by the cost of performing the fermion matrix inversion for the fermion determinant, equivalent to that for a light quark propagator. 
Previously, physical quark mass values could not be reached in practice as this cost scaled as a power law with a large exponent when using standard inversion algorithms, such as the conjugate gradient solver. Fortunately, this "Berlin wall" \cite{Ukawa:2002pc}, named after the Lattice conference where this was made clear to the community, could be ameliorated by more sophisticated solvers. Today, physical and sub-physical quark masses are accessible. However, the numerical cost is still significant. 
At the same time, fluctuations in the gauge generation process, particularly downward fluctuations of the lowest eigenmodes of the lattice Dirac operator, can significantly increase this cost or even make the calculation pathological. These fluctuations go as $\mathcal{O}(1/m_\pi, a)$, i.e., they become larger at lower input quark masses and coarser lattice spacings. In practice calculating with physical light quark parameters is consequently hard and often not possible for coarse lattice spacings.

\paragraph{Volume window.}
There are further limitations given through the size of the physical volume, which, as a rule of thumb, should be at least $3~$fm, but additionally and more concretely should satisfy $m_\pi L\gtrsim 4$. This condition ensures that volume effects are sufficiently small to observe a largely undistorted spectrum. Since the cost of the simulation scales proportionally to the lattice volume, this imposes a strong limitation on the parameter window.

\paragraph{Cut-off window.}
The last limitation considered here is of numerical origin again and concerns the smallest lattice spacings, i.e., largest cut-off values, that are accessible. As mentioned, one would like to have as small as possible lattice spacing to study heavy states, such as the doubly heavy tetraquarks. Although the volume condition imposes a significant practical constraint already, another is given through autocorrelations: The generation of gauge fields in lattice calculations proceeds via Markov-Chain Monte-Carlo techniques, and one configuration is generated from a preceding one. To sample QCD observables correctly, one needs to save a large set of configurations sufficiently apart in Monte-Carlo time for them to be decorrelated. 
Empirically, it has been found that the slowest scale to decorrelate in a lattice calculation is related to topology and the probability of jumping from one topological sector to another decreases with the lattice spacing. This means the autocorrelation time grows without bounds, and the topology freezes in practice. In turn, this induces topological contamination at the level of $\mathcal{O}(\tilde Q/V)$ to observables, where $\tilde Q$ is the topological charge and $V$ is the space-time volume. The onset of this critical slowing down, in the sense that it becomes a significant issue, depends on the lattice action used but often values of $a\lesssim 0.045$~fm are quoted.  

\begin{figure}[t!]
\centering
\includegraphics[width=0.41\textwidth]{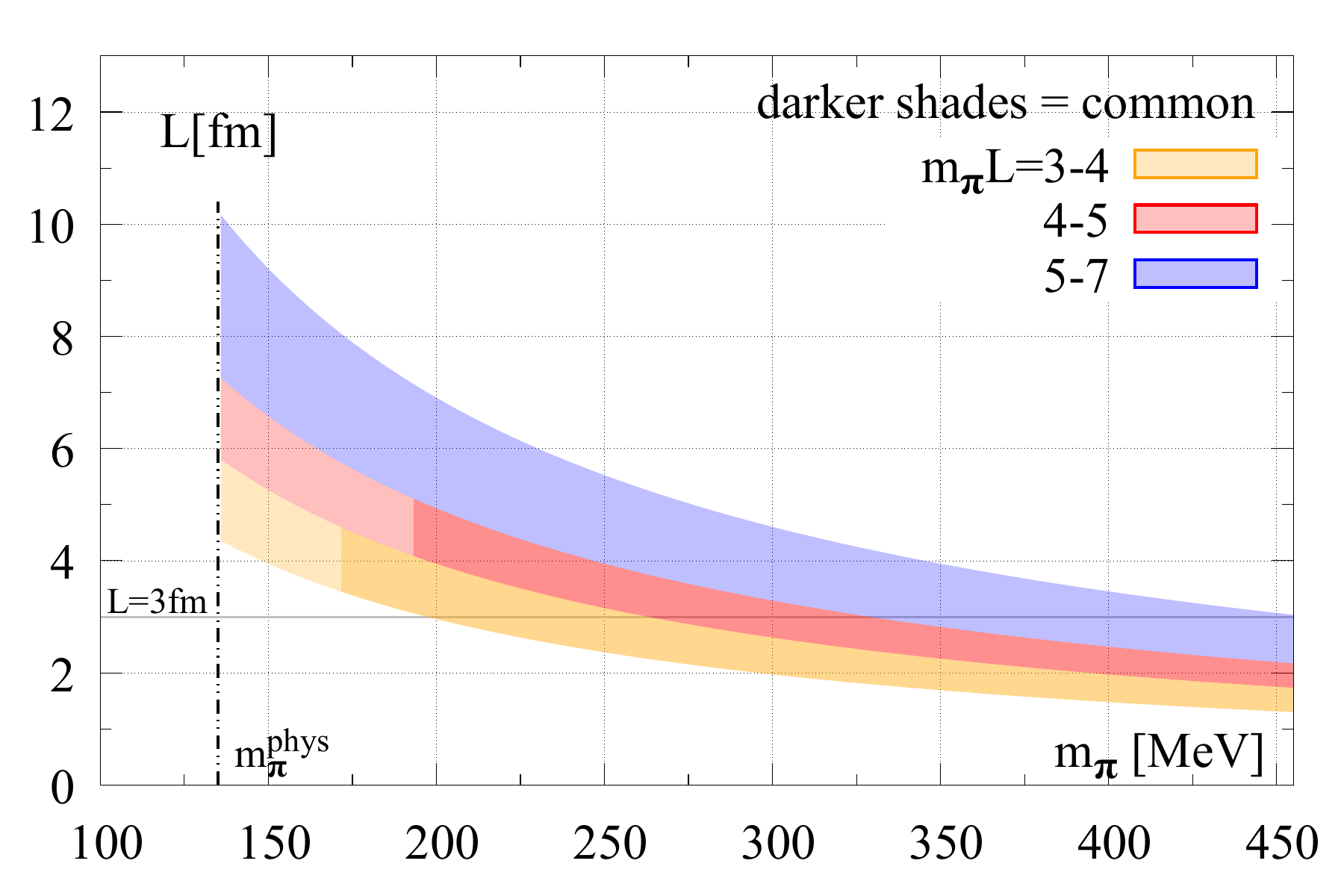}
\includegraphics[width=0.41\textwidth]{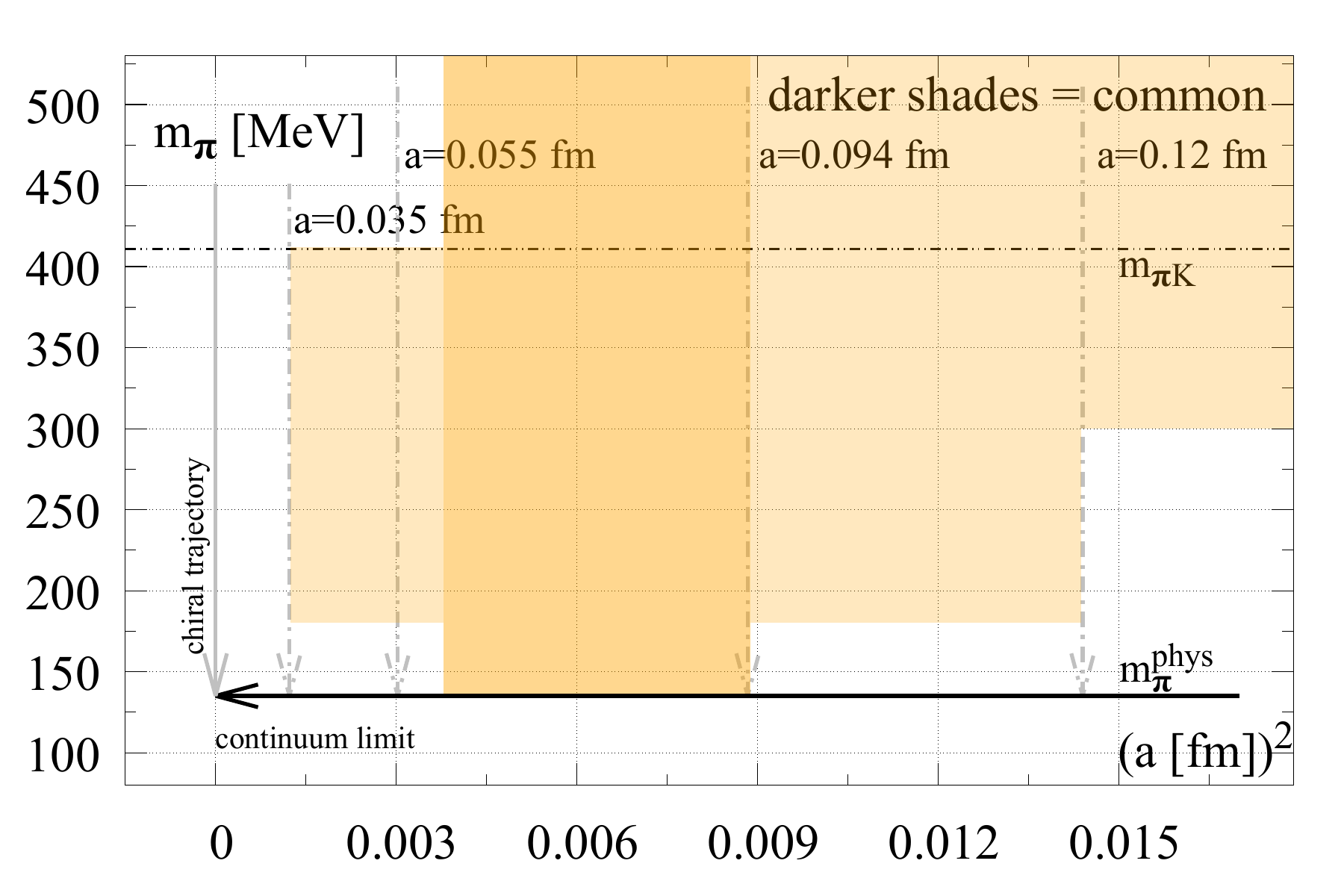}
\caption{\textit{The current parameter window of lattice calculations.
Left: Ensemble volumes over pion masses. The horizontal and vertical lines denote the $L=3~$fm bound and physical point $m_\pi=135~$MeV, respectively. The colored bands map out the regions of certain $m_\pi L$ where lattice calculations are done today. The darker shaded bands denote the most common parameter windows.
Right: Ensemble pion masses over lattice spacings.
The yellow shaded area denotes a rough representation of today's parameter window of lattice simulations. The darker area denotes the most common window.}}
\label{fig:ensemble-window}
\end{figure}

\paragraph{Current status.}
The parameter window is schematically depicted in Figure~\ref{fig:ensemble-window}. 
On the left of the figure, the ensemble volumes are shown over their pion masses. The horizontal dashed line denotes the $L=3~$fm bound, while the vertical one denotes the physical point $m_\pi=135~$MeV. The colored bands map out the regions volumes where $m_\pi L\in[3:4]$ (yellow), $[4:5]$ (red) and $[5:7]$ (blue).
The right of the figure shows the ensemble pion masses over lattice spacing.
The yellow shaded area denotes a rough, incomplete representation of today's parameter window of lattice simulations whereby the darker area depicts the most common one.

\subsubsection{Methods for heavy quarks}
\label{sec:heavyquarks}

The above discussion shows clearly the limitations for lattice calculations of the heavy hadron spectrum: With a lower bound on the lattice spacing between $a=0.035~\rm{fm}\rightarrow 5.6~\rm{GeV}^{-1}$ and $a=0.055~\rm{fm}\rightarrow 3.6~\rm{GeV}^{-1}$, we have  upper bounds on $m_Q\lesssim 3.4 (4.5)~\rm{GeV}$ or $2.2(2.9)~\rm{GeV}$, for the Wilson-type and Domain Wall (brackets) actions,  respectively.
As such, when using a fully relativistic quark action, resolving spectra including charm quarks is possible. Resolving them with bottom quarks might just be within reach, but are technically very difficult to control and extremely expensive with respects to computing resources. To avoid this the studies of doubly heavy tetraquarks have resorted to using an effective heavy quark action for the bottom quarks, for the time being. There are two main approaches to follow:

\paragraph{Non-relativistic heavy quark actions.}
The first is to perform an expansion of the QCD Hamiltonian in terms of the heavy quark mass $1/m_Q$. Starting from the static limit, describing infinitely heavy quarks, each term in the expansion then includes more and more of the mass corrections and dynamics. 
Depending on the system under study, whether it contains two heavy quarks interacting or a single heavy quark, the power counting of the terms in this expansion changes. In the first scenario, we obtain non-relativistic QCD (NRQCD), and in the second, heavy quark effective theory (HQET). Focusing on systems with two heavy quarks, the NRQCD Hamiltonian $H_{NRQCD}=H_0 + \delta H$ reads \cite{Thacker:1990bm,Lepage:1992tx,Davies:1994mp}:
\begin{align}
H_0=  -c_0 \frac{\Delta^{(2)}}{2 m_Q} ~,~~~
\delta H= & -c_1 \frac{\left(\Delta^{(2)}\right)^2}{8 m_Q^3}+\frac{c_2}{U_0^4} \frac{i g}{8 m_Q^2}(\tilde{\boldsymbol{\Delta}} \cdot \tilde{\boldsymbol{E}}-\tilde{\boldsymbol{E}} \cdot \tilde{\boldsymbol{\Delta}}) 
 -\frac{c_3}{U_0^4} \frac{g}{8 m_Q^2} \boldsymbol{\sigma} \cdot(\tilde{\boldsymbol{\Delta}} \times \tilde{\boldsymbol{E}}-\tilde{\boldsymbol{E}} \times \tilde{\boldsymbol{\Delta}}) \\\nonumber
& ~~~-\frac{c_4}{U_0^4} \frac{g}{2 m_Q} \boldsymbol{\sigma} \cdot \tilde{\boldsymbol{B}}+c_5 \frac{a^2 \Delta^{(4)}}{24 m_Q}-c_6 \frac{a\left(\Delta^{(2)}\right)^2}{16 n m_Q^2} + ...
\end{align}
where $\tilde{\boldsymbol{E}}, \tilde{\boldsymbol{B}}$ and $\tilde{\boldsymbol{\Delta}}$ denote the $O(a)$-improved color electric field, color magnetic field, and spatial lattice derivative respectively. $\Delta^{(2)}$ is the lattice Laplacian, $g$ is the bare gauge coupling and $n$ is a mode number used in the evolution equation. Often $n=4$ is chosen as a reasonable value for this stability parameter. In addition one often performs tadpole improvement through the coefficient $U_0$, which may be set either through the average gauge-fixed link or the fourth root of the plaquette.
Further, the coefficients $c_i$ may be tuned to fix physics quantities, such as the hyperfine splitting, $M_{PP}-M_{VV}$, but are typically set to their tree-level values, which are equal to one. The mass parameter $m_Q$ is then tuned to reproduce the spin-averaged mass, $3 M_{PP} + M_{VV} = 4 M_{sp.avg}$, of the target quarkonium system through the dispersion relation. That is by measuring the slope of the momentum dependence of the spin-averaged quarkonium masses determined on the lattice and tuning it to reproduce the target value in physical units. 

As part of the formulation of the non-relativistic action fermions and anti-fermions are decoupled. Computing an NRQCD quark propagator then becomes an initial value problem and can be performed by evaluating the evolution equation:
\begin{equation}
S(x, t+1)=\left(1-\frac{H_0}{2 n}\right)^n\left(1-\frac{\delta H}{2}\right) \frac{U_t^{\dagger}(x)}{U_0}\left(1-\frac{\delta H}{2}\right)\left(1-\frac{H_0}{2 n}\right)^n S(x, t),
\label{eq:heavyprop-evo}
\end{equation}
where $U_t$ is the link in time direction. This makes their determination computationally inexpensive\footnote{See \cite{ContraOb} for an implementation where NRQCD propagators can be generated on-the-fly.}. 
Additionally, in both HQET and NRQCD the constraint $am_Q\lesssim\mathcal{O}(1)$ changes into $am_Q\gtrsim\mathcal{O}(1)$. The constraint switches and coarser lattice spacings are now favored. For the bottom quark, this usually has little consequence. However, this can be a strong limitation for charm quarks as the parameter window of lattice spacings is too fine to treat them using these effective theories. 
Furthermore, in the case of NRQCD, it can be shown that a continuum limit does not exist, making a systematic study of discretization effects difficult. In contrast, for HQET, a continuum limit can be taken \cite{Gattringer:2010zz}.

Note that, as a technical aside, NRQCD correlators require the inclusion of a mass-shift $m_{0}$ to yield absolute mass values. Here, one usually sets this to be the tuned-to physical mass of a hadron. In the case of an observed meson, this is not necessarily problematic; however, when including just one quark, for example, one needs to take a fraction of this number, thereby defining a dressed quark mass parameter. This can be a source of systematic uncertainty in the study of heavy baryons with multiple NRQCD quark flavors, see e.g. \cite{Hudspith:2017bbh} or \cite{Brown:2014ena}.
A further drawback is that due to the reduction to Pauli matrices, only simple $S$ and $P$ wave operators are available without including derivatives in the operator construction. Due to this complication, the operator bases available when using NRQCD propagators within the interpolating operator for a state are often smaller than with relativistic propagators.

\paragraph{Effective relativistic heavy quark actions.}
The alternative approach is through so-called effective relativistic heavy quark actions. In practice, there are four types, whereby all rely on the derivation in \cite{El-Khadra:1996wdx}. Here, the authors could show that the pathological discretization effects could be re-absorbed into the coefficients of the fully relativistic quark action. This action has six parameters, two sets of three, which separate the temporal and spatial components. The three coefficients split in this way are related to the Wilson parameter, which ultimately relates to the quark mass parameter, an anisotropy coefficient, and the "Clover" coefficient. The latter in the relativistic action implements $\mathcal{O}(a)$ improvement \cite{Sheikholeslami:1985ij}.
The effective relativistic heavy quark Dirac operator can be generically written as:
\begin{align}
D_{x, y}= & \delta_{x y}-\kappa_f\left[\left(r_t-\gamma_t\right) U_{x, t} \delta_{x+\hat{t}, y}+\left(r_t+\gamma_t\right) U_{x, t} \delta_{x+\hat{t}, y}\right] \nonumber\\ 
&-\kappa_f \sum_i\left[\left(r_s-\nu_s \gamma_i\right) U_{x, t} \delta_{x+\hat{i}, y}+\left(r_s+\nu_s \gamma_i\right) U_{x, t} \delta_{x+\hat{i}, y}\right] 
 -\kappa_f\left[c_E \sum_i F_{i t}(x) \sigma_{i t}+c_B \sum_{i, j} F_{i j}(x) \sigma_{i j}\right] ,
\end{align}
whereby the conventions chosen here are those of the Tsukuba formulation \cite{Aoki:2001ra} and $r_t=1$. The open parameters are $\kappa_f,r_s,\nu_s,c_E$ and $c_B$. In this notation $\kappa_f$ is related to the mass, $r_s,\nu_s$ to the anisotropy and $c_E,c_B$ to the Clover term.
As such, we realize that in practice, calculating a heavy quark propagator in the valence sector may proceed by inverting the fermion matrix derived from an anisotropic quark action using these re-interpreted coefficients.
The coefficients can be perturbatively or non-perturbatively tuned by enforcing physics constraints. For example, one may choose to fix the spin-averaged mass, hyperfine splitting and dispersion relation $M_{rest}=M_{kin}$. Different choices of how to fix and choose the total six parameters yield the Fermilab \cite{El-Khadra:1996wdx}, Tsukuba \cite{Aoki:2001ra}, RHQ \cite{Christ:2006us} and OK \cite{Oktay:2008ex} effective relativistic heavy quark actions. These actions all have a continuum limit. 
Furthermore, when combined with Wilson-Clover relativistic valence and sea quarks, there is, in principle, a smooth limit between the heavy and light quark sectors. However, tuning the necessary parameters is an intensive task, e.g. \cite{Lin:2006ur,PACS-CS:2011ngu}. In its most reduced form in the RHQ action \cite{Christ:2006us}, there is a three-parameter search space to tune $m_0,\zeta,c_P$, denoting quark mass, anisotropy, and generalized Clover coefficients. Recently, the tuning of a five-parameter version was performed using machine learning techniques \cite{Hudspith:2021iqu} to handle the non-linearities and large search spaces one needs to cover.

%\newpage
\section{Overview of lattice calculations past and present}
\label{four}
\begin{figure}[t!]
\centering
\includegraphics[width=0.34\textwidth]{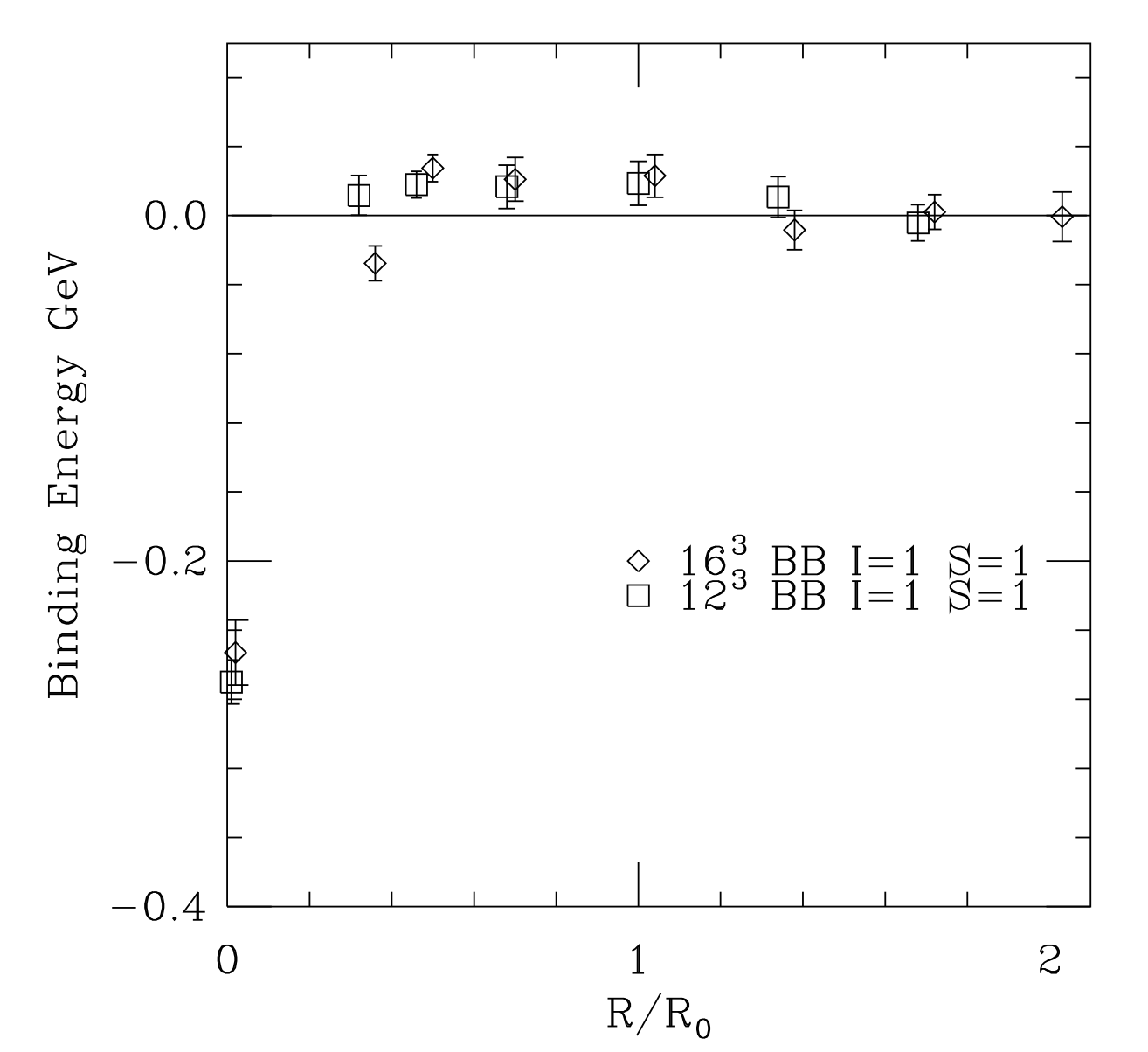}
\includegraphics[width=0.34\textwidth]{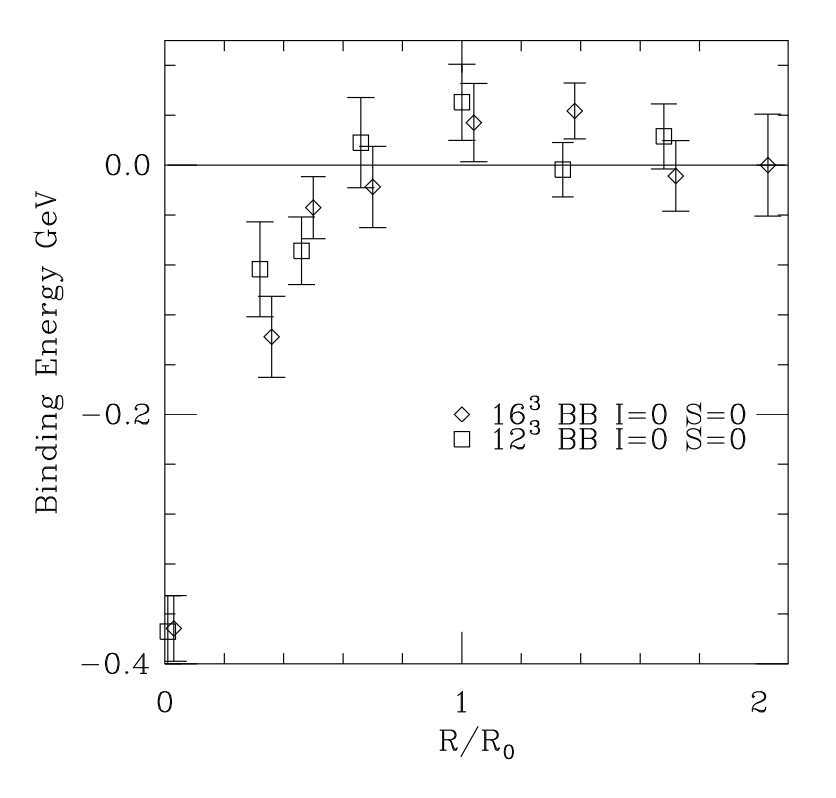}
\includegraphics[width=0.34\textwidth]{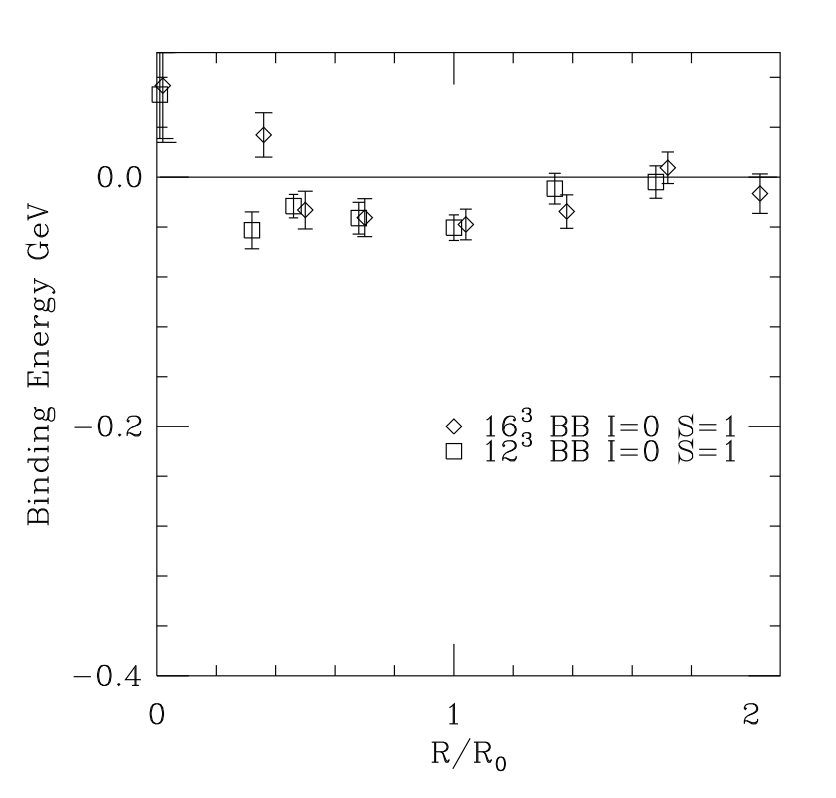}
\includegraphics[width=0.34\textwidth]{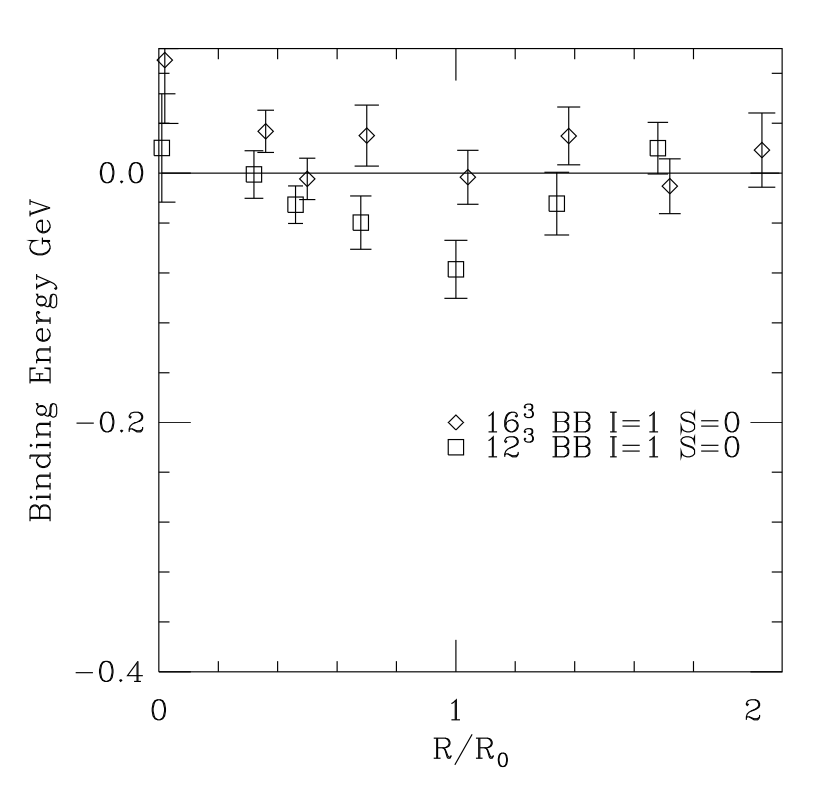}
\caption{\textit{Binding energies derived from $QQ\bar{q}\bar{q}$ potentials in \cite{Michael:1999nq}. From top left to bottom right, the channels $J^P=2^+$ (top left), $1^+,s=0$ (top right), $1^+,s=1$ (bottom left), and $0^+$ (bottom right) are shown. The $1^+,s=0$ corresponds to the modern $T_{bb}$ and already looks favored in 1999.
}}
\label{fig:historic-ukqcd}
\end{figure}

After this discussion on general methodology, next we present an overview of lattice studies at the time of writing and later discuss the emerging physics picture.
The first sections will give a rough overview of the historical development of the study of doubly heavy tetraquarks, with the later sections concentrating on the most recent work and headway made in the studies of the $I(J^P)=0(1^+)$ $T_{QQ'}$ tetraquark candidates.
Most studies in the early era violate quality criteria laid out in Sec.~\ref{sec:latlimit} for the size of the volumes, the likely finite volume effects on the spectrum, and the handling of discretization effects and autocorrelations. 
In a way, the below time history also showcases the development of lattice technology as new parameters and setups become available, significantly boosting the quality through improved statistical and systematic control over the hadronic systems studied.
In the following section, results that quote multiple errors are simplified by adding them together into a single error budget. This is done by determining the square root of the summed errors squared, and the values are first averaged in the case of upper/lower errors. This is to simplify the discussion and to make a conservative choice when representing the lattice data from multiple different efforts.

\subsection{Early and pre-QCD studies}

The earliest studies were interested in the interaction potential between two heavy-light, or actually static-light, mesons. As such, their goal often was not directly to study tetraquarks as possible heavy QCD states, but their results do still hold and have a historical relevance.
The first study \cite{Richards:1990xf}, published in 1990, was conducted on an $8^3\times 16$ quenched lattice with a staggered valence action corresponding to four staggered quark flavors. The authors' goal was to study the interaction potential of the P-V system, with the constituents called the "$\pi$" and the $"\rho$", in the approximation where one quark in each meson is static. Then, as highlighted in the previous section, studying the correlation function $W_{\rho\pi}(R,T)$ in the asymptotic $T$-limit at varying values of $R$ allows to determine the effective interaction range $\ell$ of the two constituent mesons as $W_{R,T\rightarrow\infty}\sim e^{-R/\ell}$.
Although far from today's advanced calculations, this first study already showed signs of an attractive interaction. In \cite{Mihaly:1996ue}, this type of study further advanced the determination of the potential $V(r)$ from static-static-light-light correlators. The extracted potentials show an attraction with quark exchange playing a significant role for $r\ll2a$ where $a=0.19~$fm.
At these times, the above calculations were at the cutting edge of resource requirements. Limited by computing resources, the authors of \cite{Green:1998nt}, in 1998, decided to study this system with $SU(2)$ quenched lattices with all quarks being static. Their innovation is to consider multiple basis states and to extract information on a series of potentials $V_n$ with $n=0...5$ in a type of pre-cursor to the modern GEVP analysis. In the same year, \cite{Stewart:1998hk} applied a potential type study to the system of two light and two static quarks on quenched $SU(2)$ lattices and found an attractive potential. A key innovation at this time was the lightness of the $SU(2)$ pion mass with $m_{\rho}/m_{\pi}\simeq 3$. Note at this point that the spectrum of $SU(3)$ is significantly different from that of $SU(2)$; for example, diquarks are the baryons of the theory, and the number as well as properties of the Goldstone bosons are not the same. This does not lessen the achievements but serves as a reminder of the historical nature of the results obtained.

\begin{figure}[t!]
\centering
\includegraphics[height=0.2\textheight]{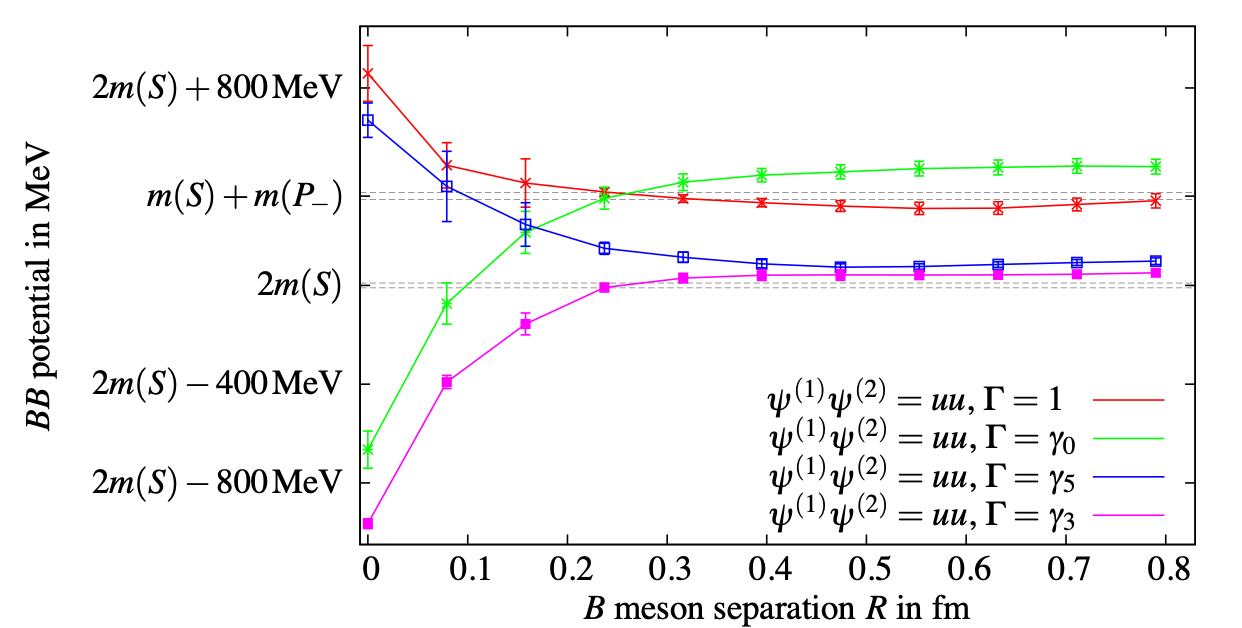}\\
\includegraphics[height=0.15\textheight]{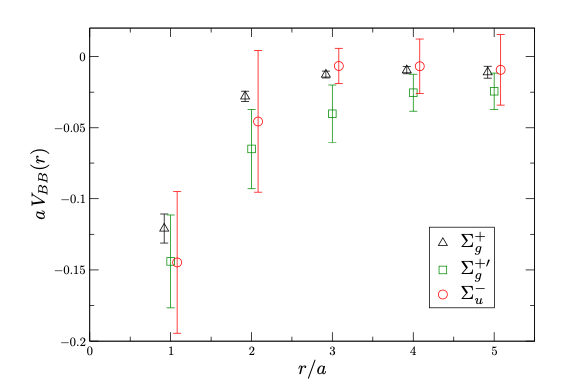}
\includegraphics[height=0.15\textheight]{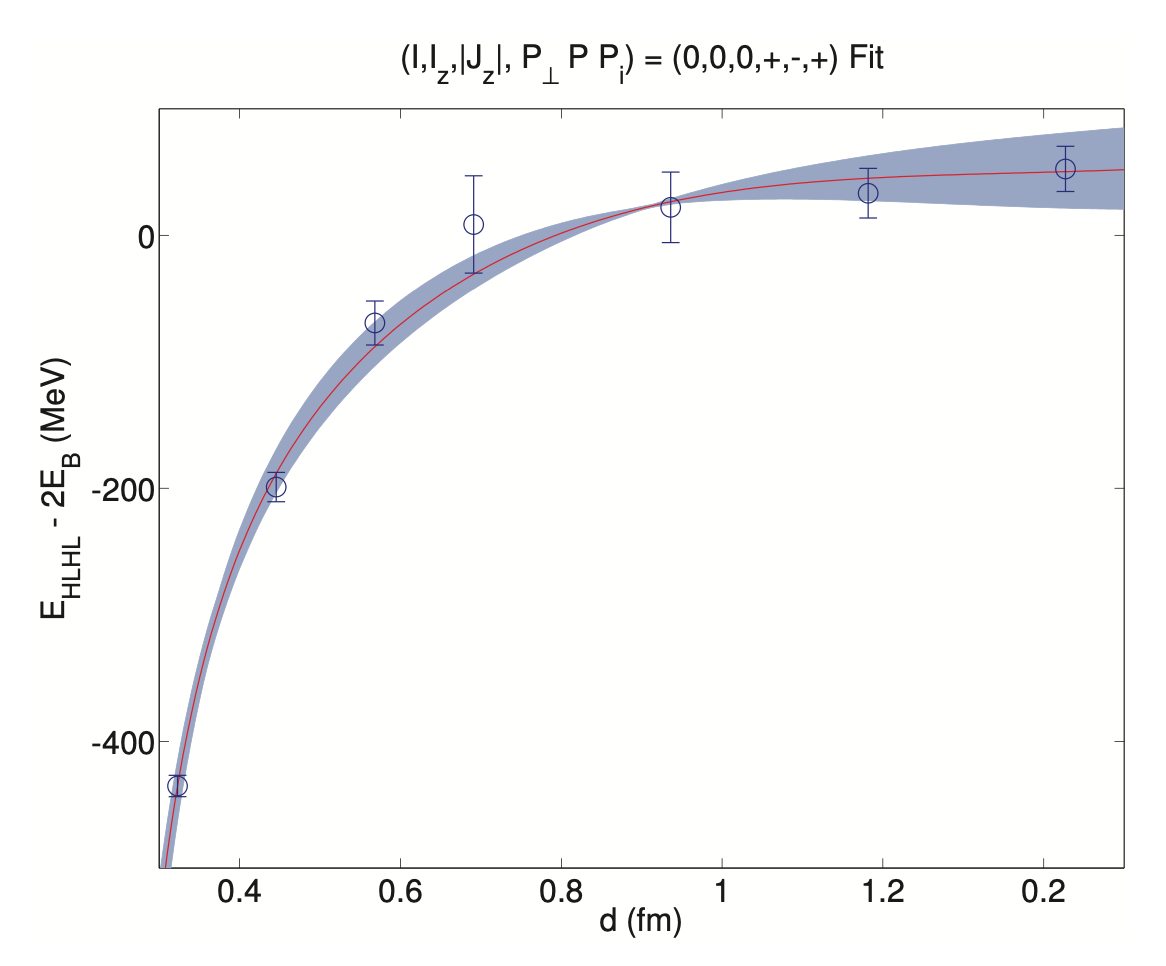}
\caption{\textit{Inter-meson potentials in the $\mathcal{B}\mathcal{B}$-system in $N_f=2$ full QCD reported in \cite{Wagner:2010ad} (top) and \cite{Bali:2010xa} (bottom left). Shortly after, a study \cite{Brown:2012tm} in $N_f=2+1$ QCD becomes available on anisotropic lattices (bottom right).
}}
\label{fig:historic-nf2QCD}
\end{figure}

A change of quality of the discussion was reached in the study of \cite{Michael:1999nq}, where, to the author's knowledge, the possibility of a bound $cc\bar{q}\bar{q}$ was explicitly mentioned for the first time as a motivation to study this system. Still in quenched QCD - but now in $SU(3)$ theory - the authors studied the static-light meson pair, which they interpret in analogy to the Born-Oppenheimer approximation as the interaction of two bottom-light mesons. Considering all possibilities, P-P, P-V, and V-V, the authors apply a variational analysis to determine the potentials. In their setup the mass of the light valence quark is around that of the strange quark and $m_{\rho}/m_{\pi}\simeq 1.5$ \footnote{The physical value is around $m_{\rho}/m_{\pi}\simeq 5.5$.} and the lattice spacing is around $a\simeq0.18~$fm.
The analysis yield attractive potentials, and "[the] results show that it is plausible that exotic $bb\bar{q}\bar{q}$ di-mesons exist as states stable under strong interactions" \cite{Michael:1999nq}. Given this remarkable conclusion, the results of the binding energies in $R$ (in dimensionless units of $R_0$) are shown in Fig.~\ref{fig:historic-ukqcd}.
The study of \cite{Pennanen:1999xi} extends this further to include dynamical fermions, here Wilson-Clover fermions, and broadly confirms the quenched results.

A slightly different perspective on these studies is given in \cite{Cook:2002am}, where a spectral reconstruction is attempted via the Maximum-Entropy method (MEM). Further, the authors construct a $2\times2$ correlation matrix, employing the previously used local di-meson operator and introducing also the notion of a nonlocal operator for the first time. They achieve this by implementing a di-meson operator separated by a distance $r$ and connected via link strings $\mathcal{O}\simeq(Q\bar{q})_x U_{x\rightarrow y} (Q\bar{q})_y$. They find the potential turns from weak repulsion to strong attraction from their extracted spectral densities $\rho(\omega,r)$ for $r\lesssim 0.2~$fm indicated through a level crossing of the first excited and the ground state at this distance. 
Interestingly, a much later study reported in 2021 \cite{Bicudo:2021qxj} that focuses on the relative overlaps between diquark-diquark operators and di-meson operators finds a crossing between the diquark-diquark and di-meson levels in this range around $r\lesssim 0.25~$fm.

%%%

As last in this section, consider the study of \cite{Detmold:2007wk}, which extends over the work in \cite{Michael:1999nq} by performing a more careful analysis as well as considering systematic effects more rigorously. In particular, the importance of finite volume and discretization effects is highlighted. The calculation employs a quenched setup with $m_\pi\simeq 400~$MeV and the lattice spacing $a\simeq0.1~$fm. The authors find "clear evidence of repulsion between the B-mesons in the isospin$\neq$spin channels and attraction in the isospin=spin channels" \cite{Detmold:2007wk}, as the authors of \cite{Michael:1999nq} did as well. 
 We mention in passing the work presented in \cite{Doi:2006kx}, which did not observe an attraction in the range of $r\gtrsim 0.2~$fm to $r\lesssim 0.8~$fm, which is generally in agreement with the studies above.
This concludes the era of early lattice work, which was dominated by the potential approach. In the next section we will see the first studies at the level of the finite volume spectrum.

\subsection{QCD potentials, finite volume spectra and the prediction of the $T_{bb}$}

\paragraph{Potentials in two-flavor QCD.}
The next era uses predominantly full QCD calculations and starts with two simultaneous efforts reported by different groups in \cite{Wagner:2010ad,Bali:2010xa} and later updated in \cite{Wagner:2011ev,Bali:2011gq}. Both pursue the goal of studying the $\mathcal{B}\mathcal{B}$-potentials, whereby the motivation of understanding tetraquarks is becoming more prominent.
In \cite{Wagner:2010ad} the calculations are performed on a lattice with $N_f=2$ twisted-mass Wilson fermions with $m_\pi\simeq 340~$MeV, $a\simeq 0.08~$fm and $L\simeq 1.92$. The study finds that "a BB potential is attractive if the trial state is symmetric under meson exchange, and repulsive if the trial state is antisymmetric under meson exchange" \cite{Wagner:2010ad}. The study of \cite{Bali:2010xa} has a very similar lattice setup and also uses $N_f=2$ dynamical quarks but of the Wilson-Clover type and for the first time on multiple lattice spacings ($a=0.084~$fm and $a=0.077~$fm). The authors do not report on conclusive findings but note possible repulsive potentials at long distances. 
The key results of both studies are shown in Fig.~\ref{fig:historic-nf2QCD} (top and bottom left).
Compared to the results of the early studies, no qualitative difference was observed; the potentials corresponding to $0(1^+)$ states 
mostly show a deeply attractive part at short distances as before. From the figure and considering both studies, the change to an attractive potential happens around $r\simeq 0.15-0.2~$fm.

\begin{figure}[t!]
\centering
\includegraphics[width=0.45\textwidth]{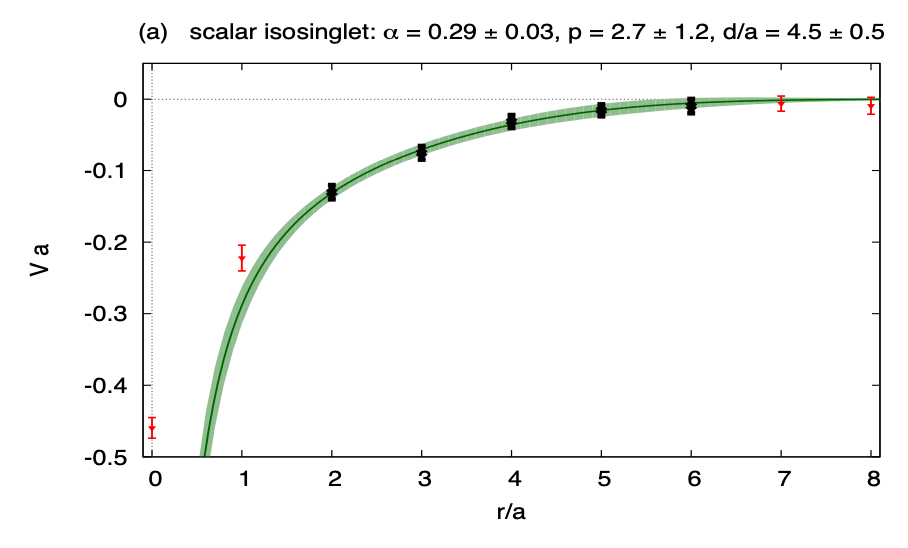}
\includegraphics[width=0.45\textwidth]{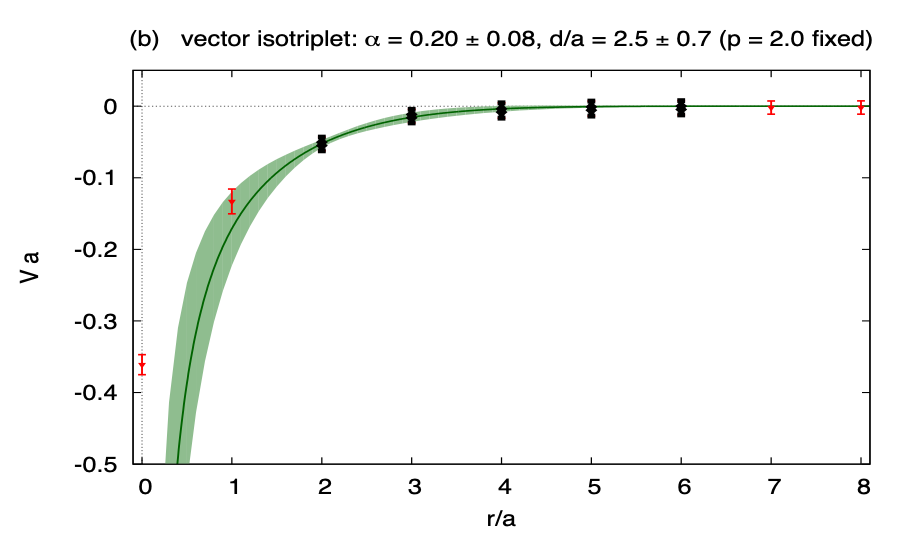}
\caption{\textit{Fitted inter-meson static-static-light-light potentials of \cite{Bicudo:2012qt} for the scalar isosinglet channel (left) and vector isotriplet channel (right). The distances are given in units of the lattice spacing $a\simeq0.079~$fm. Using these potentials the authors were the first to predict a bound $T_{bb}$ with binding energy $E_B\simeq -30(17) ~$MeV to $-57(19)~$MeV.}}
\label{fig:historic-bicudo}
\end{figure}

\paragraph{Determining binding energies from fitted lattice potentials.} A new level is reached in \cite{Brown:2012tm}. The setup of this calculation is in full QCD with $N_f=2+1$ quark flavors on an anisotropic ($\xi=3.5$) lattice with $m_\pi\simeq 380~$MeV, $a_s\simeq 0.12~$fm, $L\simeq 2.88~$fm and $m_\pi L\simeq 5.7$. For the first time, the dominant finite volume systematics are likely under good control with a large enough physical lattice volume and a sufficient number of interaction lengths in $m_\pi L$, all the while including three dynamical quark flavors. The authors employ a quark model inspired ansatz to fit the lattice potential, with the result shown in Fig.~\ref{fig:historic-nf2QCD} (bottom right). The fitted potential is then used as input into a Schr\"odinger equation to determine the ground state energy, also a first in this context to the author's knowledge. The study finds a ground state with $\simeq 50~$MeV binding energy for the $J^P=0^+$ channel, while a statement for the $J^P=1^+$ channel could not be made.

\begin{figure}[t!]
\centering
\includegraphics[width=0.35\textwidth]{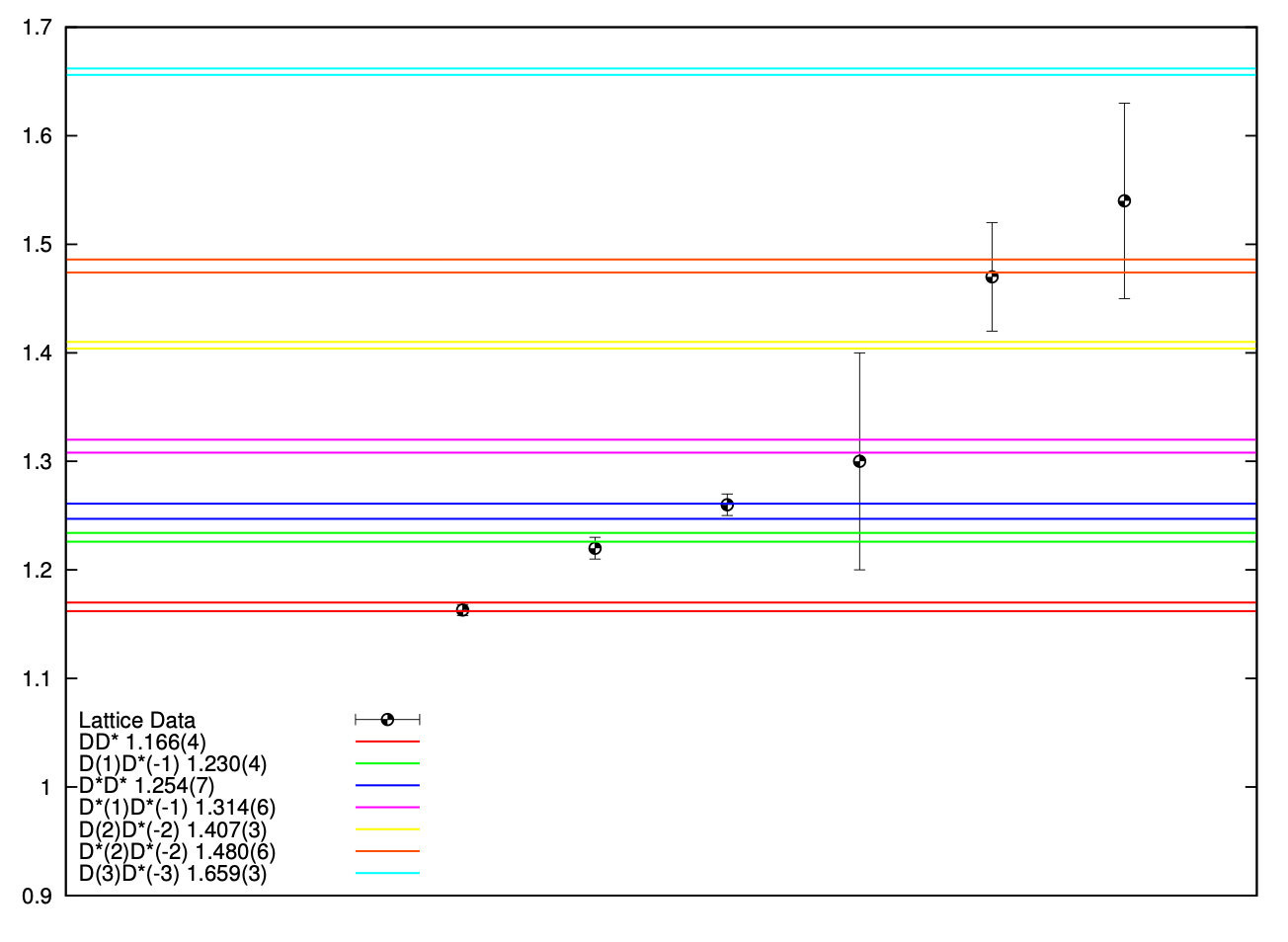}
\includegraphics[width=0.35\textwidth]{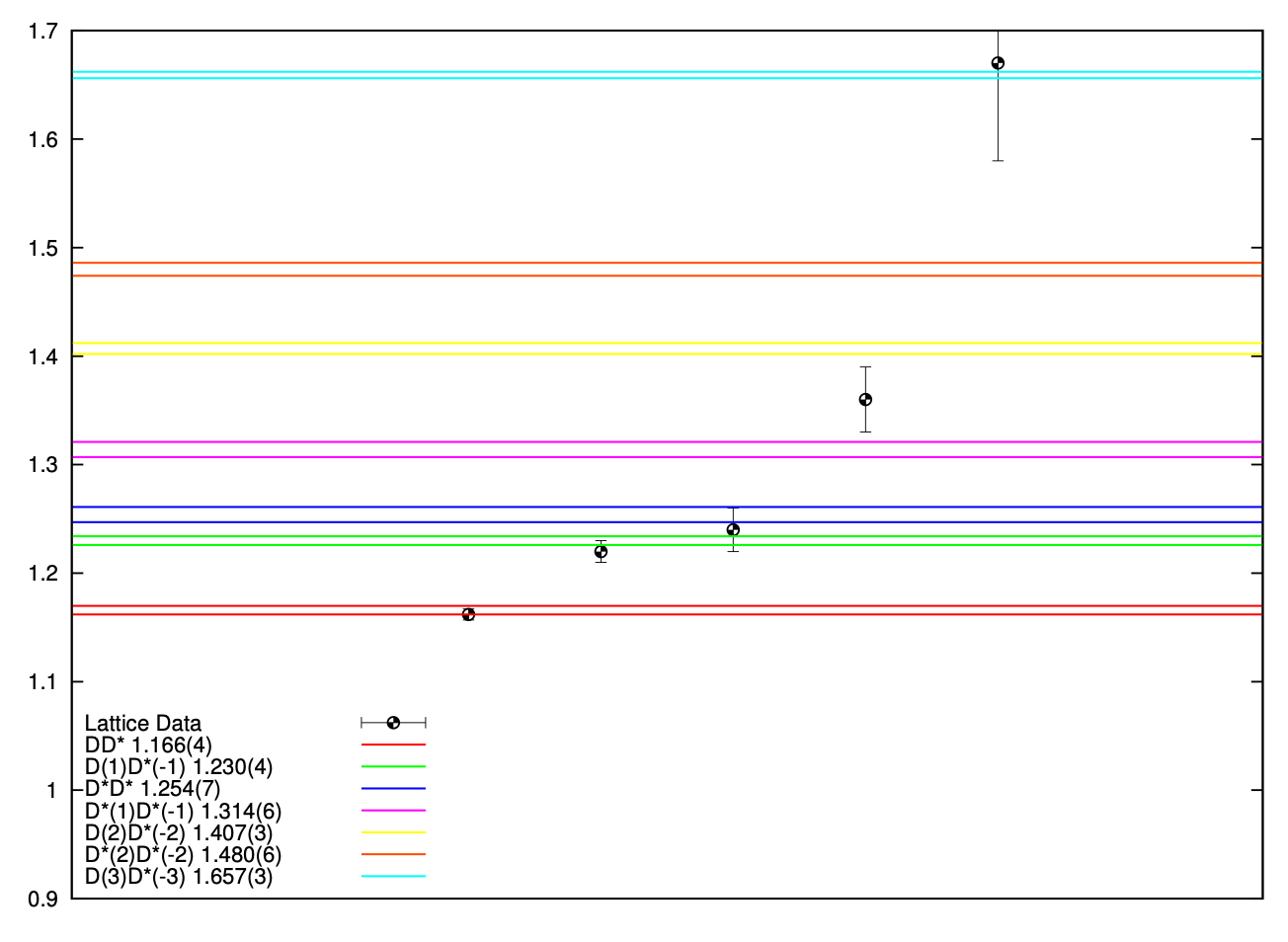}
\caption{\textit{Extracted finite volume spectra of \cite{Guerrieri:2014nxa} for the $T_{cc}^{ud}$ in the $J^P=0^+$ (left) and $J^P=1^+$ (right) channels. The spectrum was determined by estimating the splittings between excited and ground states. They are observed to be consistent with a free spectrum.}}
\label{fig:historic-tantalo}
\end{figure}

In the same year (2012), another milestone study appeared in \cite{Bicudo:2012qt} by Bicudo et al. It is the first paper clearly stating there is evidence for a bound $bb\bar{q}\bar{q}$ tetraquark. The study uses the same lattice setup as \cite{Wagner:2011ev}, on which it determines the static-static-light-light potential, similar to before. This lattice potential data is fitted to a screening potential model ansatz motivated by phenomenology.
The fitted potentials found in this way are shown in Fig.~\ref{fig:historic-bicudo}.
They are then input into a Schr\"odinger equation of the type:
\begin{equation}
H=\frac{\vec p^2}{2\mu} + 2E_0 + V(r)~~,
\end{equation}
where $E_0$ is the ground state energy of an individual $B$-type meson and $\mu$ is the energy of a heavy quark, the bottom quark in this case.
The authors find binding energies $E_B=E - 2E_0$ between $-30(17)~$MeV and $-57(19)~$MeV. Although a discussion on the likely lattice-related systematics on discretization and finite volume effects is included, the required numerical calculations were not available to confirm the discussion. The largest systematic uncertainty that is not lattice-related is the value of the reduced energy of the heavy quark, which remains a caveat of this approach, together with having to specify the fit form of the potential. 

Shortly after, in 2013, the first study of the $T_{cc}^{ud}$ system was published in \cite{Ikeda:2013vwa} using the HALQCD method to determine the inter-meson potentials. The authors use the PACS-CS ensembles \cite{Aoki:2008sm}, with $N_f=2+1$ dynamical quarks, and Tsukuba-type relativistic heavy quark action for the charm quarks \cite{Namekawa:2013vu}. However, the pion masses in this study are limited to the range $m_\pi=415...707~$MeV throughout $L=2.88~$fm and at the lowest pion mass $m_\pi L\simeq 6$. 
In the $J^P=0^+$ channels, they find a repulsive potential that is largely independent of the pion mass. In the $J^P=1^+$ channel, while repulsive in the range of masses available, they find a pion mass dependence, which could indicate a bound state at lower pion masses.
They note that "attractions become more prominent as the pion mass decreases, particularly in the $I = 0$ $D-D^*$ channel corresponding to $T_{cc}(J^P = 1^+, I = 0)$" \cite{Ikeda:2013vwa}.

\paragraph{Early spectrum results with multiple states resolved.} Next, the study \cite{Guerrieri:2014nxa} is one of the first to approach the doubly heavy tetraquark system in a spectrum-based approach and employ multiple interpolating operators to form a  GEVP. The authors use $N_f=2$ ensembles generated by the CLS collaboration with $a=0.075~$fm, $m_\pi\simeq 490~$MeV, $L=2.4~$fm and $m_\pi L\simeq 6$. Using $3\times 3$ and $4\times 4$ correlation matrices they determine the $T_{cc}^{ud}$ spectrum in the $J^P=0^+$ and $1^+$ channels,
their results are shown in Fig.~\ref{fig:historic-tantalo}.
 The observed finite volume energies and differences are consistent with free spectra in both channels.
Given the small binding energy of the $T_{cc}^{ud}$ in nature, the large pion mass and the uncertainties in the data a prediction of it at this stage would have been surprising. 
Nevertheless, the progress made in this study is impressive.

\paragraph{Consolidation of $T_{bb}$ prediction from potentials.} In \cite{Bicudo:2015vta,Bicudo:2015kna} Bicudo et al. further extend their work of \cite{Bicudo:2012qt,Wagner:2011ev} on similar gauge ensembles. Identifying the difficulty of controlling the fit of the potential, especially in the crucial short-distance regime, the studies are extended toward lower lattice spacings. Simultaneously, other flavor combinations of the light quark component are tested. Overall the signal for the $T_{bb}$ state with $I(J^P)=0(1^+)$ is confirmed and the binding energy updated to $E_B=-90(40)~$MeV. The pion mass range evaluated is $m_\pi=340~$MeV to $650~$MeV.

\begin{figure}[t!]
 \centering
 \includegraphics[width=0.48\textwidth]{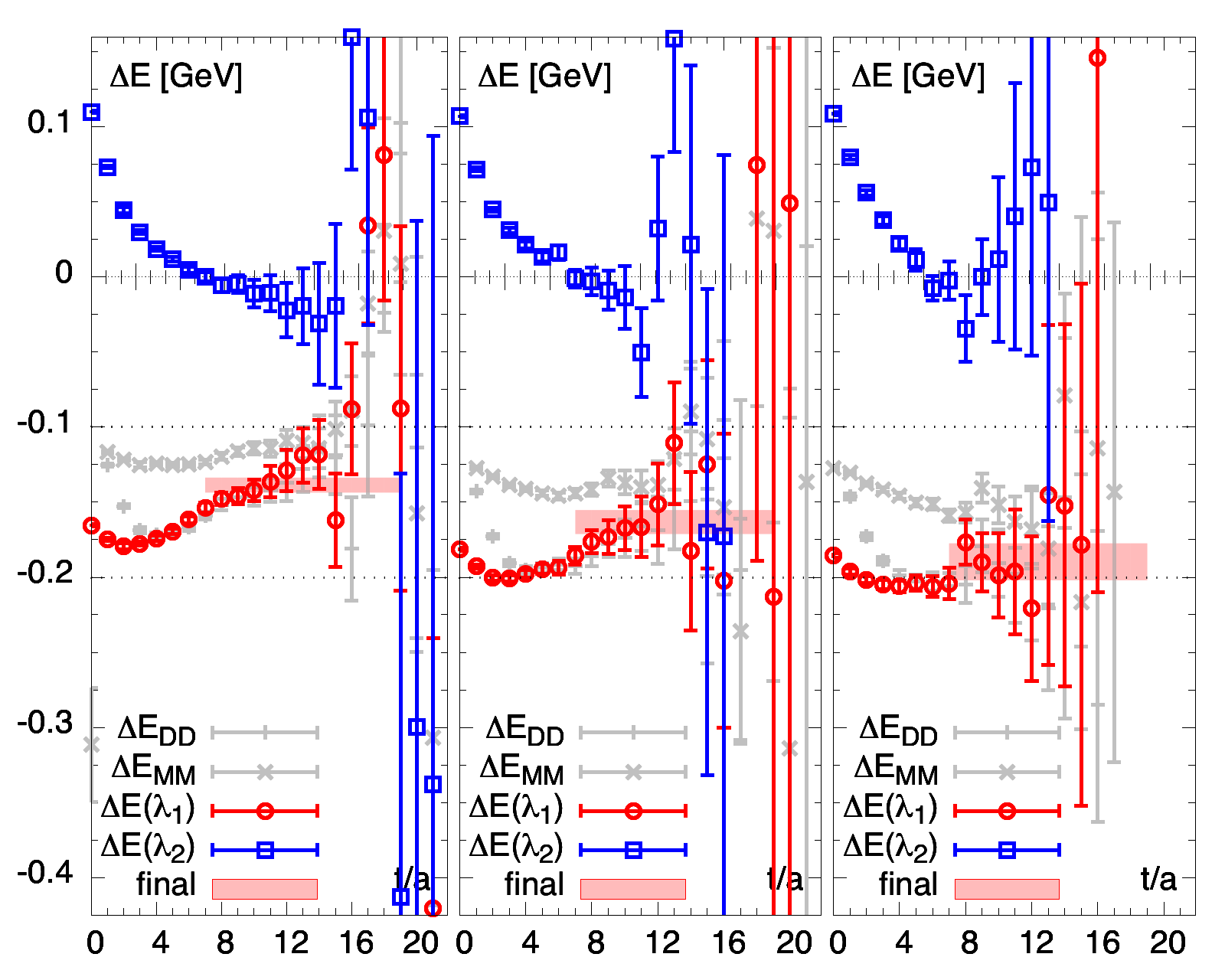}
 \includegraphics[width=0.48\textwidth]{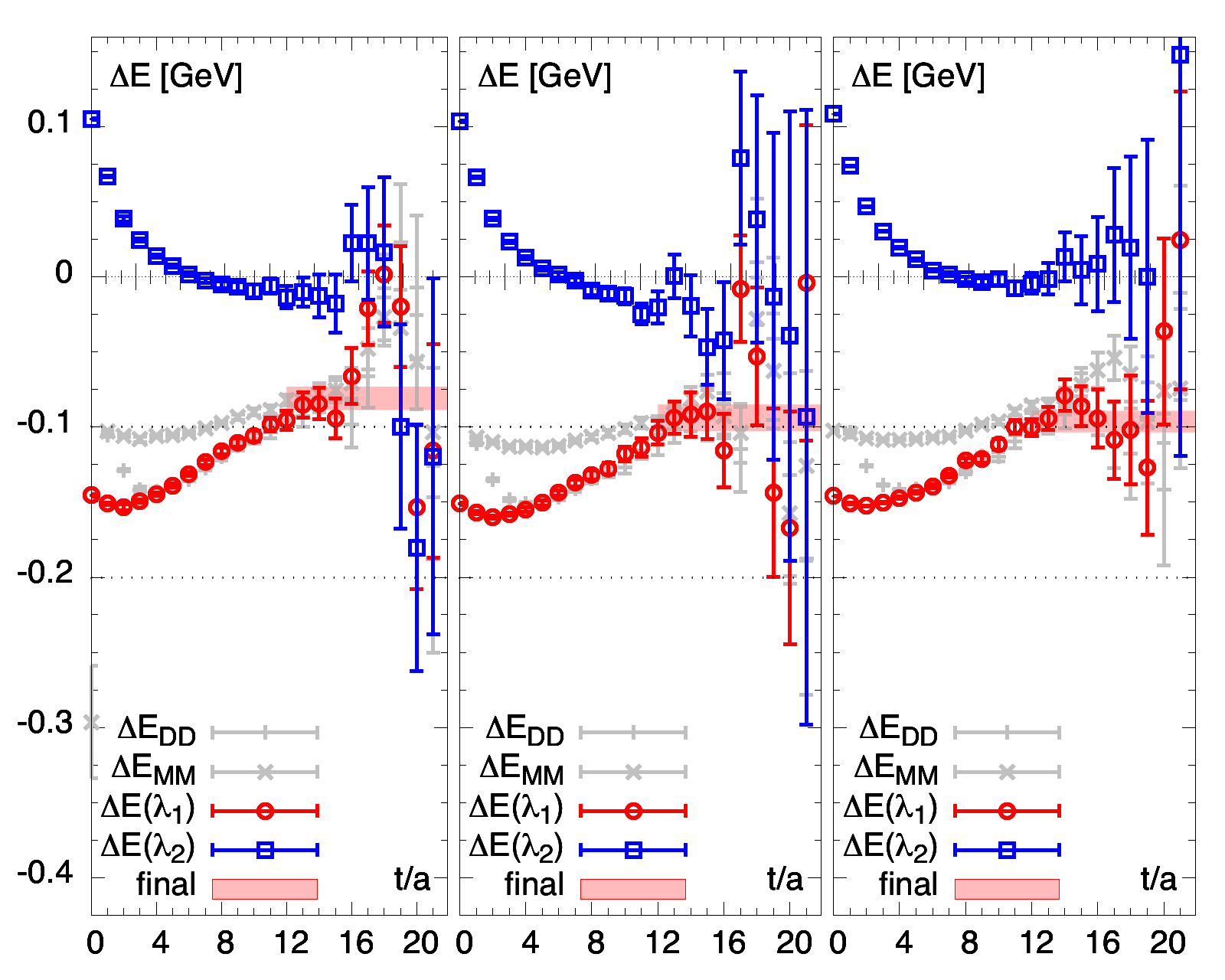}
 \caption{\textit{Effective energies of $I(J^P)=0(1^+)$ $T_{bb}^{ud}$ (left) and $T_{bb}^{us}$ (right) tetraquarks with the $B$ and $B^*$ mesons subtracted by forming a correlator ratio \cite{Francis:2016hui}. Given in red/blue are the ground/excited state energies determined via a $2\times 2$ GEVP. In grey, the results for the correlators on the diagonal of the GEVP are shown. The red band denotes the final quoted energy and the length of the fit window. The three panels of each figure show, from left to right, the results for the pion masses 415~MeV, 299~MeV, and 164~MeV. }}
 \label{fig:historic-toronto1}
 \end{figure}

\paragraph{Prediction of the $T_{bb}$ from the spectrum approach.} We continue this section with the first prediction of the $I(J^P)=0(1^+)$ $T_{bb}^{ud}$ as well as $T_{bb}^{us}$ in the spectrum based approach and at physical pion mass in \cite{Francis:2016hui} by Francis et al. This study uses the lower part of the pion range available through the PACS-CS ensembles \cite{Aoki:2008sm} with $m_\pi=164~,299$ and $415~$MeV. The lattice spacing is $a\simeq0.0899~$fm, $L\simeq 2.88~$fm and the values of $m_\pi L$ are 2.4, 4.4 and 6.1. The bottom quarks are handled through NRQCD based on the implementation developed in \cite{Wurtz:2015mqa,Lewis:1998ka,Lewis:2008fu}, see also Sec.~\ref{sec:heavyquarks}. 

\begin{figure}[t!]
\centering
\includegraphics[width=0.58\textwidth]{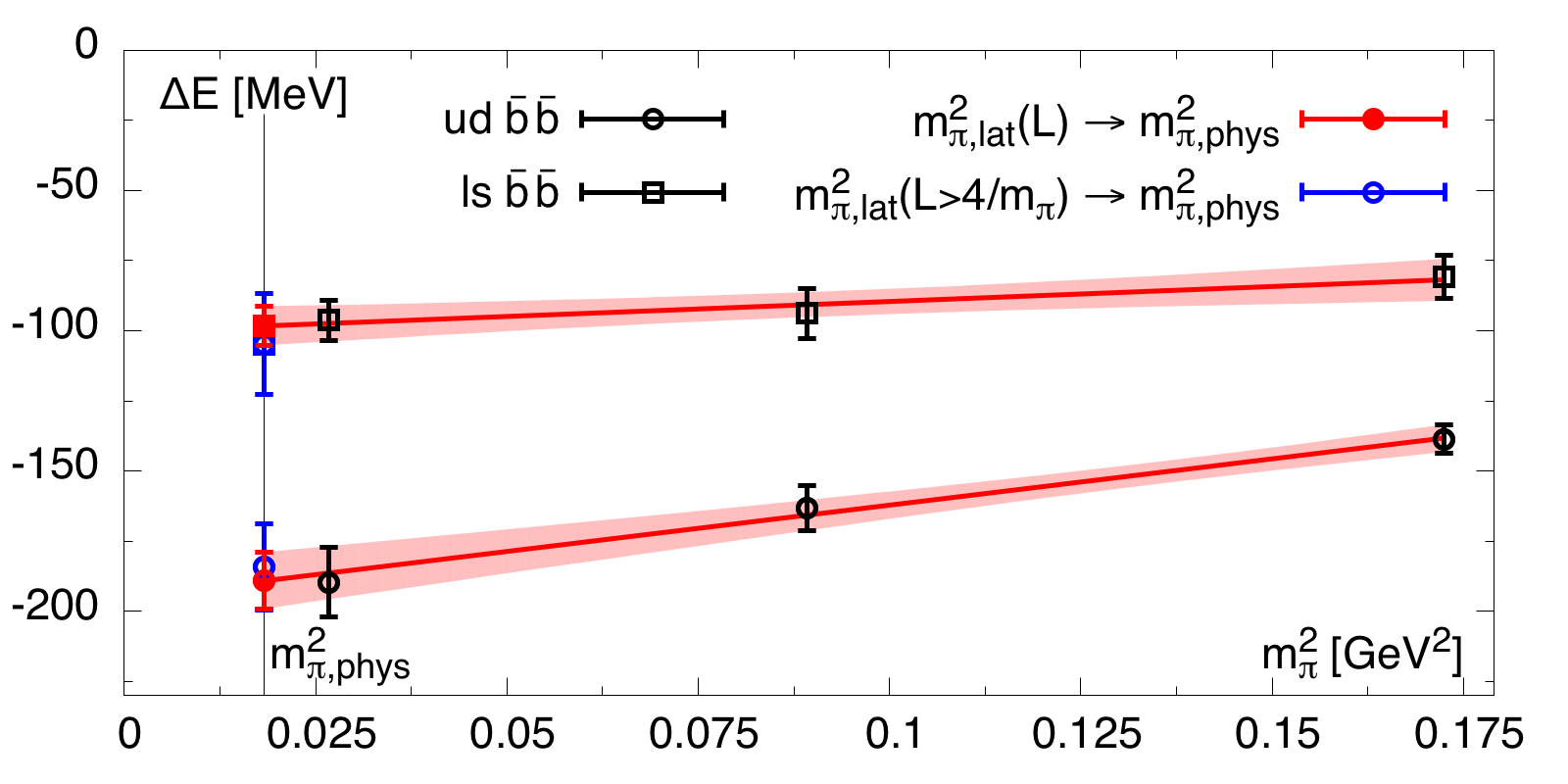}
\caption{\textit{Chiral extrapolations of the $I(J^P)=0(1^+)$ $T_{bb}^{ud}$ and $T_{bb}^{us}$ tetraquark binding energies performed in \cite{Francis:2016hui}. The physical point results using two possible cuts are given as red and blue points.}}
\label{fig:historic-toronto2}
\end{figure}

Coulomb gauge fixed wall sources are used for the propagators to enhance the signal and contracted to form wall-local correlation functions. As explained previously, this implies the asymptotic, ground state energy of the correlation functions is reached from below, introducing a possible systematic.
The results are derived from a $2\times 2$ correlation matrix with local di-meson and diquark-antidiquark interpolating operators. The two flavor channels are labeled $ud\bar{b}\bar{b}$ and $\ell s\bar{b}\bar{b}$, where the $\ell=u,d$ is introduced to highlight the fact that the lattice calculations use two mass degenerate light quarks. This is true of all lattice calculations reported in this review and is unlike in nature since this means the up and down quarks are isospin symmetric. In the following we will drop this convention and refer to the usual naming convention under the assumption of isospin symmetry.
It is reasonable to expect isospin breaking effects to be negligible compared to the dominating binding effects, with the notable exception of the $T_{cc}^{ud}$, where the shallow binding, with binding energy $E_B\simeq 300~$keV, could necessitate including isospin breaking correction. 

The results for the ground state and first excited state energies in both channels and for all three pion masses are shown in Fig.~\ref{fig:historic-toronto1}. Note that in this study, the correlator ratio of the tetraquark correlator over the combined meson correlators was analyzed. One reason this was introduced is to cancel the NRQCD mass shift at the level of the observable. Furthermore, it was hoped that this would reduce excited state contamination. As explained previously, this is not necessarily the case and can complicate the analysis. Also, the NRQCD mass shift can be canceled in post-processing since it is the same for the tetraquark and combined meson masses.
Irrespective of this complication, the authors' stability analysis gave robust indications for bound $I(J^P)=0(1^+)$ $T_{bb}^{ud}$ and $T_{bb}^{us}$ tetraquarks at physical quark masses. The results in Fig.~\ref{fig:historic-toronto2} show the chiral extrapolation of the observed binding energies to the physical point.
The final quoted binding energies are $E_B(T_{bb}^{ud})\simeq -189(13)~$MeV and $E_B(T_{bb}^{us})\simeq -98(10)~$MeV.

\begin{figure}[t!]
\centering
\includegraphics[width=0.48\textwidth]{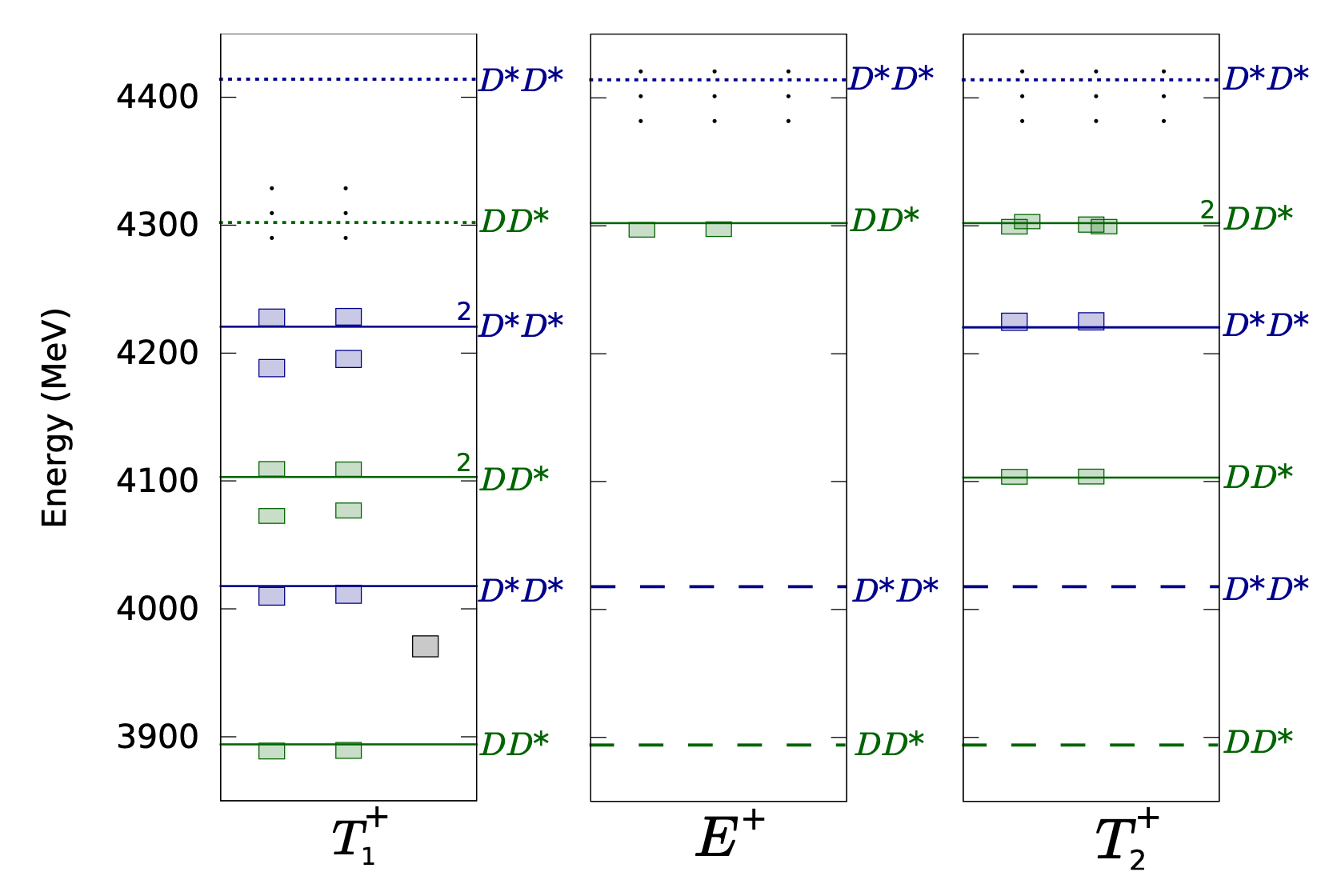}
\includegraphics[width=0.48\textwidth]{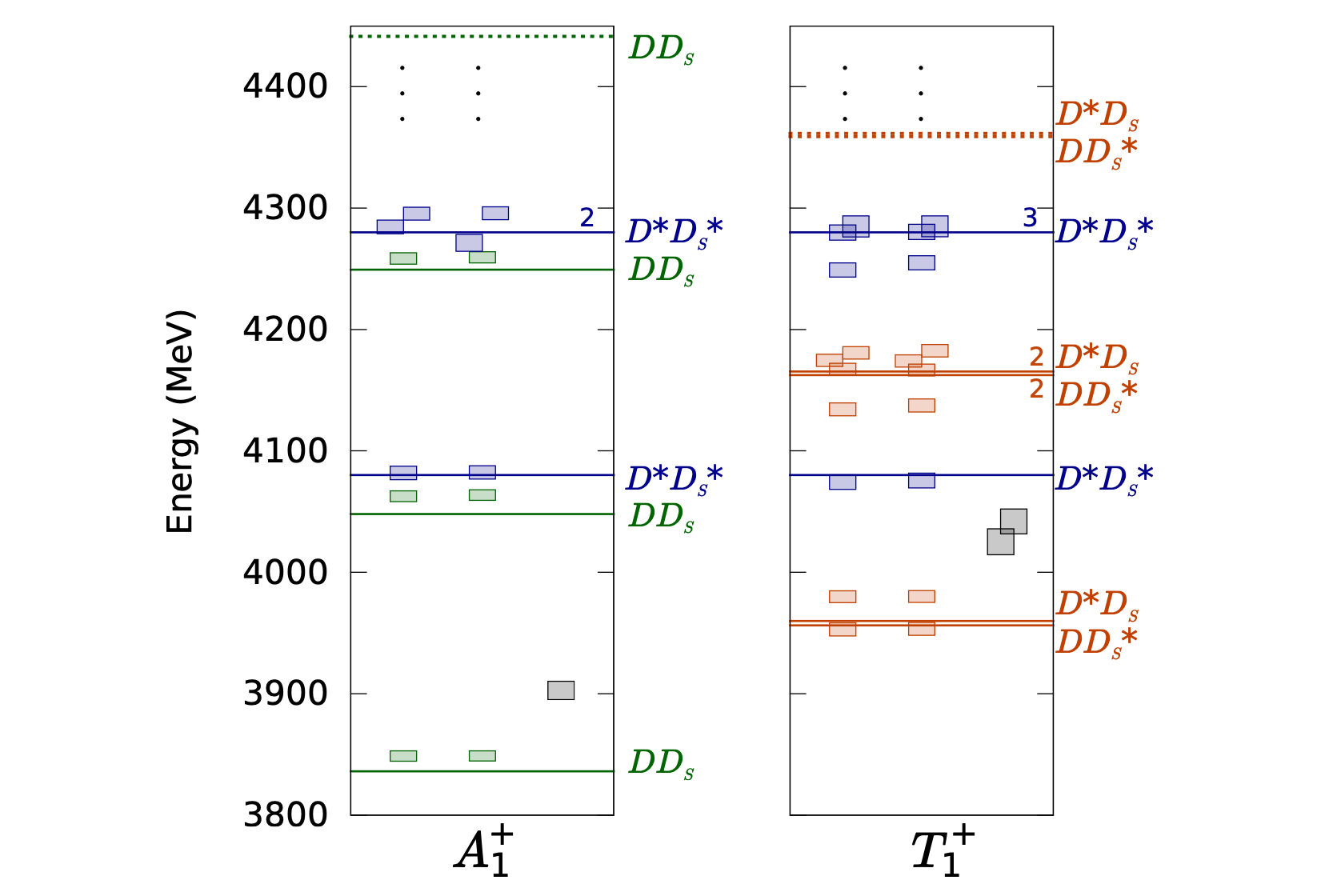}
\caption{\textit{Finite volume spectra for $I(J^P)=0(1^+)$ $T^{ud}_{cc}$(left) 
and $T^{us}_{cc}$(right) tetraquarks determined in \cite{Cheung:2017tnt}. The energies extracted from large GEVP analyses in different irreps are shown. This study represented the first extensive analysis of the $T_{cc}$ finite volume spectrum with multiple excited states on the lattice exploiting an extensive interpolating operator basis.}}
\label{fig:historic-hadspec-1}
\end{figure}

\paragraph{Extended finite volume spectroscopy for $T_{QQ}$ and results on $T_{cc}$.} Shortly after this work on $T_{bb}$-type tetraquarks also an extension  of the study of the $T_{cc}$ was published by the Hadron Spectrum collaboration \cite{Cheung:2017tnt}. In this further milestone calculation, the authors used an anisotropic lattice ($\xi=3.5$) with $m_\pi=391~$MeV, $a_s\simeq 0.12~$MeV, $L\simeq 1.92~$fm, $m_\pi L\simeq 3.8$. The key advancements are the construction of a broad basis of interpolating operators and the extensive use of the distillation method. A basis of up to a total of ten interpolating operators is made available for the study of the $I(J^P)=0(1^+)$ $T_{cc}$-type tetraquarks in a given channel and irrep. 
The results for the finite volume spectra of $T^{ud}_{cc}$ and $T^{us}_{cc}$ are shown in the left and right panels of Fig.~\ref{fig:historic-hadspec-1}. The identified levels show that an extra level is not present. At the same time, a significant distortion of the spectrum is not observed. However, as the authors conclude, "our doubly-charmed [...] spectrum is not inconsistent with there being an attractive interaction although there were no obvious signs of a bound state in this channel" \cite{Cheung:2017tnt}. As the previous discussion on the Breit-Wigner toy model highlighted, a bound state spectrum with small binding energy, and uncertainties larger than it, can be difficult to pin down even with many states available. Given the heavy quark masses involved in addition, a prediction of the $T^{ud}_{cc}$ again would have been surprising at this stage. Another outcome of this study, used later, is that including diquark-antidiquark operators did not significantly affect the determined finite volume spectrum.

\subsection{Flavor surveys, systematics improvement and first scattering phase shifts}

The combined effort and findings of the studies gathered in the previous section prepare the stage for the quantitative survey of $T_{QQ}$-type tetraquarks. The studies of \cite{Bicudo:2012qt} and \cite{Ikeda:2013vwa} pushed the boundaries of the potential-based approach by expanding it towards extracting also ground state energies or scattering parameters. 
At the same time, the studies \cite{Francis:2016hui} and \cite{Cheung:2017tnt} lifted the quality of the discussion to a new level of systematically improvable precision through the spectrum-based approach and defining a large set of interpolating operators to be used for the reliable extraction of the finite volume energy levels.
In the potential-based approach, the next step is to improve upon the systematic uncertainties, while in the spectrum-based approach, the path forward is to perform the finite volume scattering analyses to determine and interpret the scattering parameters using finite volume quantization conditions.
Both approaches should simultaneously survey and study the mass dependence of all possible flavor combinations in order to map out and give mass estimates to all thinkable doubly heavy tetraquark candidates.

\paragraph{Including spin-spin interactions and $T_{bb}$ resonances in the potential approach.} The studies of Bicudo et al. \cite{Bicudo:2016ooe,Bicudo:2017szl} followed this path and presented extensions on the tetraquark potentials, extending the work of \cite{Bicudo:2012qt,Bicudo:2015vta,Bicudo:2015kna}. After including spin-spin effects in the potential, the extended analysis confirms the $I(J^P)=0(1^+)$ $T^{ud}_{bb}$ bound state with a binding energy of $E_B\simeq -59(34)~$MeV. They further explore the possibility of tetraquark resonances and find evidence for an $I(J^P)=1(1^-)$ $T^{ud}_{bb}$ resonance that would decay into two $B$ mesons. A confirmation from a spectrum-based calculation is not yet available.

\paragraph{Flavor and mass surveys in the spectrum approach.} Continuing with the $T^{qq}_{QQ}$ tetraquarks on the spectrum approach side, the two studies by Francis et al. \cite{Francis:2018jyb} and Junnarkar et al. \cite{Junnarkar:2018twb} expand the surveyed candidates significantly. Focusing first on the latter, it follows the methodology laid out by Francis et al. \cite{Francis:2016hui} and also \cite{Francis:2018jyb}, i.e., they use Coulomb gauge fixed wall sources with local sinks to define $2\times 2$ correlation matrices and determine the ground state energy via GEVP method. Acknowledging that the GEVP is not symmetric, Junnarkar et al. use the principle correlator variant of the standard GEVP method to account for the non-Hermiticity thus introduced.
The lattice setup uses the staggered MILC ensembles \cite{MILC:2012znn}, which feature a broad range of available lattice spacings $a\simeq 0.058...0.12~$fm and $m_\pi\simeq 153...689~$MeV, whereby not all lattice spacings feature all pion masses. In the valence sector overlap quarks are used, which are automatically $\mathcal{O}(a)$ improved. The study boasts both charm and bottom quarks, where the first is handled using the Fermilab-type relativistic effective heavy quark action and the latter via NRQCD.
Overall, the authors survey different flavor combinations of candidates in the $J^P=1^+$ and $J^P=0^+$ channels with $bb$ and $cc$ heavy flavor content. They finally arrive at continuum extrapolated results for the binding energies of eleven tetraquarks.
They find none of the $0^+$ ground state energies below the non-interacting thresholds. The finalized results are gathered in Tab.~\ref{tab:nilmani-1}. Interestingly, they also find a bound $T_{cc}^{ud}$ with a binding energy of $E_B=-23(11)~$MeV.

\begin{table}
\centering
\begin{minipage}{0.49\textwidth}
\centering
\begin{tabular}{cc|cc}
\hline \hline 
State & $\Delta E^{1^+}[\mathrm{MeV}]$ & State & $\Delta E^{1^+}[\mathrm{MeV}]$ \\
\hline \hline$u d \bar{b} \bar{b}$ & $-143(34)$ & $u s \bar{b} \bar{b}$ & $-87(32)$ \\
\hline$u c \bar{b} \bar{b}$ & $-6(11)$ & $s c \bar{b} \bar{b}$ & $-8(3)$ \\
\hline$u d \bar{c} \bar{c}$ & $-23(11)$ & $u s \bar{c} \bar{c}$ & $-8(8)$ \\
\hline \hline
\end{tabular}
\end{minipage}
\begin{minipage}{0.49\textwidth}
\centering
\begin{tabular}{cc|cc}
\hline \hline State & $\Delta E^{0^+}[\mathrm{MeV}]$ & State & $\Delta E^{0^+}[\mathrm{MeV}]$\\
\hline \hline$u u \bar{b} \bar{b}$ & $-5(18)$ & $u u \bar{c} \bar{c}$ & $26(11)$ \\
\hline$s s \bar{b} \bar{b}$ & $3(9)$ & $s s \bar{c} \bar{c}$ & $14(4)$ \\
\hline$c c \bar{b} \bar{b}$ & $16(1)$ & & \\
\hline \hline
\end{tabular}
\end{minipage}
\caption{\textit{Finalized results of the first broad survey study \cite{Junnarkar:2018twb} in the $J^P=1^+$ (left) and $J^P=0^+$ (right) channels. The finalized results are extrapolated both to the physical point and continuum limit.}}
\label{tab:nilmani-1}
\end{table}

\begin{figure}[t!]
\centering
\includegraphics[width=0.58\textwidth]{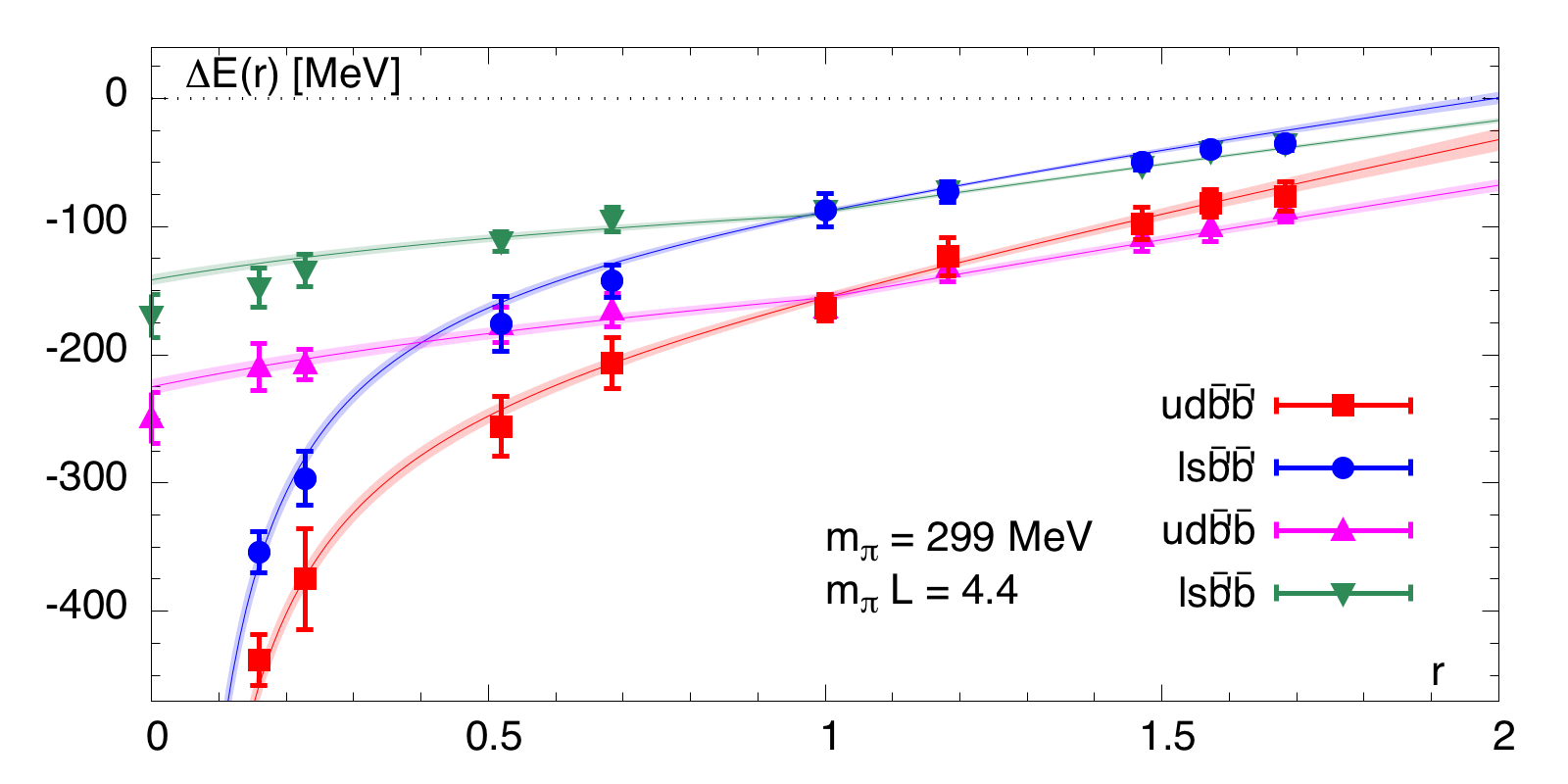}\\
\caption{\textit{The dependence on the heavy-quark mass ratio,
$r=m^b_{\rm bare}/m^{b'}_{\rm bare}$, of the binding energies for the $b{b'}\bar{u}\bar{d}$, $b{b'}\bar{u}\bar{s}$, $bb\bar{u}\bar{d}$ and $b{b'}\bar{u}\bar{s}$ channels with $J^P=1^+$ presented in \cite{Francis:2018jyb}.}}
\label{fig:francis-2}
\end{figure}

Simultaneously, Francis et al. \cite{Francis:2018jyb} expanded the survey in an orthogonal way by extending the candidates studied from $T_{QQ}$ to $T_{QQ'}$ heavy quark components. The study consists of two parts: First, the mass of the heavy quarks is varied freely within the validity bounds of NRQCD towards infinitely heavy quarks and the charm quark region. The configurations used are still the ones of the PACS-CS ensembles, i.e., there is a single lattice spacing $a=0.0899~$fm, which implies a lower bound on the NRQCD heavy quarks of $am_Q\simeq 1=2.19~$MeV. As such, the charm region cannot be reached in this setup. Within the range of validity, however, the authors survey the $J^P=1^+$ candidates with flavors $b{b'}\bar{u}\bar{d}$, $b{b'}\bar{u}\bar{s}$, $bb\bar{u}\bar{d}$ and $b{b'}\bar{u}\bar{s}$, that is they include flavor-off diagonal heavy components. 
The resulting heavy quark mass dependence is shown in Fig.~\ref{fig:francis-2}; we observe that all candidates are bound in the regime of $r=m_b/m_{b'}<1.7$. Based on this study, and assuming $m_b\simeq 4.18~$GeV, this implies all $I(J^P)=0(1^+)$ $T_{{b'}{b'}/b{b'}}^{ud/us}$ with $m_{b'}\gtrsim 2.46~$GeV are (deeply) bound. In an attempt to describe the data, a model is motivated based on the general properties of the color Coulomb potential and the diquark-HQSS picture. It has five parameters in total and permits fitting all four channels:
\begin{align}
\Delta E_{ud\bar{b}^\prime\bar{b}^\prime}&={\frac{C_0}{2r}}\, +\, C_1^{ud} \, +\, C_2^{ud}\left( 2r\right)\, +\, (23\ {\rm MeV})\, r\\
\Delta E_{\ell s\bar{b}^\prime\bar{b}^\prime}&={\frac{C_0}{2r}}\, +\, C_1^{\ell s} \, +\, C_2^{\ell s}\left( 2r\right)\, +\, (24\ {\rm MeV})\, r\nonumber\\
\Delta E_{ud\bar{b}^\prime\bar{b}}&={\frac{C_0}{1+r}}\, +\, C_1^{ud} \, +\, C_2^{ud}\left( 1+r\right)\, +\, \left( 34\ {\rm MeV} -11\ {\rm MeV}\, r\right),~~r=m_{b}/m_{b'}<1\nonumber\\
\Delta E_{\ell s\bar{b}^\prime\bar{b}}&={\frac{C_0}{1+r}}\, +\, C_1^{\ell s} \, +\, C_2^{\ell s}\left( 1+r\right)\, +\, \left( 34\ {\rm MeV} -12\ {\rm MeV} r\right),~~r=m_{b}/m_{b'}<1\nonumber\\
\Delta E_{ud\bar{b}^\prime\bar{b}}&={\frac{C_0}{1+r}}\, +\, C_1^{ud} \, +\, C_2^{ud}\left( 1+r\right)\, +\, \left( 34\ {\rm MeV}\, r -11\ {\rm MeV}\right),~~r=m_{b}/m_{b'}>1\nonumber\\
\Delta E_{\ell s\bar{b}^\prime\bar{b}}&={\frac{C_0}{1+r}}\, +\, C_1^{\ell s} \, +\, C_2^{\ell s}\left( 1+r\right)\, +\, \left( 36\ {\rm MeV}\, r -11\ {\rm MeV}\right),~~r=m_{b}/m_{b'}>1~~.\nonumber
\end{align}
The shaded lines in the figure show the results of this fit. Overall, good agreement with this model is observed, and the study demonstrates how one can combine phenomenological intuition and the capabilities of the lattice method to deviate from physical quark mass inputs in order to gain new insights into the qualitative features of the heavy hadron spectrum.

\begin{figure}[t!]
\centering
\includegraphics[width=0.78\textwidth]{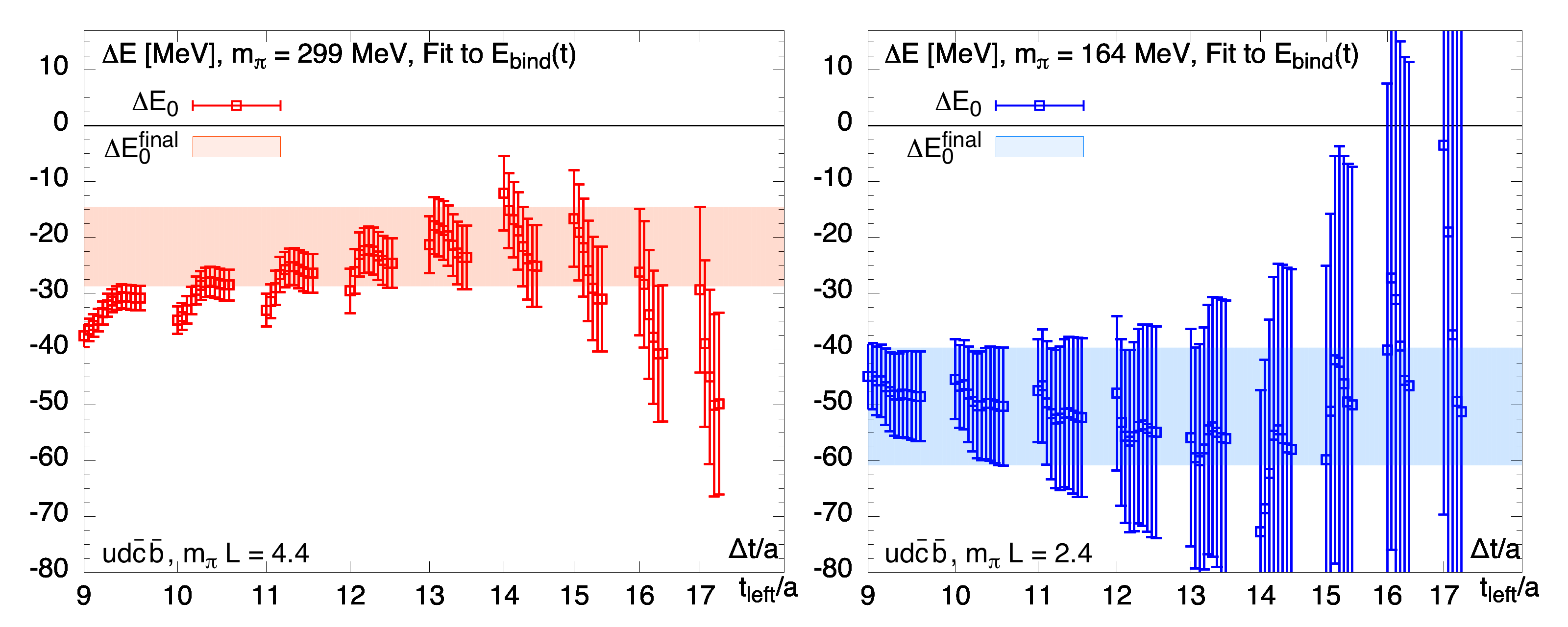}
\caption{\textit{Fit window dependence for fitting the GEVP ground state solutions of the correlation matrix made up of the ratio of the tetraquark and two meson correlator data presented in \cite{Francis:2018jyb}. All correlators were derived using wall-local correlation functions. The final results are quoted from the stable regions of this ensemble of fits.} }
\label{fig:francis-3}
\end{figure}

\paragraph{First study of $T_{bc}$.} A further innovation implemented in this study, for this group of authors, is the inclusion of a Tsukuba-type relativistic effective heavy quark action in the valence sector to determine charm quark propagators. This removes the previous limitation of not being able to probe this sector and enables the study of the $I(J^P)=0(1^+)$ $T_{bc}^{ud}$ tetraquark. Adding a new interpolating operator that becomes available due to the flavor off-diagonal heavy quark component, the authors construct a $3\times 3$ GEVP with wall-local correlation functions. The lattices used are the same as before, and the binding energies on the $m_\pi=299$ and 164~MeV ensembles are shown in Fig.~\ref{fig:francis-3} in the left and right panels, respectively. Note that the authors performed a study of both the ratio of correlators, as before, and of the tetraquark and meson energies individually. The results in the figure show the difference between the latter, where the fit window systematic shown is that of the tetraquark fit.
Later, in 2020, the study \cite{Hudspith:2020tdf} by the same authors re-examined the systematic associated with the wall-local correlation functions and their approach to the asymptotic plateau. By introducing wall-smeared correlators and after performing a new analysis on a larger volume, the authors observe a ground state energy consistent with either no bound state or a very shallow bound state within errors.

\paragraph{Systematics in the spectrum approach and wall-smeared correlators.}
\label{sec:wall-smear-1}
It was quickly realized that the wall-local correlation functions used previously and their systematic approach to the asymptotic energy plateaus from below is a significant issue. Furthermore, determining the finite volume spectra from non-symmetric GEVPs is problematic and needs to be consistently addressed. 

A potential solution to both problems is the use of distillation, especially since it was already successfully demonstrated in the $T_{cc}$ channel in \cite{Cheung:2017tnt}.
However, the method represents an advanced technology and requires a considerable investment in its development, alongside a significant up-front investment to determine the perambulators. Today, the method is becoming more widely available, but that was not the case during the period currently under review. 
Furthermore, it is most naturally applied to effective relativistic heavy quark actions over NRQCD\footnote{We note in passing unpublished work by Nilami Mathur on distillation with NRQCD for which results looked promising \cite{priv-comms-Nilmani}.}. The former would imply a new tuning of the heavy quark action parameters, which again is an investment beyond many groups at the time. 
In comparison, the wall-local approach is comparatively cheap and easy to run and implement. Furthermore, its extension to a wall-nonlocal approach is straightforward because any nonlocal sink operator construction can achieve this in principle. The problem of the non-Hermitian GEVP is not addressed in this way; however, the locality of the sink operator is. 

\begin{figure}[t!]
\centering
\includegraphics[width=0.68\textwidth]{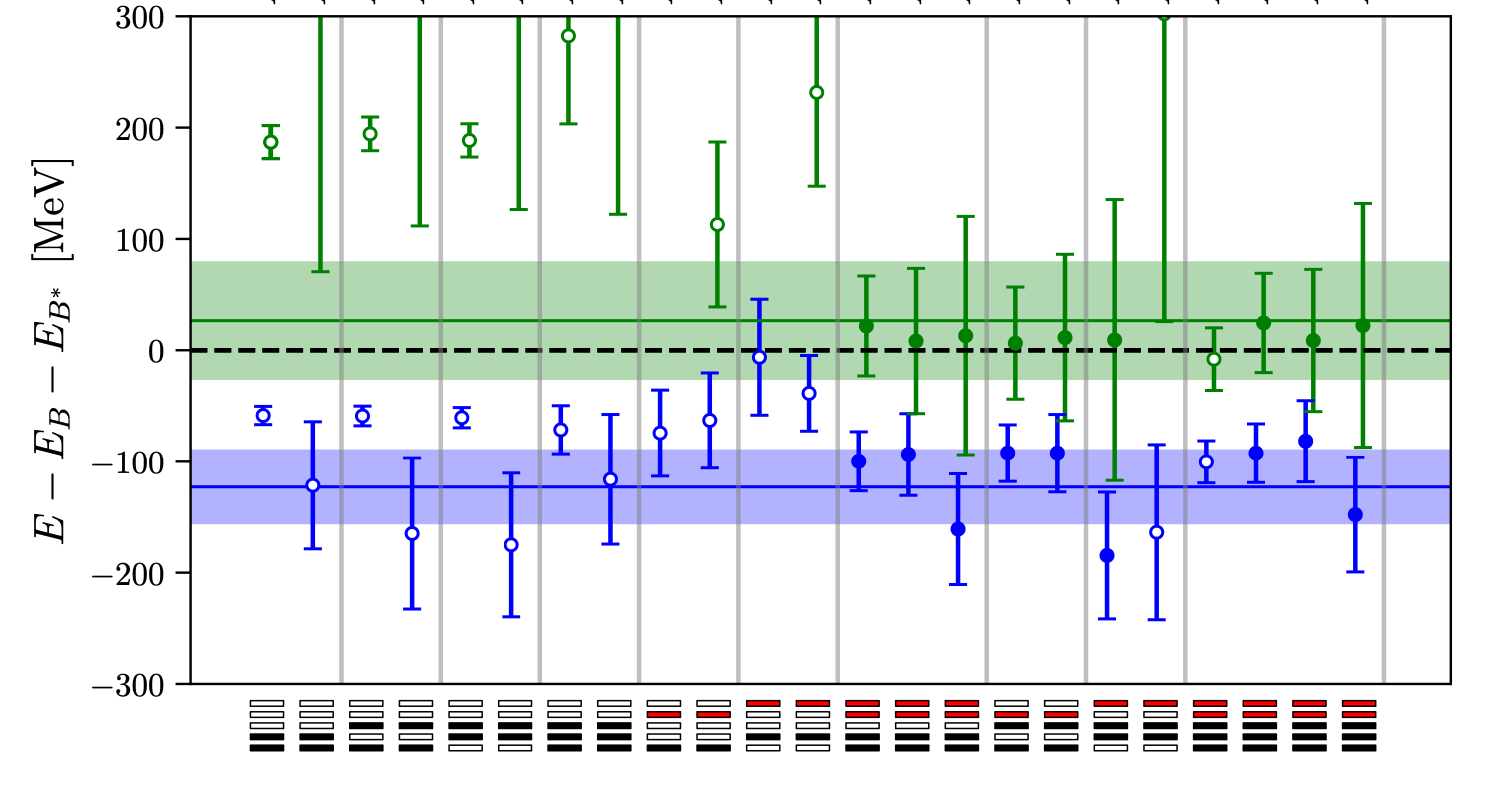}
\includegraphics[width=0.68\textwidth]{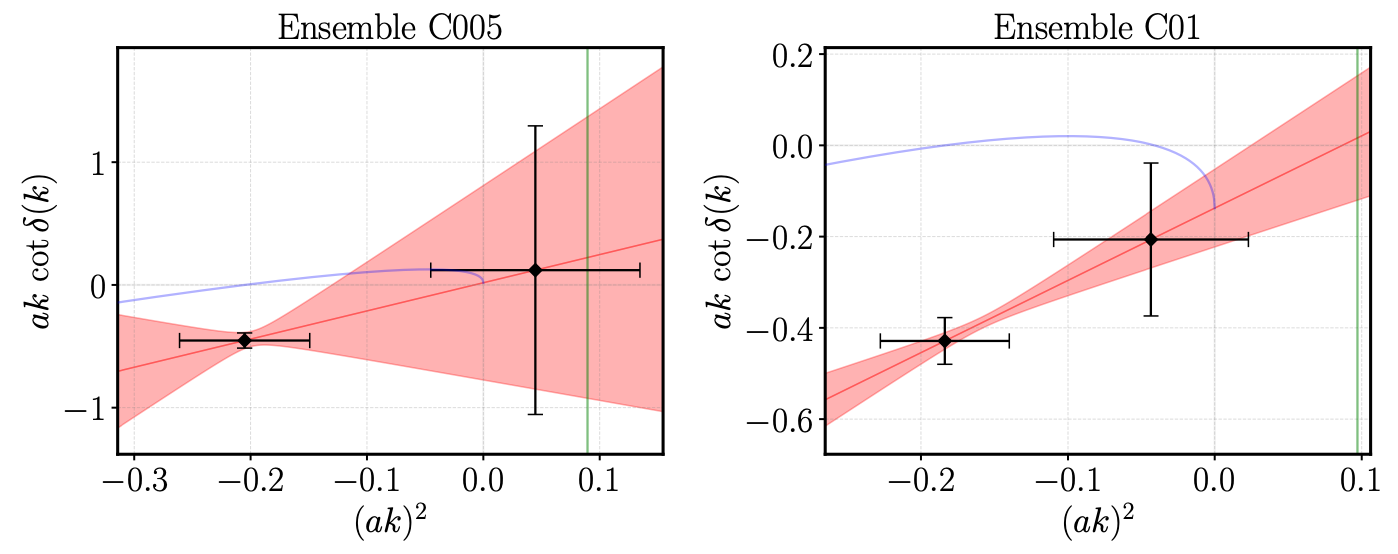}
\caption{\textit{Top: Finite volume energies of the $J^P=1^+$ $T_{bb}^{ud}$ tetraquark determined from local-local and local-smeared correlation functions using a multi-exponential fit approach on the ensemble labeled C005 ($m_\pi=340~$MeV) in \cite{Leskovec:2019ioa}. Each column represents a different fit window, the labeling of which has been removed here. The five bars below the $x-axis$ indicate which operators were used for the analysis. There are five possible sink operators, and a filled black bar denotes a local operator at the sink, while a filled red bar denotes a nonlocal one. Bottom: Extracted scattering phase shifts from the two determined finite volume energies, left on ensemble C005 ($m_\pi=340~$MeV) and right on ensemble C01 ($m_\pi=431~$MeV). }} 
\label{fig:leskovec-2019}
\end{figure}

At the same time, one way to remove the systematic from the asymptotic state being reached from below is to use local source operators instead. An approach in this direction studying the $I(J^P)=0(1^+)$ $T_{bb}^{ud}$ tetraquark was first reported on in \cite{Leskovec:2019ioa}. In this study, the authors use $N_f=2+1$ gauge configurations with Domain Wall fermions both in the sea and the valence sector for the light and strange quarks, while NRQCD is employed for the valence bottom quark. A correlation matrix that includes local and nonlocal sinks is set up using standard point-to-all propagators. Instead of solving the non-symmetric GEVP, multi-exponential fits are performed with:
\begin{equation}
G_{ij}(t) \sim \sum_n Z_i^n Z_j^n e^{-E_n\,t}
\end{equation}
where $n$ denotes the $n$-th energy level and the $Z_{i/j}^n$ the overlap factors of the interpolating operators at the source ($i$) and sink ($j$), respectively. Fitting different combinations of these operators simultaneously to constrain the first two energies in the spectrum, the authors arrive at the results of Fig.~\ref{fig:leskovec-2019}. Each column represents a different fit window, whereby the labeling has been removed for this review. The five bars below the $x$-axis indicate which operators were used for the analysis. There are five possible sink operators, and a filled black bar denotes a local operator at the sink, while a filled red bar denotes a nonlocal one. Although the results depend rather heavily on the choice of fit-window and operator combination, a robust result for a binding energy of $E_B\simeq -128(34)~$MeV is found after extrapolation to the physical point at fixed lattice spacing.\\
A further noteworthy innovation of this study is the attempt at a scattering analysis by converting the two observed energies into scattering phase shifts via the finite volume quantization condition for 2-to-2 scattering. As discussed in the Breit-Wigner toy model, it is crucial to determine the excited state energies with high precision to constrain the effective range expansion. The two examples in the bottom panels of Fig.~\ref{fig:leskovec-2019} show the situation in \cite{Leskovec:2019ioa}. Especially in the lighter ensemble, labeled C005 with $m_\pi=340~$MeV, the data cannot sufficiently constrain the effective range expansion to determine $r_0$ with significance.

In keeping with Coulomb gauge fixed wall sources, due to their advantageous signal-to-noise properties, in \cite{Hudspith:2020tdf}, the authors expanded the wall-source-local-sink approach to one in which the sinks are smeared. As such, the authors "take advantage of the gauge condition and construct propagators by summing over points lying inside a sphere [of radius $R$] around each reference sink point $x$" \cite{Hudspith:2020tdf}.
%$$
%S^B(x, t)=\frac{1}{N} \sum_{r^2 \leq R^2} S(x+r, t) .
%$$
Referring to these sinks as "box-sinks" the radius $R$ of the sphere being summed over controls the amount of smearing or nonlocality of the sink operator. In this construction the limits $R^2=0$ and $R^2=3(L/2)^2$  correspond to the standard wall-local and wall-wall correlators. Choosing a value of $R^2$ in between interpolates between these two cases, one can tune it to maximize the length of the effective mass plateau, which was done here, or to balance between approaching the asymptotic value from above and the noise.
The GEVP in this approach is still non-symmetric. However, the authors carefully monitor this systematic by evaluating the real and imaginary parts of the determined eigenvalues. In the symmetric case, the imaginary parts should be zero.
Simultaneously to this and other technical updates, the authors also carry out their analysis on a newly generated gauge ensemble with $m_\pi=192~$MeV, $a=0.0899~$fm, $L\simeq 4.3~$fm and $m_\pi L\simeq 4.2$. The setup and parameters chosen are based on those of the previously used PACS-CS ensembles, and the new configurations represent an extension of them. The results are summarized in Tab.~\ref{tab:hudspith-1} and show no further (deeply) bound states within uncertainties apart from the previously predicted $J^P=1^+$ $T_{bb}^{ud}$ and $T_{bb}^{us}$. The study confirms the results on the triply-heavy tetraquarks $T_{bb}^{uc}$ and $T_{bb}^{sc}$ of \cite{Junnarkar:2018twb} and also finds that they are not deeply bound.
In particular, in this larger volume and with better control over the systematics, the prediction of the $T_{bc}^{ud}$ could not be confirmed. Given the larger than physical pion mass and missing chiral extrapolation, this does not rule out this state per se, but a deep binding of more than $E_B \lesssim -15~$MeV already at $m_\pi=299~$MeV is not consistent with this updated analysis.

\begin{table}[t!]
\centering
\begin{tabular}{lcr}
\hline \hline Type $(\psi \phi \theta \omega)$ & $I(J)^P$ & $E_B$\\
\hline$u d c b / u d s b / u d s c$ & $0(1)^{+}$ & not bound \\
$u d b b$ & $0(1)^{+}$ & $\sim -113~$MeV\\
$l s b b $ & $\frac{1}{2}(1)^{+}$ & $\sim -36~$MeV\\
$u c b b / s c b b$ & $\frac{1}{2}(1)^{+}$ & not bound\\
$u s c b$ & $\frac{1}{2}(1)^{+}$ & not bound\\
$u d c b / u d s b / u d s c$ & $0(0)^{+}$ & not bound\\
$u s c b$ & $\frac{1}{2}(0)^{+}$ & not bound\\
\hline \hline
\end{tabular}
\label{tab:hudspith-1}
\caption{\textit{Summary of results of \cite{Hudspith:2020tdf}, in which a total of twelve tetraquark candidates across $I(J^P)=0(1^+),~\frac{1}{2}(1^+)~$ and $0(0^+),~\frac{1}{2}(0^+)~$ were studied. The quoted binding energies are mentioned to be preliminary. }}
\end{table}

Finally, we note in passing a study on the $T^{bb}_{bb}$ \cite{Hughes:2017xie} that was published in 2017. It uses NRQCD for the four bottom quarks involved and determines individually di-meson and diquark-antidiquark tetraquark correlation functions. These are compared to the non-interacting thresholds without a GEVP analysis.
In both channels, the asymptotic state agrees with that of two non-interacting mesons, and the study subsequently does not observe a bound state in contrast to expectations. These expectations were recently reiterated using pNRQCD methods \cite{Assi:2023dlu}. A renewed study of this channel using a different lattice approach could illuminate this discrepancy.

This concludes the era of flavor surveys, as to the author's knowledge no further broad searches of multiple channels have been performed\footnote{A new study by the author and collaborators that will study the $J^P=0^+$ channels in addition to the $1^+$ is in preparation \cite{TO2024paper}. It was not ready for this review but will further update the heavy quark mass dependence.}. From now on, studies tend to hone in on one or two channels with a focus on determining scattering phase shifts, more systematic control, or attempting an interpretation in terms of the dominant structure of the bound tetraquark candidates.

% -----------------------------
\subsection{First attempts at studying the substructure of the $T_{bb}$ tetraquark}
\label{sec:structure}

The previous studies revealed deeply bound tetraquarks in the $I(J^P)=0(1^+)$ $T_{bb}^{ud}$ and $T_{bb}^{us}$ channels that were furthermore confirmed by multiple groups. 
The observed properties also seemingly confirmed the binding mechanism predictions based on the good diquark and heavy quark spin symmetry. As such, the predictions that the lighter the light quarks and the heavier the heavy quarks, the deeper the binding were all consistently observed. This raises the question of whether one can infer from these findings that it is indeed the diquark-HQSS picture that drives this binding and that the associated substructure of the deeply bound doubly heavy tetraquarks is dominantly of this type.
Especially which role the invoked diquark plays and whether it indeed acts as an effective QCD degree of freedom are interesting questions to answer.%\footnote{Note that the experimentally observed $T_{cc}$ are more difficult terms of the outlined binding mechanism, as the heavy quark spin symmetry in the charm quark case is a much less good approximation than for the bottom and residual interactions are not as heavily suppressed.}.

\begin{figure}[t!]
\centering
\includegraphics[height=0.18\textheight]{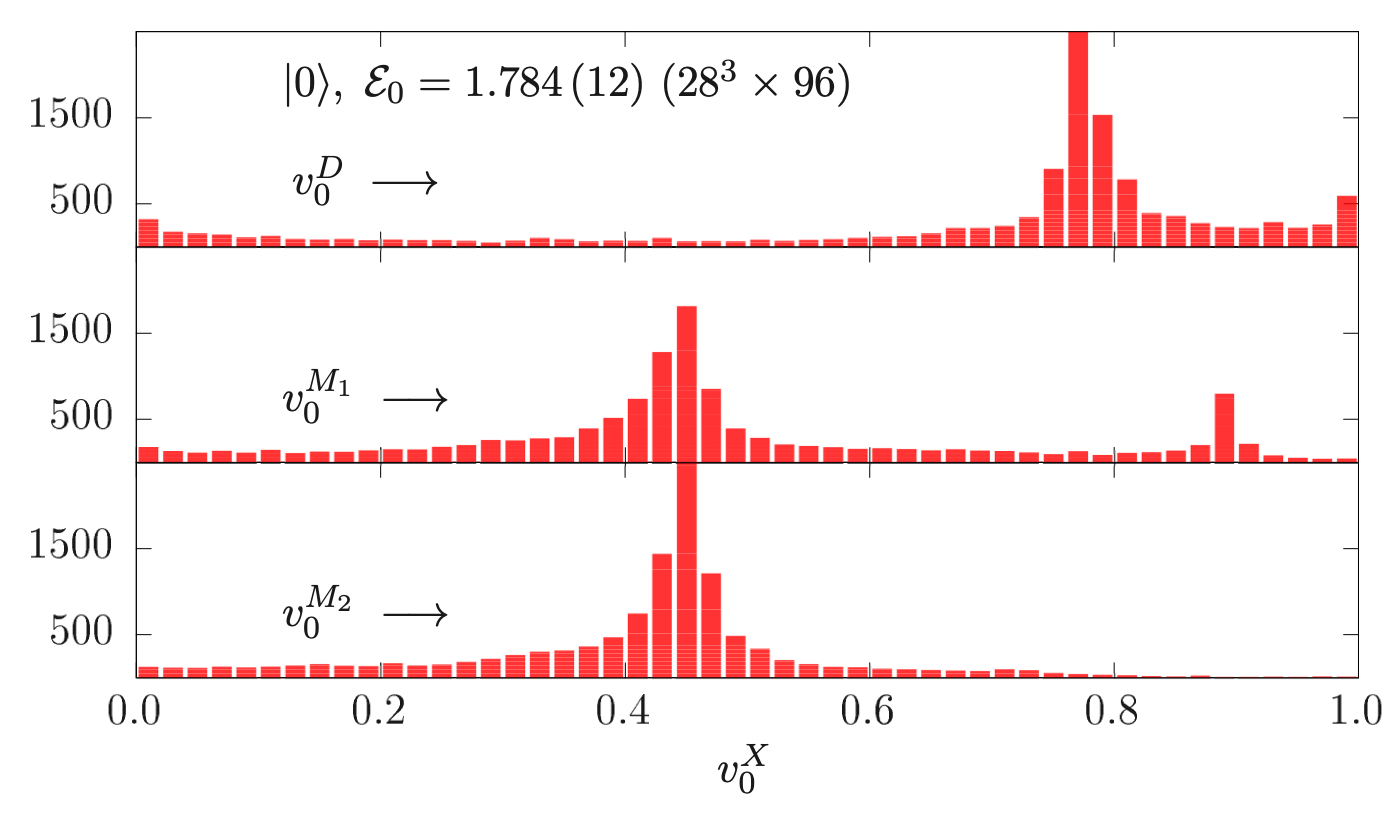}\\
\includegraphics[height=0.18\textheight]{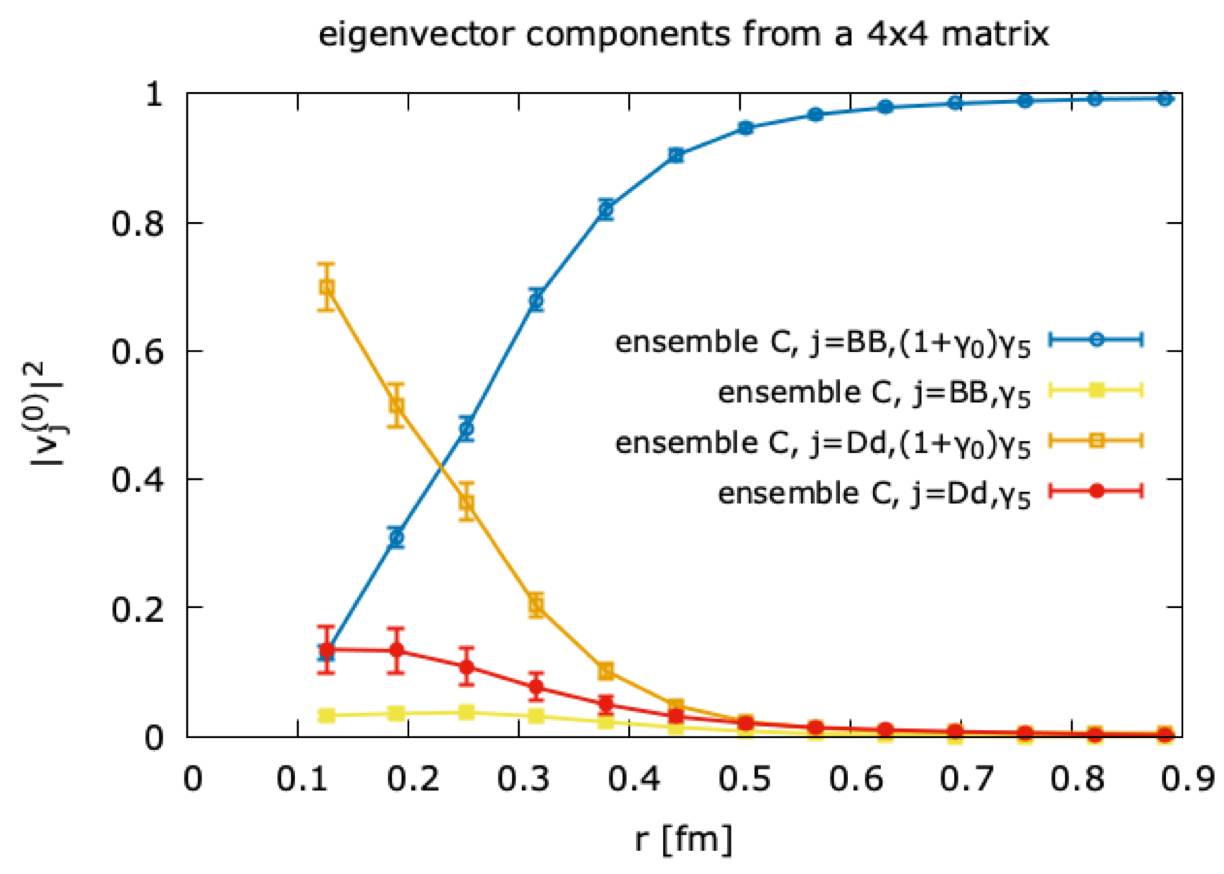}
\includegraphics[height=0.19\textheight]{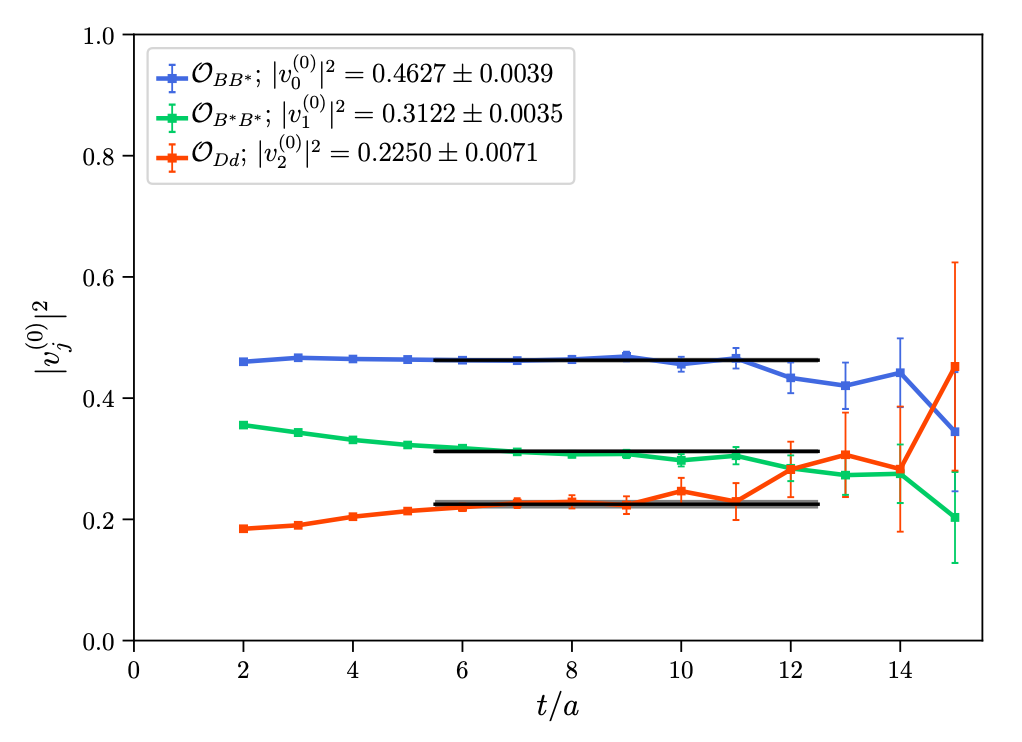}
\caption{\textit{Contradicting results from trial state overlap analyses for $T_{bb}^{ud}$ at similar physical lattice parameters. Top: Eigenvector component analysis for the ground state of the tetraquark system of \cite{Mohanta:2020eed} from a local-source-local-sink $3\times 3$ correlation matrix. The peak at 0.8 indicates diquark-antidiquark dominance. Bottom left: Inter-meson potential analysis with optimized trial states \cite{Bicudo:2021qxj} over distance. The authors find a 60/40 percentage for the di-meson/diquark-antidiquark contributions to the tetraquarks. Bottom right: Eigenvector component analysis of \cite{Pflaumer:2021ong} using a local-source-nonlocal-sink correlation matrix. The $BB^*$ structure in blue is seen to be dominant.
} }
\label{fig:structure-1}
\end{figure}

A first notable attempt in the direction of understanding the dominant structure of the $I(J^P)=0(1^+)$ $T_{bb}^{ud}$ doubly heavy tetraquark was made in \cite{Mohanta:2020eed}.
This study is unique in a few ways, one of which is that the authors use the MILC ensembles, i.e., using staggered quarks, with the HISQ action for the light quarks and NRQCD for the bottom quarks. In the used ensembles $m_\pi\simeq 323~$MeV, $a\simeq 0.09...0.15~$fm, $L\simeq 2.45~$fm and $m_\pi L\simeq 4$. Taking the continuum limit, they find a binding energy of $E_B\simeq -189(18)~$MeV. %, which is in contrast with other determinations, including those by the authors of \cite{Francis:2016hui} which have been consistently seeing smaller binding energies. 
The results were derived from a $3\times 3$ correlation matrix of local-local lattice correlation functions with operators of the diquark-antidiquark and di-meson types. Further analyzing their GEVP results, the eigenvector components are also evaluated. The argument being made is that they give insights into the dominant structure of the observes states: "the components of the eigenvectors provide the relative contribution of the trial states, corresponding to the lattice operators, to the energy eigenstates" \cite{Mohanta:2020eed}. The results are shown in Fig.~\ref{fig:structure-1} (top) and depict histograms of normalized eigenvector components for the ground state. The finding is that the ground state is dominated by the diquark-antidiquark structure, indicated by the peak at $v_0^D=0.8$, with a significant admixture of the di-meson structure. To compare with the results below this implies a dominant fraction of $0.64\%$ (via $ (v_0^D)^2$) of the diquark-antidiquark structure.

As mentioned, this requires extensive control over the operator structures being probed. In particular, having non-symmetric correlation matrices or an operator basis that is too small can cause problems. Although this study does not suffer from the first issue, it does from the second since extended structures are not considered. To highlight this further, we consider two studies published not long after. 
The first is the study by Bicudo et al. \cite{Bicudo:2021qxj},
where the authors also focus on $I(J^P)=0(1^+)$ $T_{bb}^{ud}$ doubly heavy tetraquarks and they extend their previous work in determining lattice potentials. They generate a set of $2\times 2$ and $4\times 4$ correlation matrices with different operator structures.
Resolving their relative weights to the optimized tetraquark operators at all available distances, they arrive at the results shown in Fig.~\ref{fig:structure-1} (bottom left). They find an inversion of the dominant structure around $r\simeq 0.25~$fm, compare also the results of \cite{Cook:2002am} mentioned previously. For the ratio of the different structures to the tetraquark, they find 60\% for the di-meson and 40\% for the diquark-antidiquark structures. 
A related study from an overlapping group of authors is \cite{Pflaumer:2021ong}. It is an extension of their effort using Domain Wall fermions \cite{Leskovec:2019ioa} and recall that in the setup of this group, nonlocal sinks are available. As a technical update, instead of using multi-exponential fits to determine the different finite volume energies, the authors shifted towards a mixed approach: By solving a larger non-symmetric GEVP, they reduce the size of the correlation matrix and perform multi-exponential fits on those. 
This proceedings article shows results on an eigenvector component analysis from a $3\times 3$ correlation matrix for the $J^P=1^+$ $T_{bb}^{ud}$. The key plot is shown in Fig.~\ref{fig:structure-1} (bottom right). At a similar pion mass of $m_\pi=340~$MeV, they find that the di-meson contribution dominates with $\simeq 77\%$ and the diquark-antidiquark contribution at $\simeq 23\%$.
Note that this data did not enter the authors' later publication \cite{Meinel:2022lzo}, which instead focuses on $1^+$ $T_{bb}^{us}$ and the $0^+$ as well as $1^+$ $T_{bc}^{ud}$.

While the lattice parameters are similar for all three studies, or at least are so on a subset of the gathered data, a key difference is the operator basis, with local and nonlocal structures being the most prominent. That they reach contradictory statements on the dominant substructure of the doubly bottom tetraquark highlights the difficulties of eigenvector component analysis on the lattice, i.e., the study of trial state overlaps. It is not clear how to effectively control the introduced systematics through the choice of basis and to make robust statements. An interpretation based on the scattering parameters derived from finite volume energies seems the better way forward, even though, as explained later on, this comes with its own difficulties. 
Note that this issue is of a different flavor than the difficulty in determining ground state energies and connected statements on binding energies, for example, in the $T_{bc}^{ud}$. Here, larger statistics, better-suited lattice parameters, and a more careful study of the approach to the asymptotic plateau are systematical paths to improving the situation. 

%----------------------------------
\subsection{Scattering analyses and focus on $I(J^P)=0(1^+)$ $T_{cc}$, $T_{bc}$ and $T_{bb}$ channels}

In today's era, approximately after 2020, studies have started to see results from the completed multi-step recipe of the spectrum approach, which is aimed at determining the scattering parameters of the main doubly heavy tetraquark candidates. The previous successes and especially the experimental discovery of $T_{cc}^{ud}$ \cite{LHCb:2021vvq,LHCb:2021auc} have catalyzed the interest in studying doubly heavy tetraquarks on the lattice, and many new lattice groups have begun working in this direction.
The most recent results are reviewed in the following, organized by their main search channel.

\begin{figure}[t!]
\centering
\includegraphics[width=0.495\textwidth]{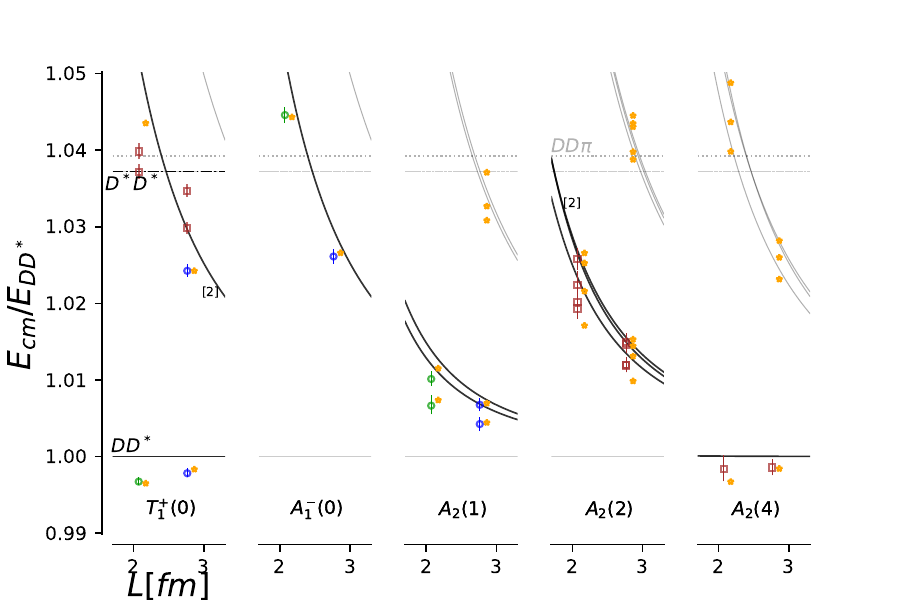}
\includegraphics[width=0.495\textwidth]{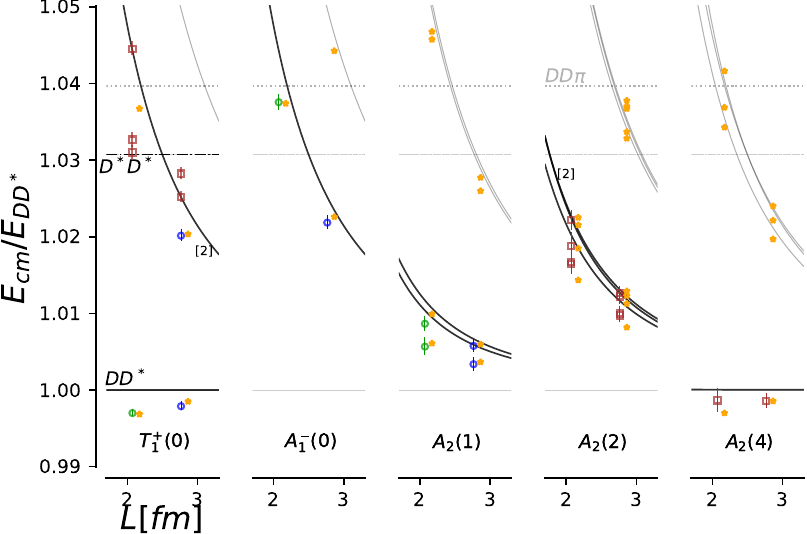}
\caption{\textit{Finite volume energies for the $J^P=1^+$ $T_{cc}^{ud}$ tetraquark determined in \cite{Padmanath:2022cvl}. Energies were extracted from symmetric GEVPs across a large basis of operators in five irreps, at multiple momenta and in two lattice volumes at $m_\pi\simeq 280~$MeV and at slightly lower (left) and slightly larger (right) charm quark mass. The figures were taken from the supplementary material of the publication.
} }
\label{fig:sasa-1}
\end{figure}

%----------------------------------
\subsubsection{The $T_{cc}$}

The $T_{cc}^{ud}$ channel is particularly difficult to study using lattice QCD since it is so shallowly bound in nature. In this channel, a careful scattering analysis is mandatory and the idea of tracking the trajectory of the particle pole to potentially interpolate its properties at the physical point comes into its own. In the following, we detail the milestone calculations of the $T_{cc}^{ud}$ demonstrating the full program that can be used as blueprints, with improvements possible, for future studies.

\paragraph{Finite volume energies and scattering parameters.}

\begin{figure}[t!]
\centering
\includegraphics[width=0.45\textwidth]{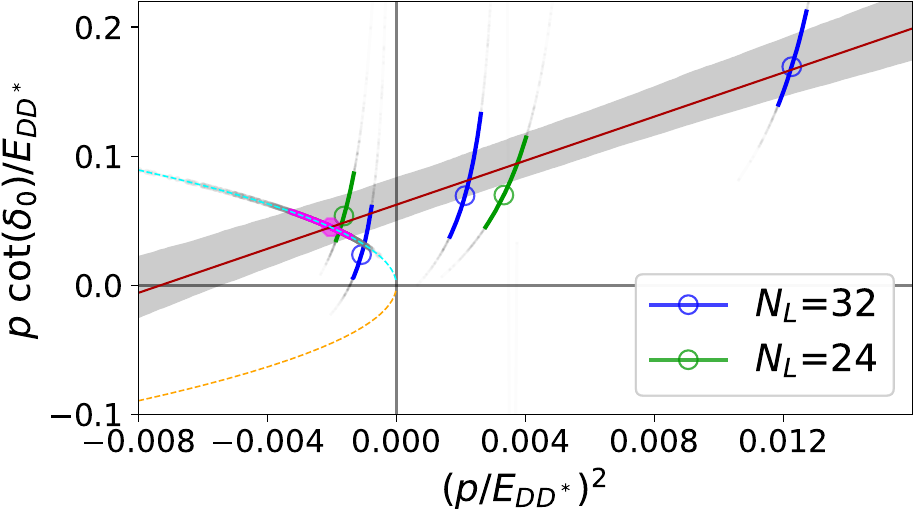}
\includegraphics[width=0.45\textwidth]{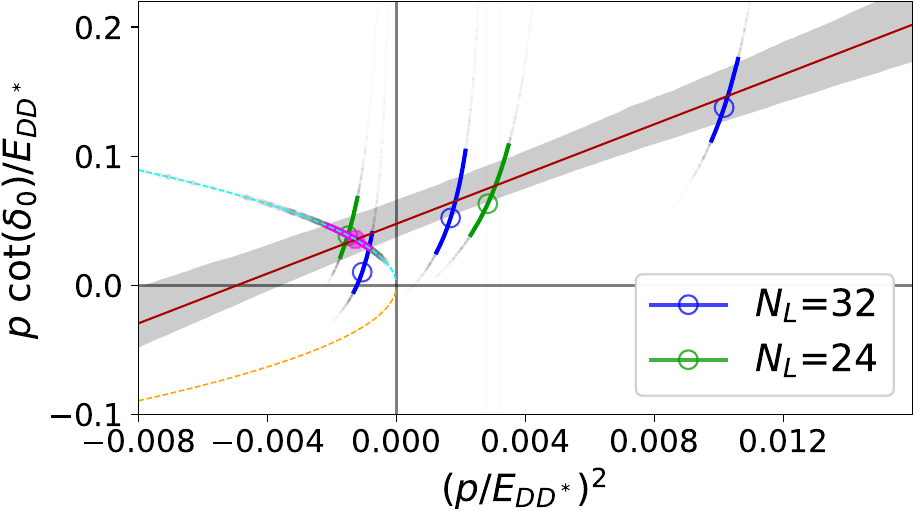}
\includegraphics[width=0.25\textwidth]{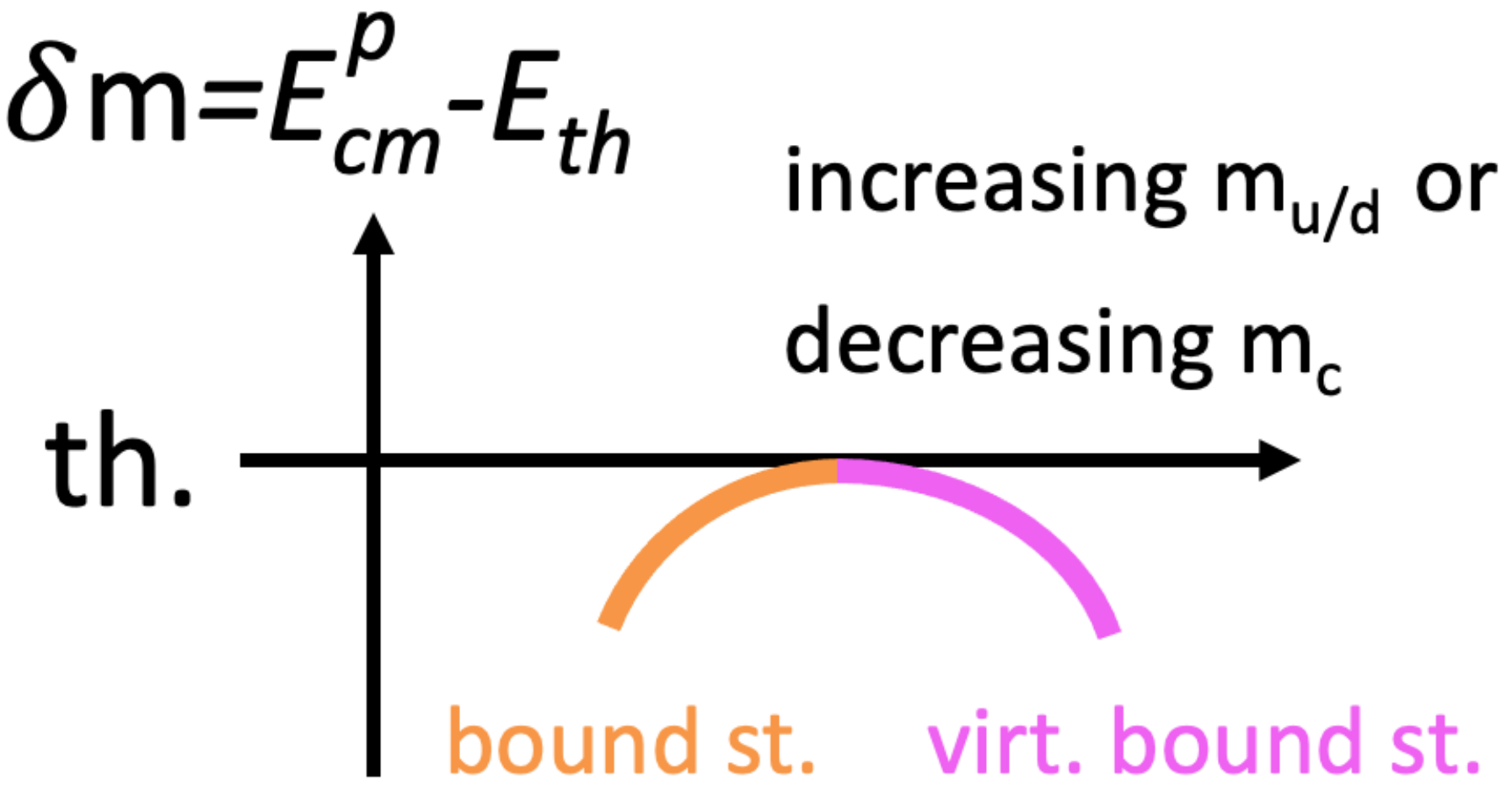}
\caption{\textit{Scattering phase shifts $p \cot \delta$ and fitted effective range expansion of \cite{Padmanath:2022cvl} for the lighter than physical charm quark mass (top left) and the heavier (top right), whereby the pion mass is $m_\pi=280~$MeV. The figures were taken from the supplementary material of the publication. The figure in the bottom panel depicts the possible trajectory of the pole with quark mass.
} }
\label{fig:sasa-2}
\end{figure}

The study of \cite{Padmanath:2022cvl} represents a breakthrough because, for the first time, a complete analysis was achieved in the $1^+$ $T_{cc}^{ud}$ channel; this is in the sense of the determination of multiple finite volume energies at sufficient precision to convert them to scattering phase shifts and to determine with statistical significance the scattering parameters. In their study the authors use $N_f=2+1$ gauge ensembles provided by the CLS effort \cite{Bruno:2014jqa,Bali:2016umi,Bruno:2016plf}, the lattice parameters are $m_\pi\simeq 280~$MeV, $a\simeq0.086~$fm, $L\simeq 2.1,~2.8~$fm and $m_\pi L\simeq 2.9,~3.9$. The charm quark parameters were previously determined in \cite{Piemonte:2019cbi} and set to two values, such that the spin-averaged masses are just below ($M_l\simeq 2820~$MeV) and just above ($M_h\simeq 3103~$MeV) their physical value.
The authors make use of the distillation method. Based on the previous observation of \cite{Cheung:2017tnt} that diquark-antidiquark operators do not significantly contribute to the resolving power they decided to study di-meson type operators only. Nevertheless, as suggested in the same study \cite{Cheung:2017tnt}, the authors here implement a large basis of these operators in the different irreps. Furthermore, they consider operators with relative momenta, further boosting their resolution capabilities. 
The determined finite volume energies are shown in Fig.~\ref{fig:sasa-1}, throughout a high precision for each state resolved is observed. This makes it possible in the next step to convert the energies to scattering phase shifts via the finite volume quantization condition for 2-to-2 scattering and to determine the scattering length and effective range. The results are shown in Fig.~\ref{fig:sasa-2}, and the determined scattering parameters and predicted binding energy are gathered in Tab.~\ref{tab:sasa-1}.

One thing to note is that although the quoted results in the table are derived from an effective range expansion of the type discussed in the toy model before, the authors consider the partial waves $\ell=0$ and also $\ell=1$. That is, they explicitly consider the energy dependence of the scattering amplitude $t$ in $S=$ $e^{2 i \delta}=1+i \frac{4 p}{E_{c m}} t$ and approximate the dependence near threshold with more terms in the effective range expansion:
\begin{equation}
t_\ell^{(J)}=\frac{E_{c m}}{2} \frac{1}{p \cot \delta_\ell^{(J)}-i p}~, ~~p^{2 \ell+1} \cot \delta_\ell^{(J)}=\frac{1}{a_\ell^{(J)}}+\frac{r_\ell^{(J)}}{2} p^2~~.
\end{equation}
In the heavier-than-physical charm quark case they find $a_1^{(0)}=0.076\left({ }_{-0.009}^{+0.008}\right) \mathrm{fm}^3, r_1^{(0)}=6.9(2.1) \mathrm{fm}^{-1}$ for the $\ell=1$ scattering case, while for $\ell=0$ they find the parameters given in Tab.~\ref{tab:sasa-1}.
The authors' final result is that they determine scattering parameters consistent with a virtual bound state interpretation for both the heavier-than- and lighter-than-physical charm quark masses. They expect the pole trajectory's dependence on the quark mass as is sketched out in Fig.~\ref{fig:sasa-2} (bottom), where the bound state observed in experiment lies between two virtual bound state regimes. Note that the figures have been taken from the supplementary material of the publication.

\begin{table}
\centering
\begin{tabular}{l|cc|c|c}
\hline \hline & $a_{l=0}^{(J=1)}[\mathrm{fm}]$ & $r_{l=0}^{(J=1)}[\mathrm{fm}]$ & $\delta m_{T_{c c}}[\mathrm{MeV}]$ & $T_{c c}$ \\
\hline \hline lat. $\left(m_\pi \simeq 280 \mathrm{MeV}, m_c^{(h)}\right)$ &  $1.04(29)$ & $0.96\left({ }_{-0.20}^{+0.18}\right)$ & $-9.9_{-7.2}^{+3.6}$ & virtual bound st. \\
lat. $\left(m_\pi \simeq 280 \mathrm{MeV}, m_c^{(l)}\right)$  & $0.86(0.22)$ & $0.92\left({ }_{-0.19}^{+0.17}\right)$ & $-15.0\left({ }_{-9.3}^{+4.6}\right)$ & virtual bound st. \\
$\quad \operatorname{exp.}[2,38]$ &  $-7.15(51)$ & {$[-11.9(16.9), 0]$} & $-0.36(4)$ & bound st. \\
\hline \hline
\end{tabular}
\caption{\textit{Collected results for the scattering parameters and their interpretation of \cite{Padmanath:2022cvl}.}}
\label{tab:sasa-1}
\end{table}

Since this study came out, a few new discussions started. One is the validity of the effective range expansion since in the $T_{cc}^{ud}$ there is a potential left-hand cut problem. This will be discussed further later.
Here, we note that the discussion of the point of diquark-antidiquark operators was picked up very recently by the group around the authors \cite{Ortiz-Pacheco:2023ble}. This new study finds that while including diquark-antidiquarks does not affect the ground state energy level it does affect the second energy level in their test.
Consequently, the omission of these operators can contribute a systematic that should be quantified. 

A further study on $DD^*$ scattering was published in \cite{Chen:2022vpo} using anisotropic, $N_f=2$ Wilson-Clover fermion gauge ensembles with a pion mass of $m_\pi\simeq 349~$MeV and $m_\pi L\simeq 3.9$. The temporal lattice spacing is fine enough to use a standard relativistic quark action for the charm quarks.
The authors aim to compare the $I=0$ and $I=1$ channels and learn something about the attractive or repulsive character of the $DD^*$ interaction. Indeed, they find the $I=0$ channel is attractive for a broad range of momenta, which is not the case for the $I=1$ channel.

\paragraph{Potential approach.}

\begin{figure}[t!]
\centering
\includegraphics[width=0.6\textwidth]{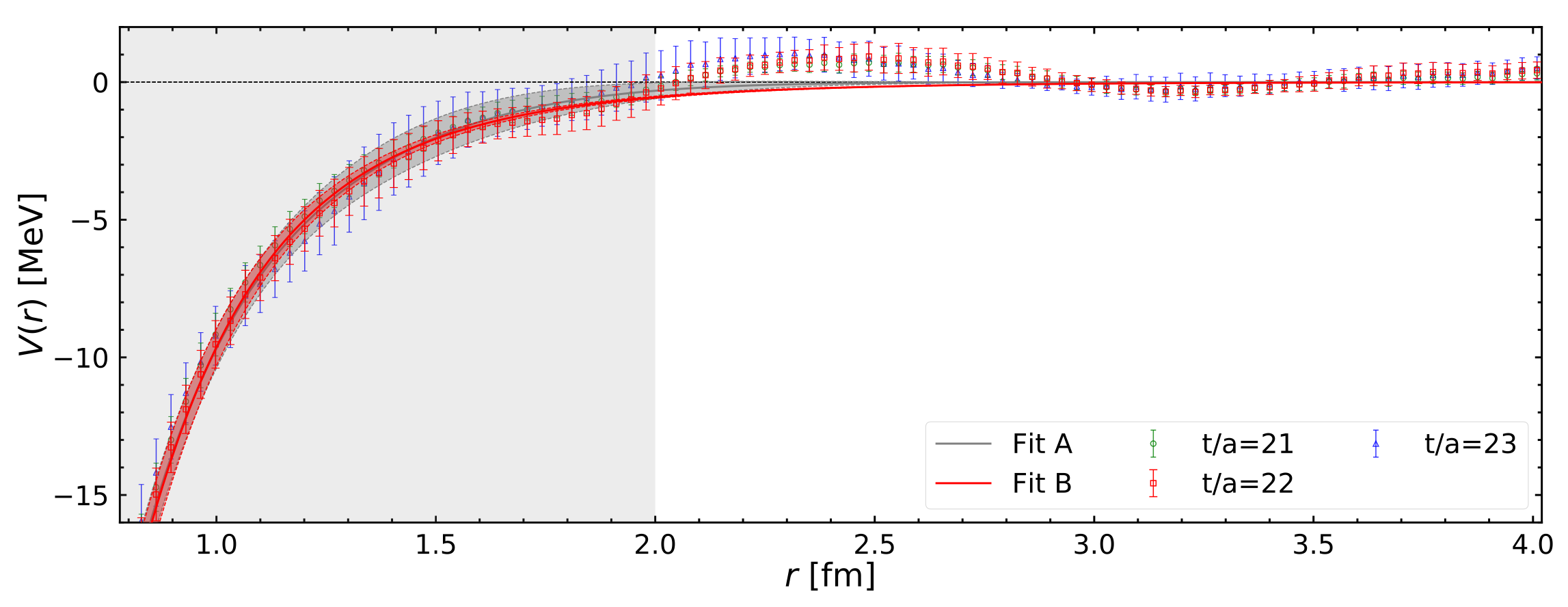}
\includegraphics[width=0.32\textwidth]{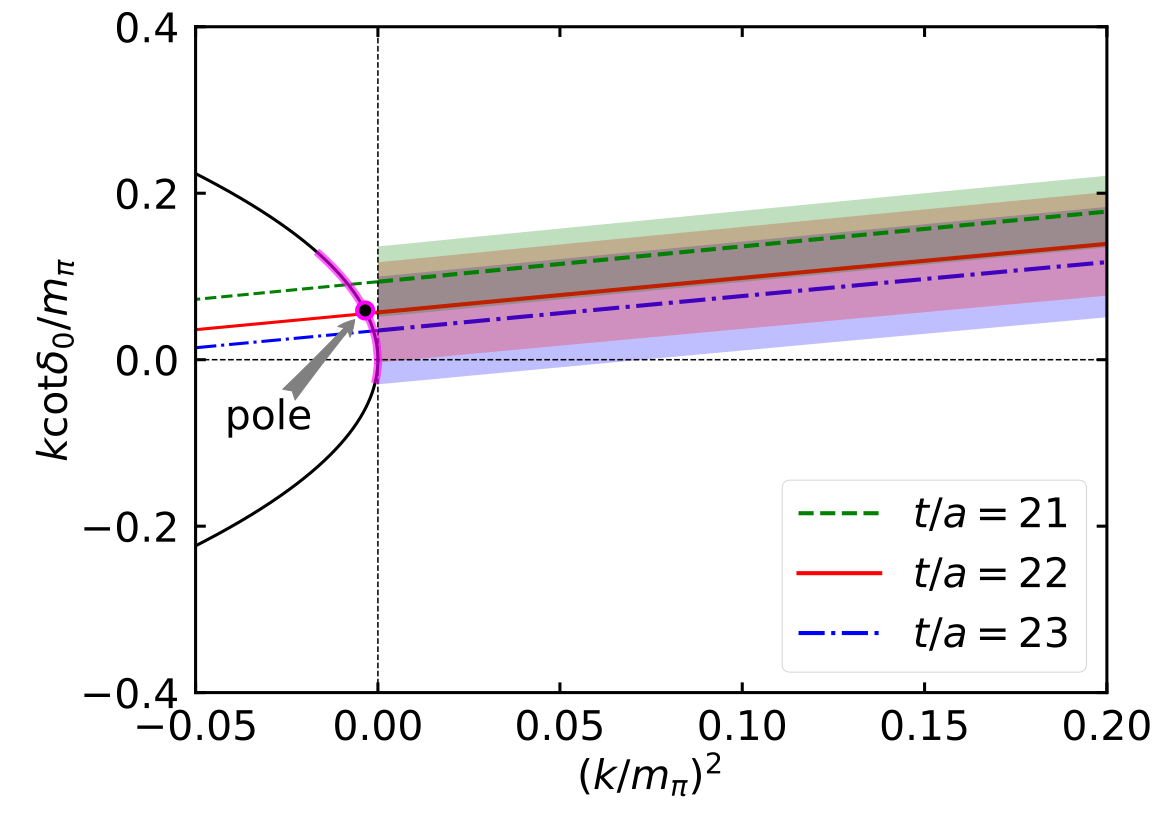}
\caption{\textit{The left panel shows results on the $DD^*$ inter-meson lattice potential determined in \cite{Lyu:2023xro} using the HALQCD method. The scattering phase shift $\delta$ can be determined by fitting a potential ansatz and solving the resulting Schr\"odinger equation. Then $p\cot \delta$ can be evaluated via an effective range expansion to yield the scattering length and effective range. This result is shown in the right panel.
}}
\label{fig:hal-tcc-1}
\end{figure}

The alternative to studying the $T_{cc}^{ud}$ using the spectrum approach is to study it via the HALQCD method to determine the inter-meson potential.  
In \cite{Lyu:2023xro}, the authors report on results using this method in the first major update since \cite{Ikeda:2013vwa} in 2013. 
The authors use a new generation of the PACS-CS ensembles \cite{Ishikawa:2015rho} and employ configurations with $m_\pi\simeq 146~$MeV, $a\simeq 0.085~$fm, $L\simeq 8.1~$fm and $m_\pi L\simeq 6$. They use Tsukuba-type effective relativistic heavy quarks for the charm quark in the tuning of \cite{Namekawa:2017nfw}.
Recall that in the HALQCD method, one determines the unprojected correlation function and the NBS wave function through it, as outlined in Sec.~\ref{sec:potentials}.
%\begin{equation}
%\begin{aligned}
%\Phi(\vec{r}, t) & =\sum_{\vec{x}}\left\langle 0\left| \mathcal{O}\Big(D^*(\vec{x}+\vec{r}, t), local\Big) \mathcal{O}\Big(D(\vec{x}, t), local\Big) \bar{\mathcal{O}}\Big(DD^*(0), wall\Big)\right| 0\right\rangle / e^{-\left(m_{D^*}+m_D\right) t} %\\
%& =\sum_n A_n \psi_n(\vec{r}) e^{-\left(\Delta E_n\right) t}+O\left(e^{-\left(\Delta E^*\right) t}\right)
%\end{aligned}
%\end{equation}
%which can be related to a leading order local potential via:
%\begin{equation}
%V(r)=\Phi^{-1}(\boldsymbol{r}, t)\left[-\partial_t + \frac{\vec\nabla^2}{2\mu}\right] \Phi(\boldsymbol{r}, t)
%\end{equation}
%where $\mu$ is the reduced mass of the $D$ and $D^*$ mesons. %Further, a term proportional $\partial_t^2$ has been dropped as it was %observed to be consistent with zero in this study. 
The resulting $DD^*$ inter-meson lattice potential $V(r)$ is shown in Fig.~\ref{fig:hal-tcc-1} for multiple values of $t$. In principle, the potential should be $t$-independent, but often contaminations are observed, and the values used for fitting are chosen such that these are as small as possible.
In the next step, the potential is fitted to two forms:
\begin{align}
V_{\text {fit}}^A(r)=\sum_{i=1}^4 a_i e^{-\left(r / b_i\right)^2}~,~~
V_{\text {fit}}^B\left(r ; m_\pi\right)=  \sum_{i=1,2} a_i e^{-\left(r / b_i\right)^2} +a_3\left(1-e^{-\left(r / b_3\right)^2}\right)^2 (e^{-m_\pi r} / r )^2
\label{eq:hal-tcc-fit}
\end{align}
From the two fitted forms, given in Fig.~\ref{fig:hal-tcc-1} by the grey and red bands, the $S$-wave scattering phase shifts can be determined via the Schr\"odinger equation. The authors find a scattering length $1/a_0\simeq 0.05(7)~\rm{fm}^{-1}$ and effective range $r_0\simeq 1.12(8)~$fm, where we have combined two quoted errors into one. Since the value of $m_\pi$ enters the second fit form, the authors also consider changing it to $135~$MeV while keeping all other parameters fixed. In this case the scattering length changes $1/a_0\simeq 0.05(7)~\textrm{fm}^{-1}\rightarrow -0.03(4)~\textrm{fm}^{-1}$. Furthermore, the intersection of the determined effective range expansion with $\pm \sqrt{-p^2}$ gives information on the pole location of a possible bound state. Converting this into a binding energy the authors find $E_B=-0.59(83)~$MeV and for the mass shifted case $E_B=-0.45(78)~$MeV.\\%, again after combining the quoted errors into one. \\
It is important to remember that the HALQCD method relies on the validity of the local approximation of the potential and the potential ansatz. This is better fulfilled the heavier the quarks involved. 
However, controlling this systematic is difficult. 
Taking the two pion mass shifted results of the second fit form together, they would imply the switch from a virtual bound state to a bound state; however, although it is interesting to see this, there is no strong reason why the potential should not change more when one repeats the analysis at the target shifted mass. Further, the change of interpretation could be worrying since it is unclear whether a similar effect cannot be achieved by simply changing the potential fit ansatz. Future work is eagerly awaited on what is an impressive study.

\subsubsection{The $T_{bb}$ }

\paragraph{Spectrum approach.} Already mentioned briefly before, \cite{Meinel:2022lzo} studies the $T_{bb}^{us}$ channel on $N_f=2+1$ Domain Wall fermion gauge ensembles. One of the ensembles is at physical pion mass with $m_\pi\simeq 139~$MeV, $a\simeq 0.11~$fm, $L=5.5~$fm and $m_\pi L\simeq 3.9$, while the others are mostly around $m_\pi\sim 300~$MeV with some having a lower lattice spacing $a\simeq 0.08~$fm. For the heavy quark flavors, the authors use NRQCD to handle the bottom quarks and the RHQ effective relativistic heavy quark action for charm quarks, whose parameters are tuned as part of this study. Recall that the authors employ nonlocal sink operators in their setup. Dubbed scattering operators in this work, their structure is of the nonlocal di-meson type. Overall, they confirm that the previously predicted $J^P=1^+$ $T_{bb}^{us}$ tetraquark is indeed bound and give the physical point binding energy $E_B=-86(32)~$MeV.

A next extensive study of both the $J^P=1^+$ $T_{bb}^{ud}$ and the $T_{bb}^{us}$ channels was presented in \cite{Hudspith:2023loy}. The calculational setup is similar to that of \cite{Hudspith:2020tdf} using wall-local and wall-smeared correlation matrices. For the heavy bottom quarks it uses NRQCD, and the parameters were tuned using a machine-learning approach.
Furthermore, it uses a large set of $N_f=2+1$ Wilson-Clover fermion ensembles generated within CLS \cite{Bruno:2014jqa,Bali:2016umi,Bruno:2016plf}. This enables combined chiral-volume extrapolations, thereby achieving a new level of control over some of the main lattice systematics. 
For the $T_{bb}^{ud}$ they find the binding energy $E_B=-112(13)~$MeV, while for the $T_{bb}^{us}$ they quote $E_B=-46(12)~$MeV.
These findings are significantly lower than those of the study just above. However, they agree well with those preliminarily reported previously in \cite{Colquhoun:2022dte,Colquhoun:2022sip}.

\begin{figure}[t!]
\centering
\includegraphics[width=0.45\textwidth]{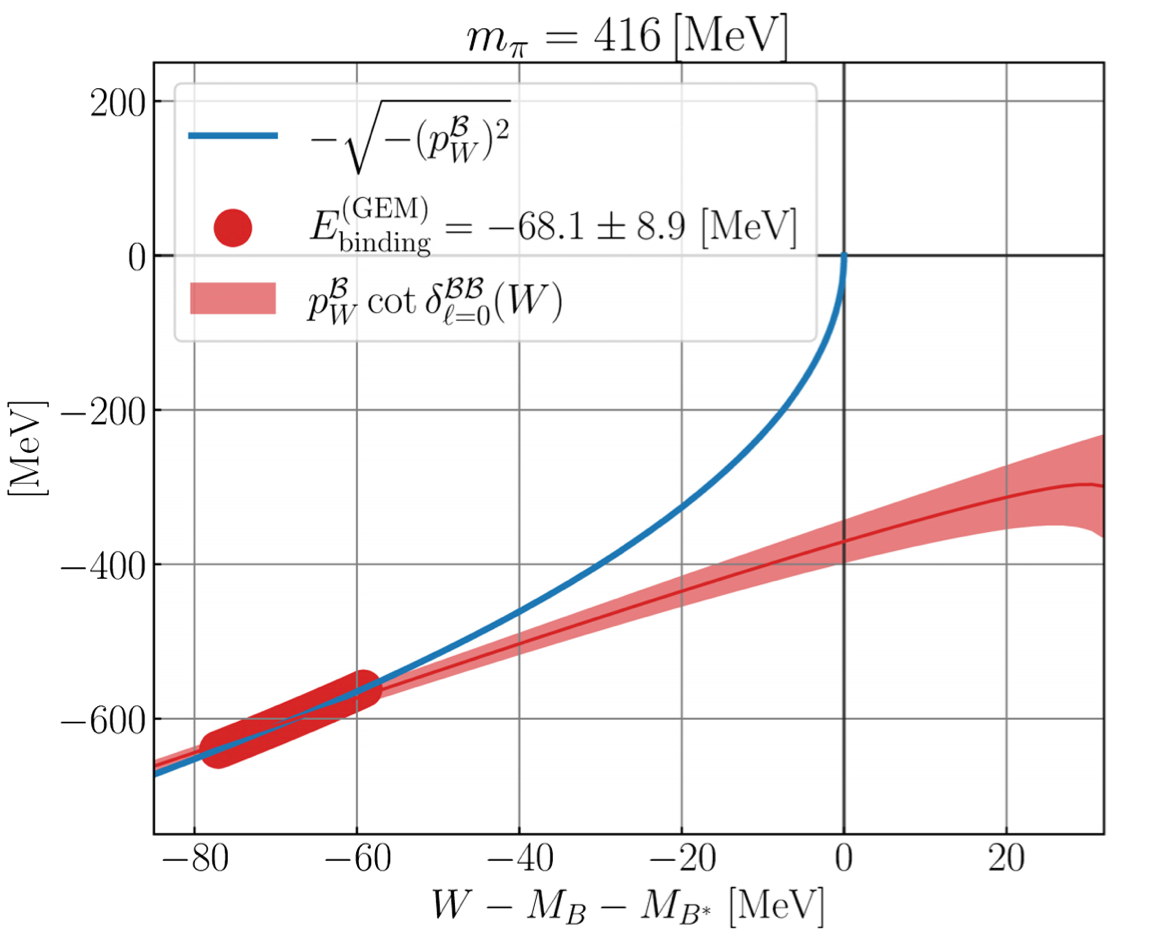}
\includegraphics[width=0.45\textwidth]{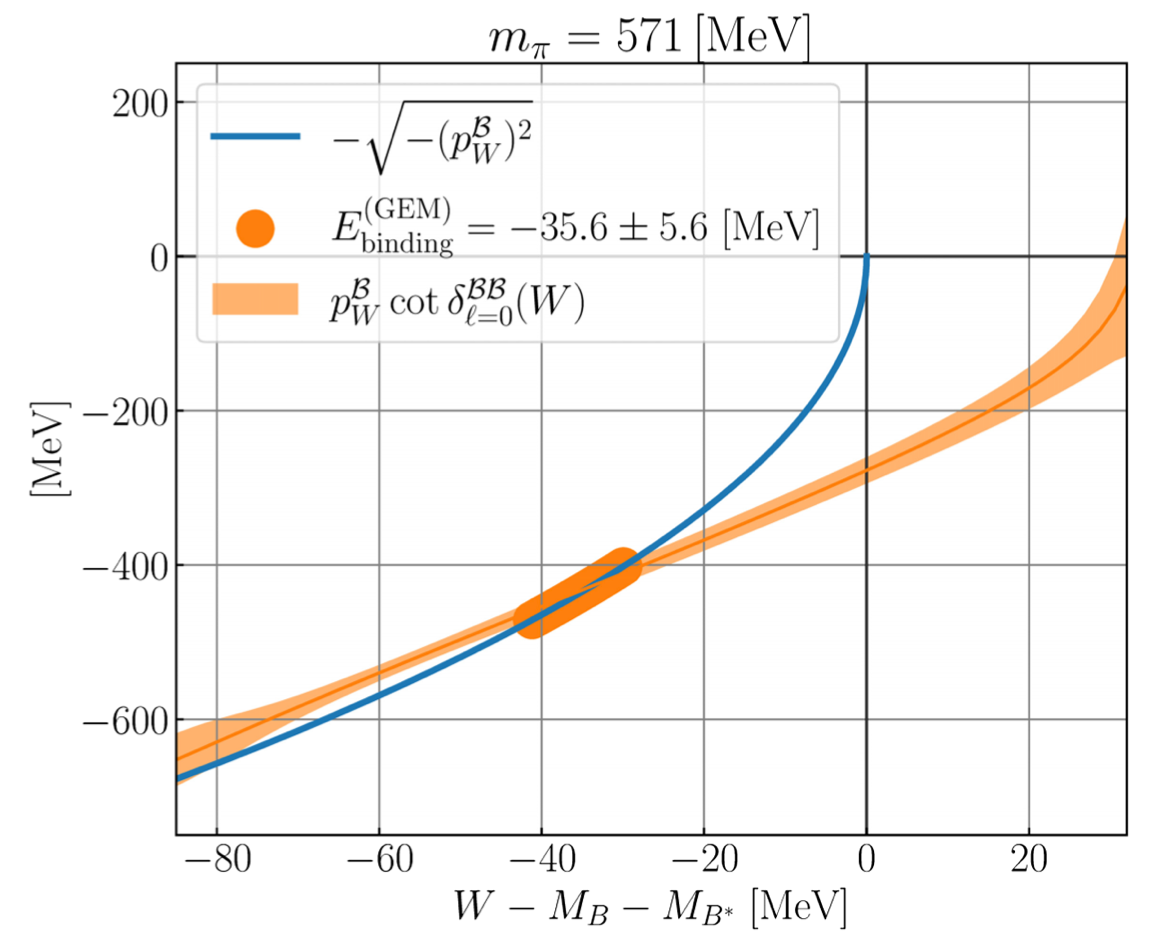}
\includegraphics[width=0.45\textwidth]{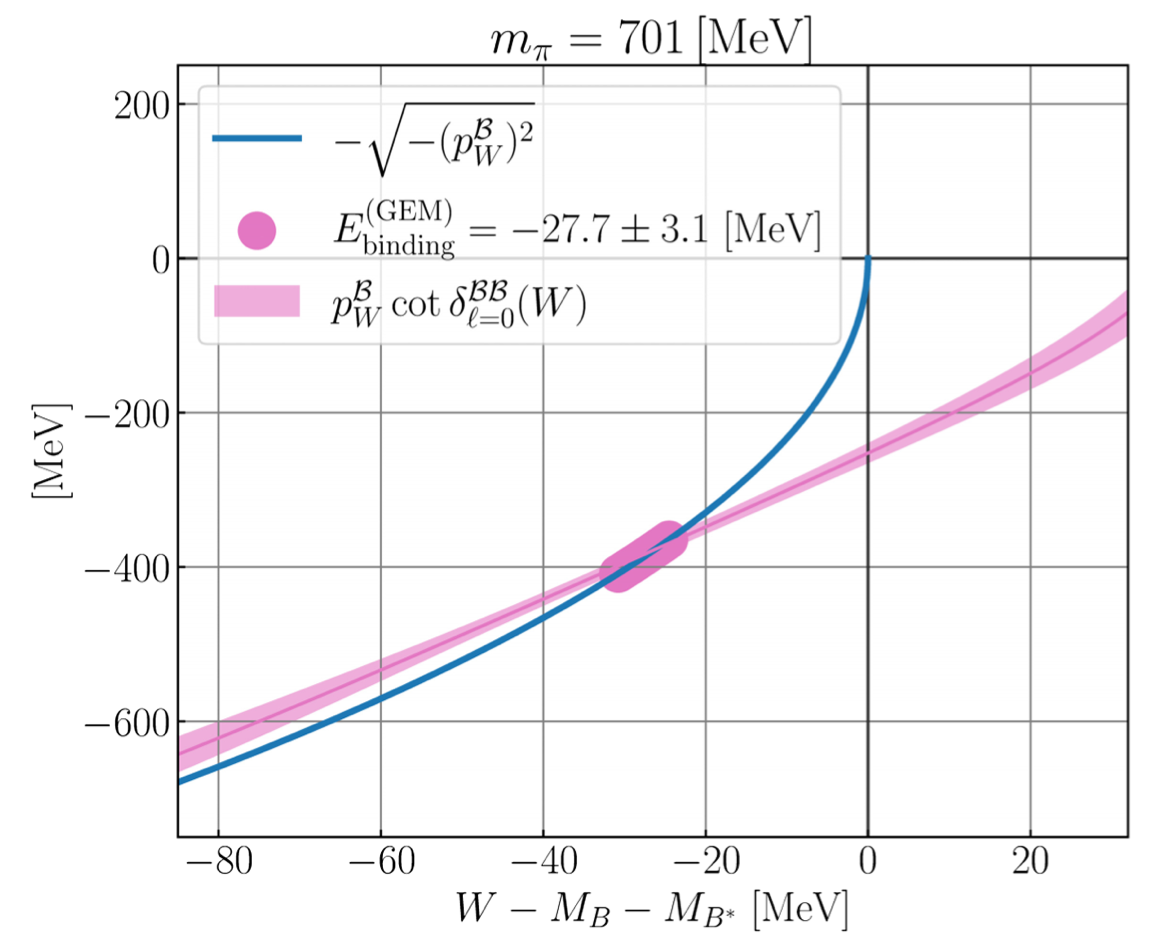}
\includegraphics[width=0.45\textwidth]{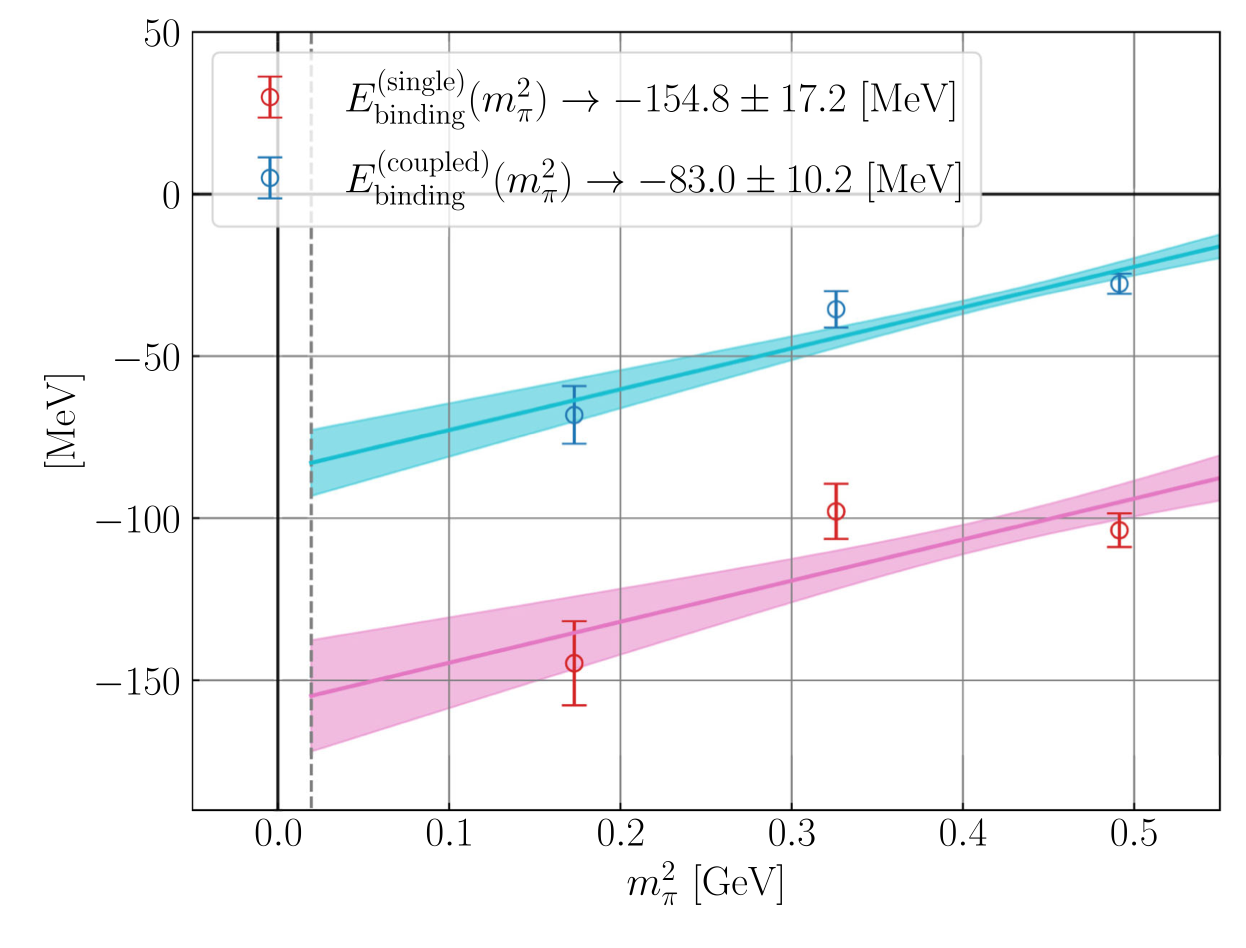}
\caption{\textit{Results on $p\cot\delta$ for the $J^P=1^+$ $T_{bb}^{ud}$ tetraquark from a coupled-channel potential analysis using the HALQCD method presented in \cite{Aoki:2023nzp}. The results are shown in descending pion mass from top left to right to bottom left. The shaded bands denote the effective range expansions while the blue curves denote the pole condition $-\sqrt{-p^2}$. Their intersection gives the binding energy, which is shown at the bottom right, together with an extrapolation to the physical pion mass (shaded bands). The authors observe a significant shift in the binding energy between the single $E_B^s\simeq -155(17)~$MeV and the coupled channel $E_B^c\simeq -83(10)~$MeV approach.
}}
\label{fig:hal-tbb-1}
\end{figure}

\paragraph{Potential approach.} 

Shortly after the study in the $T_{cc}^{ud}$ channel in \cite{Lyu:2023xro} also a study on the $J^P=1^+$ $T_{bb}^{ud}$ tetraquark using the HALQCD method was published \cite{Aoki:2023nzp}. Unlike the former study, the current one uses the same PACS-CS ensembles as other groups have previously, e.g. in \cite{Francis:2016hui}. The authors select only the three heavier ensembles with $m_\pi\simeq 416,~571, 701~$MeV and the lattice spacing $a=0.0907~$fm\footnote{Although they are the same ensembles, these values do not match with those of for example \cite{Francis:2016hui} and other studies. This is due to the redetermination of the lattice spacing and pion masses in these efforts. It is not entirely clear why the groups' results do not agree throughout.}. A key innovation of this study considers a coupled-channel potential between $BB^*$ and $B^*B^*$. 
As mentioned previously, the authors of \cite{Cheung:2017tnt} worked out a large basis of interpolating operators on the lattice for the study of doubly heavy tetraquark states; the vector-vector dimeson channel was part of this and has been used in, for example, \cite{Hudspith:2023loy,Hudspith:2020tdf}. Given the closeness of the thresholds in both channels, the authors argue that a coupled-channel potential is important to resolve the physics correctly.\\
The results are fitted to an ansatz of three Gaussians with six parameters. Since, in this study, a coupled-channel potential is determined, potentials are fitted for a $2\times 2$ correlation matrix between the $BB^*$ and $B^*B^*$ channels.
Evaluating the Schr\"odinger equation with this matrix of potentials adds a layer of complication. Nevertheless, the authors succeed in determining $p\cot \delta$ and performing an effective range expansion. The results for the coupled channel analysis for all three available pion masses are shown in Fig.~\ref{fig:hal-tbb-1}. The analysis is also repeated for the single channel case, and both scenarios are extrapolated to the physical pion mass, as shown in Fig.~\ref{fig:hal-tbb-1} (bottom right). 
The authors observe a significant shift in the binding energy between the single $E_B^s\simeq -155(17)~$MeV and the coupled channel $E_B^c\simeq -83(10)~$MeV approach. 
The current binding energies determined from the spectrum approach lie between these two values. However, a scattering analysis has not been performed yet, so it will be interesting to compare those results in the future. Furthermore it is interesting to note that both in the studies using static potentials and HALQCD adding in the $B^*B^*$ channel and performing a coupled channel analysis reduces the binding energy.

\subsubsection{The $T_{bc}$ }

Finally, the already mentioned work of \cite{Meinel:2022lzo} studies also the $I(J^P)=0(1^+)$ and $I(J^P)=0(1^+)$ $T_{bc}^{ud}$ channels. The setup of the calculation was outlined previously. For the two mentioned channels, the authors observe finite volume ground state energies compatible with the di-meson threshold. However, given the statistical uncertainties, they cannot rule out a shallow bound state at the $20-50~$MeV level. 

\begin{figure}[t!]
\centering
\includegraphics[width=0.495\textwidth]{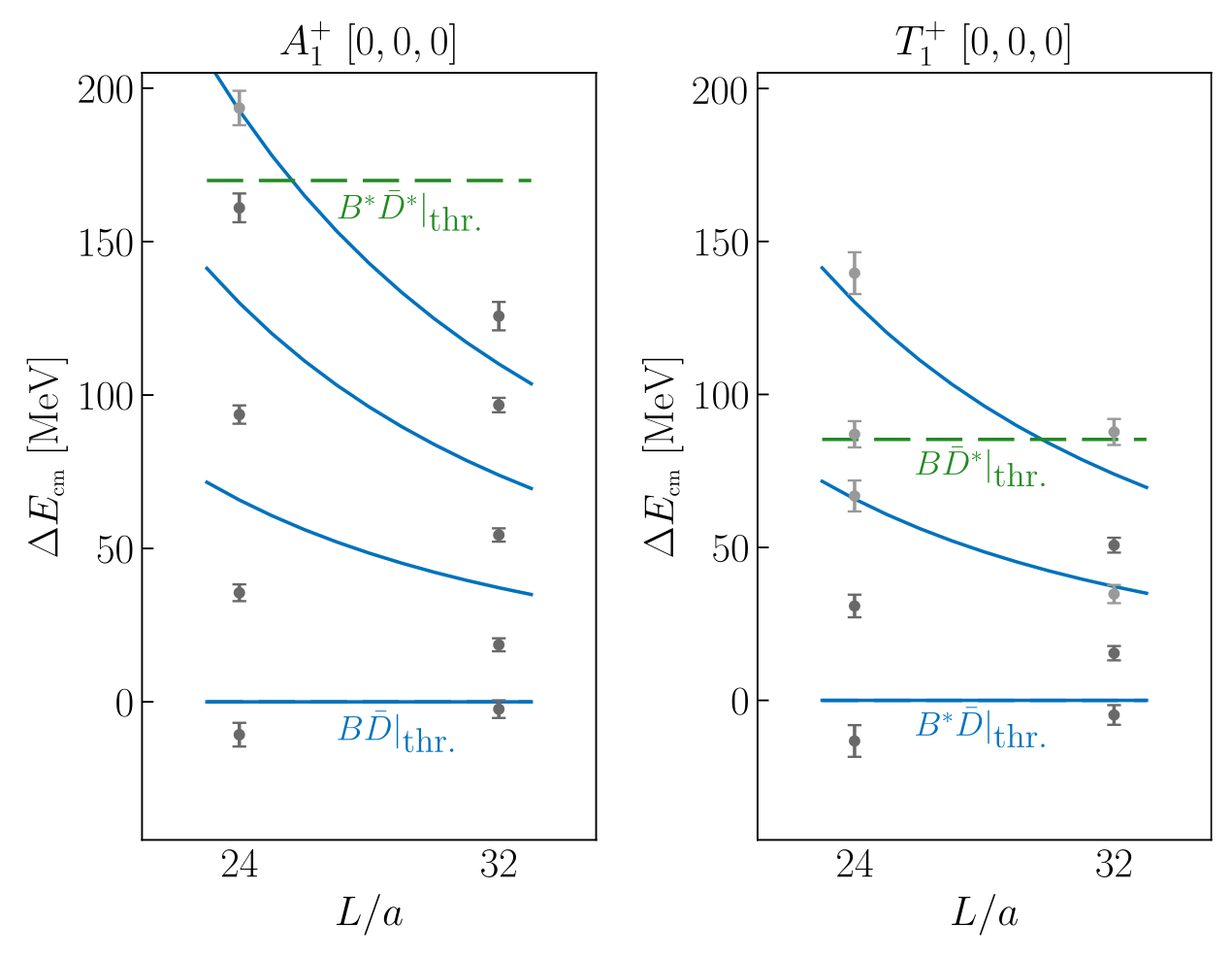}
\includegraphics[width=0.4\textwidth]{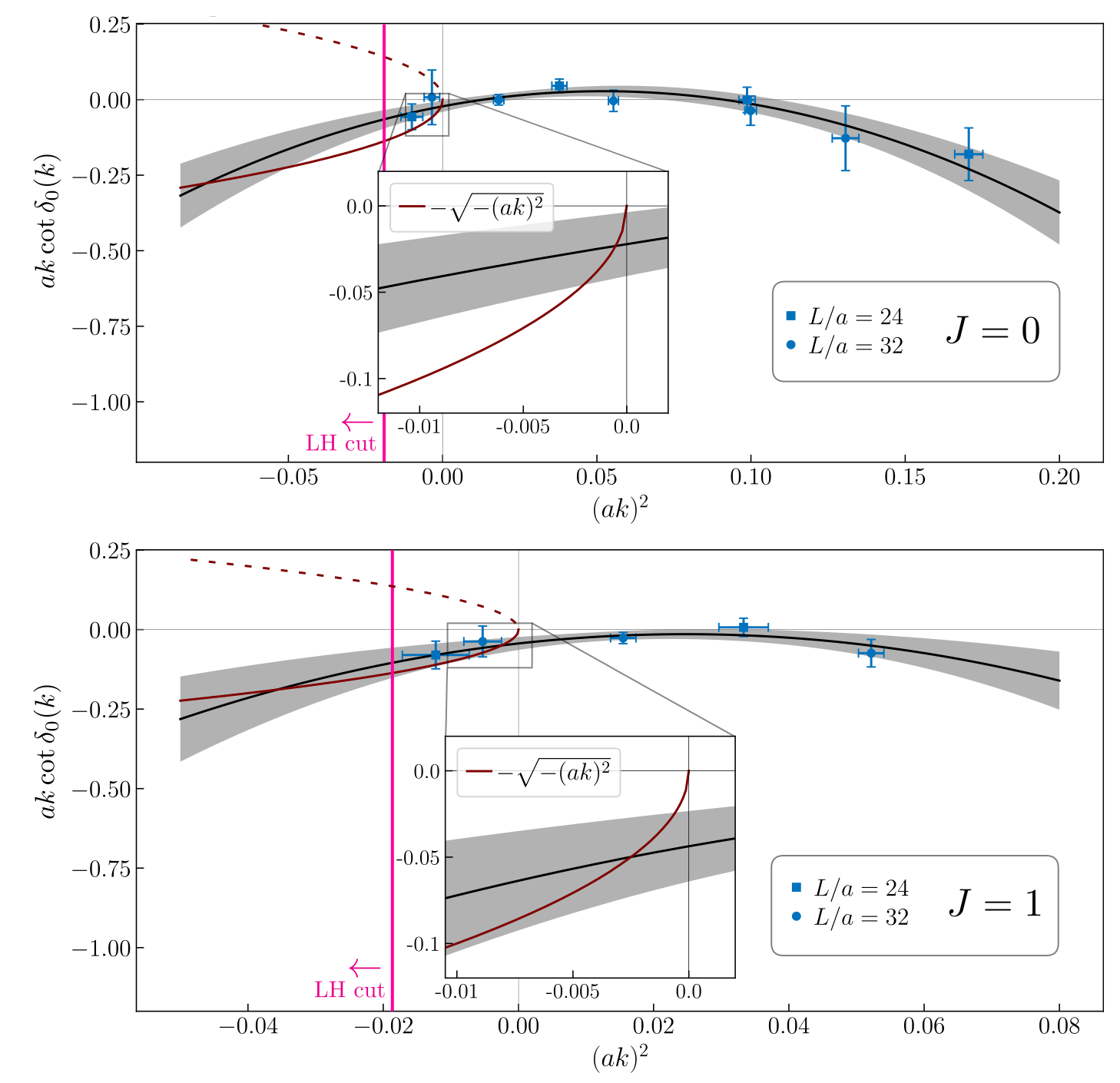}
\caption{\textit{Left: Finite volume energies determined in \cite{Alexandrou:2023cqg} the left part of the figure shows the $I(J^P)=0(0^+)$ $T_{bc}^{ud}$ spectrum while the right shows the $I(J^P)=0(1^+)$. Right: The corresponding scattering phase shifts and subsequent fits to the effective range expansion. The top part shows the $I(J^P)=0(0^+)$ and the bottom the $I(J^P)=0(1^+)$.
}}
\label{fig:wagner-2024-1}
\end{figure}

In their newest study \cite{Alexandrou:2023cqg}, the researchers around the previous authors could significantly update their calculational setup. Using ensembles with $m_\pi\simeq 220~$MeV, $a\simeq 0.12~$fm, $L\simeq 2.88,~3.8~$fm and $m_\pi L\simeq 3.2,~4.3$, they used a combination of Gaussian smeared point sources and stochastic wall sources to form $7\times 7$ and $8\times 8$ symmetric correlation matrices for the study of $DB$ and $DB^*$ scattering, i.e. the tetraquark channels $I(J^P)=0(0^+)$ and $I(J^P)=0(1^+)$ $T_{bc}^{ud}$. The standard GEVP approach is used to determine the finite volume energies, shown in Fig.~\ref{fig:wagner-2024-1}, and the scattering parameters are given in Tab.~\ref{tab:wagner-2024}. 
Although a conclusive statement cannot be made due to the statistical uncertainties, the results hint at a shallowly bound ($E_B\gtrsim -3MeV$) $I(J^P)=0(1^+)$ $T_{bc}^{ud}$ tetraquark, in addition to a new $I(J^P)=0(0^+)$ candidate. 

\begin{figure}[t!]
\centering
\includegraphics[width=0.43\textwidth]{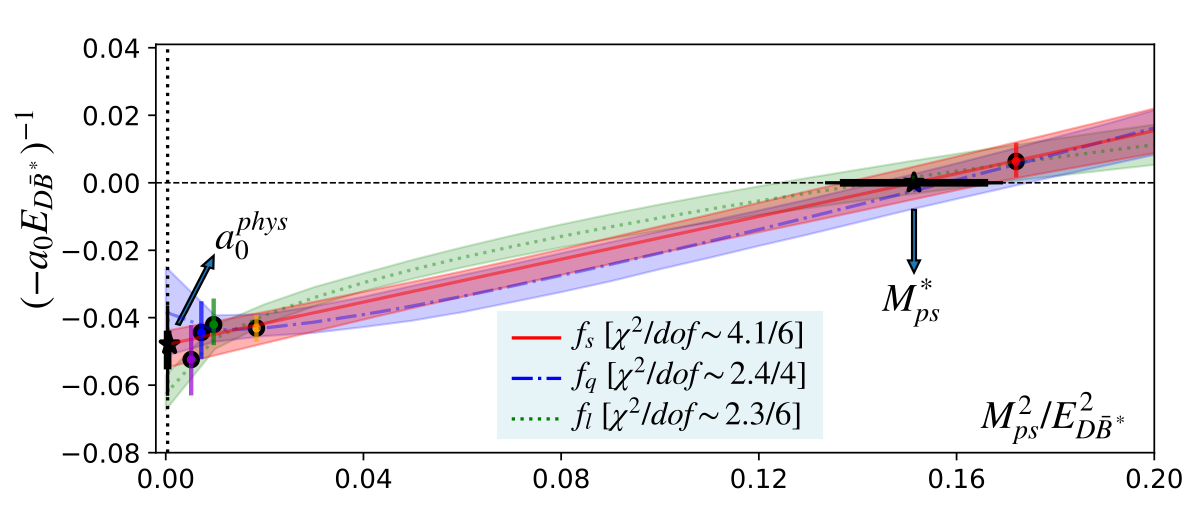}
\includegraphics[width=0.42\textwidth]{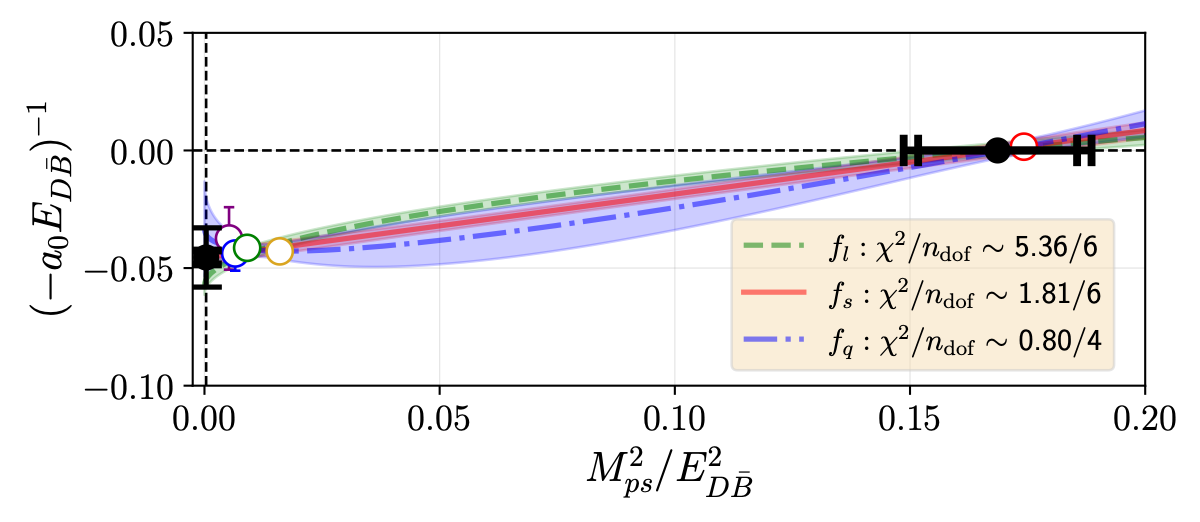}
\caption{\textit{Left: Continuum extrapolated results for $1/a_0$ in the $I(J^P)=0(1^+)$ $T_{bc}^{ud}$ channel \cite{Padmanath:2023rdu}. Right: The same for the $I(J^P)=0(0^+)$ $T_{bc}^{ud}$ channel \cite{Radhakrishnan:2024ihu}. The authors observe bound states in both cases.
}}
\label{fig:padmanath-2024}
\end{figure}

Both candidates are also observed in \cite{Padmanath:2023rdu,Radhakrishnan:2024ihu}, which represent the newest updates by the authors Mathur, Padmanath and others. They use staggered sea quark ensembles provided by the MILC collaboration \cite{MILC:2012znn} and construct their correlation matrix with overlap light and charm quarks in addition to NRQCD bottom quarks. Performing a scattering phase shift analysis, they find binding energies of $E_B\simeq -43(20)~$MeV and $E_B\simeq -39(18)~$MeV for the $I(J^P)=0(1^+)$ and $I(J^P)=0(0^+)$ candidates respectively. An interesting aspect about these studies is that they also provide the scattering lengths from a 1-parameter fit form, both results are shown in Fig.~\ref{fig:padmanath-2024} and their quoted numbers are tabulated in Tab.\ref{tab:summary-table-ot}.

\begin{table}
\centering
\begin{tabular}{cccc}
\hline \hline & $1 / a_0\left[\mathrm{fm}^{-1}\right]$ & $r_0[\mathrm{fm}]$ & $b_0\left[\mathrm{fm}^3\right]$ \\
\hline$J=0$ & $-0.19(16)$ & $0.46(16)$ & $-0.031(8)$ \\
$J=1$ & $-0.37(17)$ & $0.56(30)$ & $-0.080(41)$ \\
\hline \hline
\end{tabular}
\caption{\textit{Results on the scattering parameters determined by fitting an effective range expansion ansatz of \cite{Alexandrou:2023cqg}.}}
\label{tab:wagner-2024}
\end{table}

% -----------------------------
\subsection{Incremental updates through proceedings}

A unique feature of lattice QCD studies is their multi-year maturation times. As a result, it is common that incremental updates are published, especially at the yearly Lattice conference \cite{Bietenholz:2021unf}, which are combined into a full publication after possibly multiple iterations. 
Parts of these updates often contain extra checks, pathological directions explored, and explanations of technological developments that are seen as too detailed for a journal publication. For this review, we attempt to gather all such proceedings as a means to provide a more or less complete resource on what has been done on the lattice connected to the study of doubly heavy tetraquarks. \\
As preparation for an extension to the charm sector, the group of authors around those of \cite{Francis:2016hui} presented updates on the $T_{bc}$ and $T_{cc}$, \cite{Hudspith:2017bbh,Francis:2016nmj}. Further updates on the study of \cite{Francis:2018jyb} with wall-smeared sources have been reported \cite{Colquhoun:2022dte,Colquhoun:2022sip}. These studies also indicated a much lower binding energy for the $J^P=1^+$ $T_{bb}^{ud}$ and $T_{bb}^{us}$ channels than previously reported.
Around the group of authors of \cite{Bicudo:2012qt}, the development towards the spectrum-based approach was documented in \cite{Pflaumer:2021ong,Peters:2016isf,Peters:2017hon,Pflaumer:2020ogv,Wagner:2022bff,Pflaumer:2022lgp} starting from a local-local setup towards the local-nonlocal one and currently towards a full scattering phase shift analysis. Here, we also found the eigenvector component analysis on the $T_{bb}^{ud}$ shown previously. The newest results were recently published as preprint \cite{Alexandrou:2024iwi}.
Simultaneously \cite{Bicudo:2016jwl,Pflaumer:2018hmo,Hoffmann:2022jdx,Mueller:2023wzd} show the incremental advancement in improving the inter-meson potential studies.
Prior to their first work, the authors of \cite{Junnarkar:2018twb} developed their approach in \cite{Junnarkar:2017sey}. Their ongoing study of the $T_{bc}^{ud}$ was reported in \cite{Mathur:2021gqn} and its finalized results were recently submitted as preprint \cite{Radhakrishnan:2024ihu} and quoted above.
Interesting new updates were also reported in \cite{Ortiz-Pacheco:2023ble,Collins:2024sfi,Green:2023es,Baeza:2023es}, whereby the re-opened discussion on including diquark-antidiquark operators in \cite{Ortiz-Pacheco:2023ble} was already mentioned.

%\input{diquarkreview.tex}

%\newpage
\section{Challenges and opportunities}\label{second}
\label{five}

% -----------------------------
\subsection{Emerging physics picture}

\subsubsection{Synopsis of the technical progress in lattice studies so far}
In the preceding section, 30+ years of research on doubly heavy tetraquarks have been summarized. The research started with studying inter-meson potentials and slowly evolved towards dedicated tetraquark studies in the spectrum approach and updated methods to determine potentials. Today, the first lattice determinations of the binding energies and their scattering parameters, i.e., the scattering lengths and effective ranges, are available. 
The evolution of this research program also coincides with the lattice technological advancements in generating better lattice configurations and new methods to determine spectra through lattice correlation functions.

In the first studies, small volumes, often around $L\sim 1-2~$fm and coarse lattice spacings and everything done in the quenched approximation with static and heavy valence quarks were the norm, with some notable exceptions where heavy dynamical quarks were available. Statements that were possible and found consensus were that the $I(J^P)=0(1^+)$ inter-meson potential is attractive for short distances.

With the wide-spread inclusion of dynamical quarks at intermediate pion masses, often around $m_\pi\sim 300-400~$MeV, and larger volumes $L\gtrsim 2~$fm, a new quality of discussion was reached. Frequently still in the potential approach, the studies were extended towards determining ground state energies from fitted model ans\"atze to the lattice data. This led to the first prediction of the $I(J^P)=0(1^+)$ $T_{bb}^{ud}$ with a binding energy $E_B\sim  -90(40)$MeV.
Pushing the envelope further, by managing to obtain signal for almost physical pion masses, the first calculation of this channel in the spectrum approach showed strong evidence for bound $I(J^P)=0(1^+)$ $T_{bb}^{ud}$ and $T_{bb}^{us}$ tetraquarks with binding energies around $E_{B}=-189(13)~$MeV and $-98(10)~$MeV, respectively. 
Around the same time, studies of the $T_{cc}$ using both spectrum and potential approaches, the latter in the form of the HALQCD method, at heavier-than-physical pion masses found a spectrum consistent with a free spectrum, ruling out a resonance, but with insufficient precision to rule out a shallow bound state. 

These developments led to a broad effort to scan for more tetraquark candidates across all possible flavor channels and quantum numbers $I(J^P)$. At the same time, studies were expanded to include multiple pion masses and lattice spacings whenever ensembles were available. Further advances in determining more finite volume energies than just the ground state also enabled first analyses in terms of scattering phase shifts, although the precision was not yet high enough to be strongly constraining. 
To approach this difficulty, studies considered probes of tetraquark structure from alternative avenues, in particular through trial state overlaps. However, those studies proved difficult to interpret and have so far led to contradictory statements on the structure of doubly heavy tetraquarks.

With wide-spread availability of multiple large lattice volumes at low pion masses $m_\pi\sim 200-350~$MeV the program reached the next level, and scattering analyses became possible. This was first shown in the study of the $T_{cc}^{ud}$, which had recently been observed experimentally, where the first full analysis in spectrum approach was carried through and led to the prediction that the $J^P=1^+$ $T_{cc}^{ud}$ is a virtual bound state at slightly below-physical and slightly above-physical charm quark masses and heavier-than-physical light quark masses. Since this study, new results determining scattering phase shifts in various doubly heavy tetraquark results have been coming in and continue to do so. The most current status is that $J^P=1^+$ $T_{bc}^{us}$ is likely also a shallow bound state in nature. At the same time, binding energies for the $J^P=1^+$ $T_{bb}^{ud}$ and $T_{bb}^{us}$ are settling into a lower value than first predicted by the spectrum approach but larger than in the potential approach. It is still too early to tell what the final result will be as new studies are being produced; however it seems safe to say that $-190 \lesssim E_B \lesssim -70~$MeV ($T_{bb}^{ud}$) and $-100 \lesssim E_B \lesssim -40~$MeV ($T_{bb}^{us}$).

\subsubsection{Physics results summary and overview tables}

After summarizing the technical evolution of the research program, we gather and repeat the main insights gained from a physics perspective here.

\paragraph{Stable or not stable?} Note that all studies reviewed here do not consider QED effects and use a pair of mass-degenerate light fermions to describe the up and down quarks.  Often, this is highlighted in notation by using $\ell=u/d$. The focus on QCD also implies a more narrow concept of bound or stable states: In all cases, stability with respect to QCD processes is implied. From an experimental point of view, it can be interesting to consider whether they are also stable under electroweak decays since that would significantly alter the experimental signature. However, this must be evaluated separately based on the QCD binding energy reported in a given lattice study.

\paragraph{Deep or shallow?} Lattice studies often distinguish between a deeply bound state and a shallowly bound state. This is driven by the idea that if the observed binding energy is large enough, there is little chance for the observed ground state to be merely a scattering state induced by the finite lattice volume. 
According to Eq.~\ref{eq:fvol-exp-formula}, the finite volume dependence of a stable state is proportional to an exponential with its binding momentum in the argument. As such, finite volume corrections are exponentially suppressed, and the state should move relatively little from one volume to the next. This relies on the hope that the precision reached and prefactors are such that the exponential behavior can be cleanly resolved from the power law of a scattering state. %, see Eq.~\ref{eq:fvol-plaw-formula}. However, when the binding energy is deep, one may hope to see this.
Of course, what is deep enough is a subjective statement. However, based on the experiences with the finite volume quantization conditions and in other areas of lattice QCD spectroscopy, one could put $E_B\lesssim -50~$MeV as a rule of thumb bound.

\paragraph{Survey and precision results.}  Together, the lattice community has surveyed a large mass parameter space region and many doubly heavy tetraquark flavor combinations. Focusing on the current situation and the overview of lattice results, we give a complete table of studied doubly heavy tetraquark candidates after 2012, in Tab.~\ref{tab:summary-table-bb} for the $T_{bb}^{qq'}$ and Tab.~\ref{tab:summary-table-ot} for the $T_{bc}^{qq'}$, $T_{cc}^{qq'}$ and other channels. Since many studies continue to refine their calculations and scattering analyses have not been achieved for all main channels, it is too early to attempt an averaging to quote a lattice community mean result. It may be interesting to do so in the near future, however.

Together, the tables and discussions above permit a few insights. Firstly, we observe there is strong evidence for deeply bound $I(J^P)=0(1^+)$ doubly heavy tetraquarks with flavor content $bb\bar{u}\bar{d}$ and $bb\bar{u}\bar{s}$. The binding energy of these tetraquarks increases with increasing heavy quark mass $m_Q$. From \cite{Francis:2018jyb}, we see that it grows as $m_Q$ approaches the static limit, i.e. $m_Q\rightarrow \infty$, in the spectrum approach. Furthermore, the binding energy increases as the light quark masses decrease. The deepest binding energies are observed for $\bar{u}\bar{d}$ at physical pion masses, and binding energies with $\bar{u}\bar{s}$ components are consistently shallower.
All three of these predictions are in accordance with the good diquark heavy quark spin symmetry picture. However, some important details need to be clarified. Studies of the mass difference between $\bar{u}\bar{d}$ and $\bar{u}\bar{s}$ good diquarks predict a splitting around $\sim 53~$MeV; for some studies, this is the right ballpark, e.g. \cite{Hudspith:2023loy}, but further work is necessary especially through more studies that treat both channels at the same time in a consistent way. Another aspect is regarding the effectiveness of heavy quark spin symmetry. As such, we can hope the picture describes the $T_{bb}$ situation, but for $T_{cc}$, the dynamics are likely more complicated, with the proposed binding mechanism playing just a minor role, if even. 
For example, the picture also runs into trouble for non-generate heavy quark components in the $T_{bc}$. Considering the general $T_{QQ'}$ case, the $I(J^P)=0(0^+)$ opens up as an interesting new candidate which would be based on a good diquark picture that does not necessitate the effectiveness of heavy quark spin symmetry per se. There are some indications that there are bound states here, both in direct studies \cite{Alexandrou:2023cqg,Radhakrishnan:2024ihu}, and in a study smoothly varying the heavy quark masses \cite{TO2024paper}, similar to \cite{Francis:2018jyb}, see below.

The tables also show that lattice QCD calculations are capable of determining the properties of the $I(J^P)=0(1^+)$ $T_{cc}^{ud}$, the candidate observed in experiment, with good precision. In this channel a scattering analysis is key to success since the binding energy is too shallow to reliably make conclusions otherwise. For example, \cite{Cheung:2017tnt} already studied the finite volume spectrum of this state and could conclude that it did not coincide with that corresponding to a resonance. However, at the level of precision reached, a bound state or no binding at all could not be distinguished. Using multiple masses around the physical parameters and performing a scattering analysis, i.e., determining scattering phase shifts and performing an effective range expansion, is a way to identify this state. 

Here, the appearance of a left-hand cut complicates the scattering analysis significantly. At the same time, this has boosted further development in handling these effects with solid interest from multiple communities converging on the $T_{cc}^{ud}$ as an ideal test-bed system, not least because well-controlled and fairly precise lattice data is available.

With both $0(1^+)$ $T_{cc}^{ud}$ and $T_{bb}^{ud}$ bound, the next interesting candidate is clearly $T_{bc}^{ud}$. Here, the situation is less decided; for one, there is no experimental observation at this time, but further, there is a lively discussion among lattice calculations. 
To some extent, the question is what the appropriate formalism is to handle this state. The scattering analysis will ultimately be the best, but including a bottom quark, the go-to method of distillation for determining the many finite volume states at high precision required for this is not straightforwardly available. If the binding energy of this state were deep, then one might get away with just determining the lowest one, two, or three states like in the $T_{bb}$ case. 
The first study, \cite{Francis:2018jyb} saw a relatively deep binding energy. However, further studies by the group around these authors revealed this to be overestimated and quoted an unbound result in \cite{Hudspith:2020tdf}. Since then, some studies found binding energies that were in the previously discounted range, with the newest result indicating $E_B\simeq-43(20)~$MeV and the scattering length $a_0=0.57(17)$~fm \cite{Padmanath:2023rdu}. The most complete scattering analysis of this system was performed in \cite{Alexandrou:2023cqg}. The authors find scattering parameters that make this state very shallowly bound and similar to the $T_{cc}^{ud}$. 

New results on the $0(0^+)$ $T_{bc}^{ud}$ candidate \cite{Alexandrou:2023cqg,Radhakrishnan:2024ihu,TO2024paper} show indications that there is a bound state in this channel. It is too early to draw firm conclusions, but this is an interesting new development for understanding the binding mechanism as explained above.

Apart from these channels, no further bound doubly heavy tetraquarks have been observed and confirmed at the time of writing. Studies varying the heavy quark masses find more candidates, but at close-to-physical flavor contents the $I(J^P)=0(1^+)$ $T_{bb}^{ud}$, $T_{bb}^{us}$, $T_{cc}^{ud}$ and $T_{bc}^{ud}$ as well as $I(J^P)=0(0^+)$ $T_{bc}^{ud}$ are the only candidates observed by multiple groups. Previously mentioned studies on doubly heavy tetraquark resonances, e.g., \cite{Bicudo:2017szl}, are not further discussed here. The results are listed in the overview table to encourage further work in this direction.

% BB
\begin{table}
\centering
\begin{tabular}{ll|r|cc|c|c|c}
\hline \hline Type & $I(J^P)$ & $E_B[\mathrm{MeV}]$ & $a_0[\mathrm{fm}]$ & $r_0[\mathrm{fm}]$ &  Study & Method & Comments \\
\hline \hline 
\multicolumn{8}{c}{$T_{bb}$ candidates}\\
\hline \hline 
$T_{bb}^{bb}$ & $0(0^+,1^+,2^+)$ & not bound & & &  \cite{Hughes:2017xie}& spectrum &\\
\hline 
$T_{bb}^{cc}$ & $0(0^+)$ & not bound & & &  \cite{Junnarkar:2018twb}& spectrum &\\
\hline 
$T_{bb}^{sc}$ & $\frac{1}{2}(1^+)$ & -8(3) & & & \cite{Junnarkar:2018twb}& spectrum & \\
& & not bound & & & \cite{Hudspith:2020tdf}& spectrum & \\
\hline 
$T_{bb}^{uc}$ & $\frac{1}{2}(1^+)$ & -6(11) & & &  \cite{Junnarkar:2018twb}& spectrum &\\
&& not bound & & &  \cite{Hudspith:2020tdf}& spectrum &\\
\hline 
$T_{bb}^{ss}$ & $0(0^+)$ & not bound & & &  \cite{Junnarkar:2018twb}& spectrum &\\
\hline 
$T_{bb}^{us}$ & $\frac{1}{2}(1^+)$ & -98(10) &  &  &  \cite{Francis:2016hui} &spectrum &\\
&& -87(32) &  &  &  \cite{Junnarkar:2018twb} & spectrum &\\
&& -36 &  &  &  \cite{Hudspith:2020tdf} & spectrum & preliminary\\
&& -86(32) &  &  &  \cite{Meinel:2022lzo} & spectrum &\\
&& -46(12) &  &  &  \cite{Hudspith:2023loy}&spectrum &\\
&& $-30(3)\left({ }_{-31}^{+11}\right)$ &   &  & \cite{Alexandrou:2024iwi} & spectrum & \\
\hline 
$T_{bb}^{ud}$ & $0(1^+)$ & -57(19) to -30(17) &  &  & \cite{Bicudo:2012qt} & static potential &\\
&& -90(40) &  &  &  \cite{Bicudo:2015vta,Bicudo:2015kna} & static potential & single channel\\
&& $-59^{+30}_{-38}$ &  &  &  \cite{Bicudo:2016ooe} & static potential & coupled channel\\
&& -189(10) &  &  &  \cite{Francis:2016hui} & spectrum &\\
&& -143(34) &  &  &  \cite{Junnarkar:2018twb} & spectrum &\\
&& -128(34) &  &  &  \cite{Leskovec:2019ioa} & spectrum &\\
&& -189(18) &  &  &   \cite{Mohanta:2020eed} & spectrum &\\
&& -113 &  &  &  \cite{Hudspith:2020tdf} & spectrum & preliminary\\
&& -112(13) &  &  &  \cite{Hudspith:2023loy} & spectrum &\\
&& -155(17) &  &  & \cite{Aoki:2023nzp} & HAL potential & single channel\\
&& -83(10) &  -0.43(5) &  0.18(6) & \cite{Aoki:2023nzp} & HAL potential & coupled channel\\
&& $-100(10)\left({ }_{-43}^{+36}\right)$ &   &  & \cite{Alexandrou:2024iwi} & spectrum & \\
\hline 
$T_{bb}^{uu}$ & $0(0^+)$ & -5(18) &  &  &  \cite{Junnarkar:2018twb} & spectrum &\\
\hline \hline
\end{tabular}
\caption{\textit{Collected results on all studied doubly heavy tetraquarks after 2012. They are given in descending expected absolute mass order. Only studies with quoted parameters are listed. We refer to the publication for results for the smoothly varying heavy quark mass of \cite{Francis:2018jyb}.}}
\label{tab:summary-table-bb}
\end{table}

% BC
\begin{table}
\centering
\begin{tabular}{ll|r|cc|c|c|c}
\hline \hline Type & $I(J^P)$ & $E_B[\mathrm{MeV}]$ & $a_0[\mathrm{fm}]$ & $r_0[\mathrm{fm}]$ &  Study & Method & Comments \\
\hline \hline 
\multicolumn{8}{c}{$T_{bc}$ candidates}\\
\hline \hline 
$T_{bc}^{us}$ & $\frac{1}{2}(1^+)$ & not bound & & &  \cite{Hudspith:2020tdf}& spectrum &\\
\hline 
$T_{bc}^{us}$ & $\frac{1}{2}(0^+)$ & not bound & & &  \cite{Hudspith:2020tdf}& spectrum &\\
\hline 
$T_{bc}^{ud}$ & $0(1^+)$ & -50 to -15 & & &  \cite{Francis:2018jyb}& spectrum &\\
&& not bound & & &  \cite{Hudspith:2020tdf} & spectrum &\\
&& -50 to -20 & & &  \cite{Meinel:2022lzo} & spectrum &\\
&& $-2.4_{-0.7}^{+2.0}$ & $-2.7(12) $&$ 0.56(30)$ &  \cite{Alexandrou:2023cqg} & spectrum & $J=1$\\
&& $-43\left({ }_{-7}^{+6}\right)\left({ }_{-24}^{+14}\right)$ & $0.57\left({ }_{-5}^{+4}\right)(17)$ & &  \cite{Padmanath:2023rdu} & spectrum & 1-parameter ERE\\
\hline
$T_{bc}^{ud}$ & $0(0^+)$ & not bound & & &  \cite{Hudspith:2020tdf}& spectrum &\\
&& $-0.5_{-1.5}^{+0.4}$ & $-5(4) $&$ 0.46(16)$ &  \cite{Alexandrou:2023cqg} & spectrum & $J=0$\\
&& $-39\left({ }_{-6}^{+4}\right)\left({ }_{-18}^{+8}\right)$ & $0.61\left({ }_{-4}^{+3}\right)(18)$ & &  \cite{Radhakrishnan:2024ihu} & spectrum & 1-parameter ERE\\
\hline \hline 
%
% CC
\multicolumn{8}{c}{$T_{cc}$ candidates}\\
\hline \hline 
$T_{cc}^{ss}$ & $0(0^+)$ & not bound & & &  \cite{Junnarkar:2018twb}& spectrum &\\
\hline 
$T_{cc}^{us}$ & $\frac{1}{2}(1^+)$ & -8(8) & & &  \cite{Junnarkar:2018twb}& spectrum &\\
\hline 
$T_{cc}^{ud}$ & $0(1^+)$ & not resonant & & &  \cite{Cheung:2017tnt}& spectrum &\\
& & -23(11) & & &  \cite{Junnarkar:2018twb}& spectrum &\\
& & $-9.9_{-7.2}^{+3.6}$ & $1.04(29)$ & $0.96\left({ }_{-0.20}^{+0.18}\right)$ &  \cite{Padmanath:2022cvl}& spectrum & heavy charm\\
& & $-15.0_{-9.3}^{+4.6}$ & $0.86(0.22) $&$ 0.92\left({ }_{-0.19}^{+0.17}\right)$ &  \cite{Padmanath:2022cvl}& spectrum & light charm\\
& & $-59\left({ }_{-99}^{+53}\right)\left({ }_{-67}^{+2}\right)$ & $20(20)\left({ }_{-8}^{+8}\right)$ & $1.12(3)\left({ }_{-8}^{+3}\right)$ &  \cite{Lyu:2023xro}& HAL potential & lattice $m_\pi$\\
& &  $-45\left({ }_{-78}^{+41}\right)$ & $-33(44)$ & $1.12(3)$ &  \cite{Lyu:2023xro}& HAL potential & rescaled $m_\pi$\\
\hline
$T_{cc}^{uu}$ & $0(0^+)$ & not bound & & &  \cite{Junnarkar:2018twb}& spectrum &\\
\hline \hline 
% 
% others table
\multicolumn{8}{c}{Other candidates}\\
\hline \hline 
$T_{bb}^{ud}$ & $0(1^-)$ & & Re$(E)=17^{+4}_{-4}$ & $-2\textrm{Im}(E)=112^{+90}_{-103}$ &  \cite{Bicudo:2017szl}& static potential & resonance\\ \hline
$T_{bs}^{ud}$ & $0(1^+)$ & not bound & & &  \cite{Hudspith:2020tdf}& spectrum &\\
\hline 
$T_{bs}^{ud}$ & $0(0^+)$ & not bound & & &  \cite{Hudspith:2020tdf}& spectrum &\\
\hline 
$T_{cs}^{ud}$ & $0(1^+)$ & not bound & & &  \cite{Hudspith:2020tdf}& spectrum &\\
\hline 
$T_{cs}^{ud}$ & $0(0^+)$ & not bound & & &  \cite{Hudspith:2020tdf}& spectrum &\\
\hline \hline
\end{tabular}
\caption{\textit{Continued collected results on all studied doubly heavy tetraquarks after 2012. They are given in descending expected absolute mass order. Only studies with quoted parameters are listed. We refer to the publication for results for the smoothly varying heavy quark mass of \cite{Francis:2018jyb}.}}
\label{tab:summary-table-ot}
\end{table}

\begin{figure}[t!]
\centering
\includegraphics[width=0.38\textwidth]{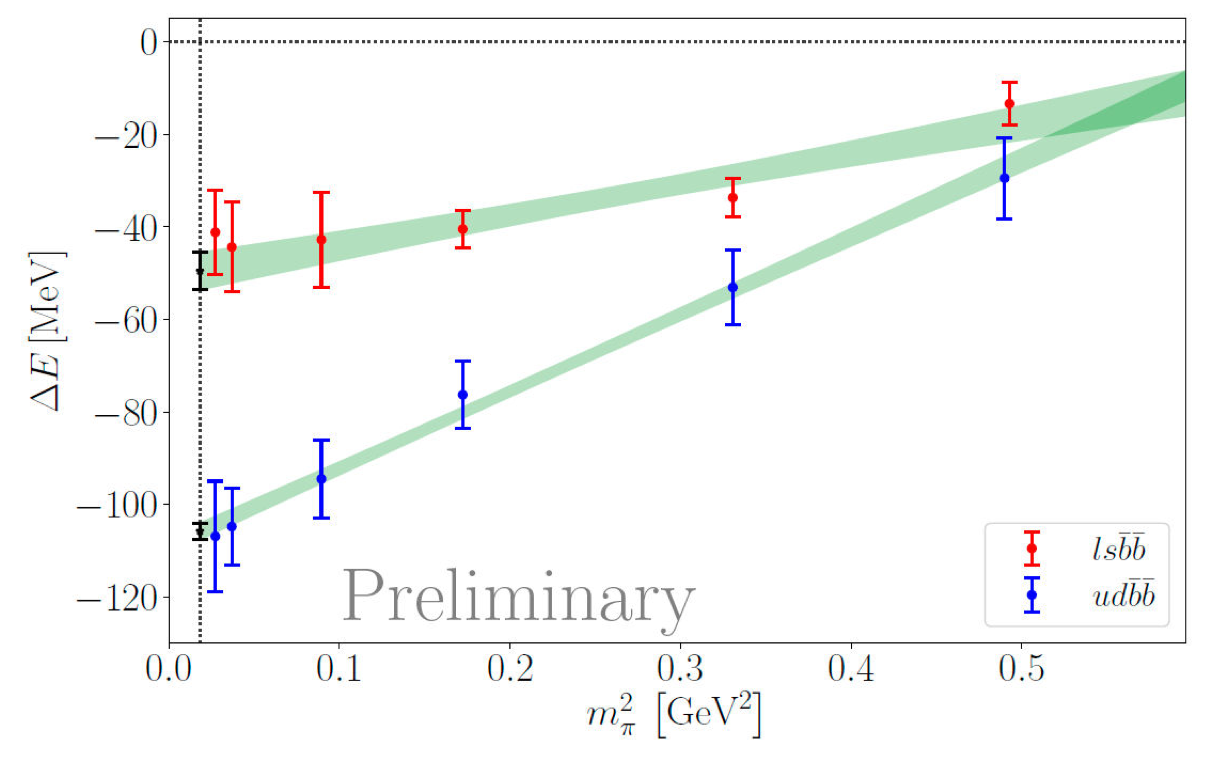}\\
\caption{\textit{Preliminary pion mass dependence of the binding energies of the $J^P=1^+$ $T_{bb}^{ud}$ and $T_{bb}^{us}$ in \cite{Colquhoun:2022sip}.}}
\label{fig:brian-1}
\end{figure}

\subsubsection{Limitations and avenues for advancements of current studies}

With the progress and positive developments in the last few years, it is worth looking at current limitations to see where possible future advancements might come from.
On the side of lattice setups, current calculations are still suffering from not having enough ensembles at the physical pion mass. Especially the study of doubly heavy tetraquarks shows a dependence on the details of the light quarks that is not mild. To illustrate we show the pion mass dependence of the $1^+$ $T_{bb}^{ud}$ and $T_{bb}^{us}$ from a current, yet preliminary, study presented in \cite{Colquhoun:2022sip} in Fig.~\ref{fig:brian-1}.
Non-physical pion masses seem sufficient for qualitative statements, but quantitative precision determinations require better control over the chiral behavior.
In the case of the $T_{cc}^{ud}$, this becomes even more important as the physical binding energy is $E_B\sim -0.3~$MeV. Here also the study using the HALQCD method \cite{Lyu:2023xro}, with its external rescaling of the pion mass in the potential ansatz, showed a change in the interpretation of the state and suggested that the system is sensitive to the light quarks, see the discussion around Eq.~\ref{eq:hal-tcc-fit}.
At the same time, while the light quark details are important, the heavy quark details are equally so. This is intuitively clear and was shown numerically, e.g. in \cite{Francis:2018jyb}, however especially in the $T_{cc}^{ud}$ this can change the interpretation, as shown in \cite{Padmanath:2019ybu}.
From the point of view of tuning the heavy quarks more accurately, new methods were presented in \cite{Hudspith:2023loy,Hudspith:2021iqu}.
The study \cite{Hudspith:2023loy} finds that the discretization effects are the leading systematic and need to be addressed. This means ideally, the use of effective relativistic heavy quark actions or non-relativistic heavy quark actions should be phased out. This is only possible through using ensembles at finer lattice spacings, which are difficult to obtain due to topological freezing. An alternative, stop-gap solution is using higher improved quark actions, such as Domain Wall and Overlap. However, these are also limited and today can only reach the charm quark regime with a high level of confidence\footnote{Staggered quark actions in the valence sector are not mentioned here since the operator construction and evaluation is often more difficult due to some ambiguity as to the lattice quantum numbers. The mentioned actions do not have this feature and are, therefore, easier to use.}.
A further limitation are the available volumes. For lattice studies in the spectrum approach, volumes with $m_\pi L\gtrsim 4$ and $\lesssim 6$ are ideal since then finite volume effects should be parametrically under control on the one hand and the spectrum not too dense for the GEVPs to reliably extract finite volume energies on the other. Many studies do not have access to multiple volumes in the required range. Note that at $m_\pi=135~$MeV the required physical volume is $L\simeq 6~$fm, which at current lattice spacings mandates volumes with $L/a\gtrsim 64$ or 92 if the charm quark is to be resolved with a fully relativistic quark action. This adds a layer of complication for the distillation approach.
As mentioned, these points are, in the end, requests to the lattice parameters of the gauge configurations used and require developments in generating lattice QCD gauge ensembles.\\
On the side of spectral analysis, to make progress, a better understanding of how to analyze the scattering phase shifts is required. As highlighted below, the nearby left-hand cut in the $T_{cc}^{ud}$ poses a significant challenge on the side of the formalism. 
New pathways to control the systematics in the potential approach also need to be developed to boost trust in the findings derived from the fitted potential ans\"aetze. 
To determine many finite volume energies, a consensus is emerging on using the GEVP method for the best results. However, the problem of non-symmetric correlation matrices persists. Furthermore, the symmetric matrices that are being employed today are often determined via the distillation approach. Since this relies heavily on smearing, i.e., on inducing a nonlocal component in the operator this introduces a bias
towards di-meson interpolating operators from the point of view of trial state selection.
The study of the inclusion of diquark-antidiquark operators mentioned above contributes and connects to this discussion.
Furthermore, determining finite volume energies and scattering phase shifts is an advanced application in lattice spectroscopy that is challenging at each step and brings new challenges with each channel considered.

%---------------------------------
\subsection{Molecules, compact states and binding mechanisms in lattice QCD}
\label{sec:weinberg}

One question consistently being asked, especially about the $T_{bb}$ tetraquarks, is whether or not they are compact or molecular states and whether it really is the good diquark heavy quark spin symmetry picture that is driving the binding as its mechanism. 
%In the GEVP analysis, it is tempting to see the operators as trial state selections and to interpret the result of the variational analysis in terms of overlap factors. %As discussed in Sec.~\ref{sec:structure}, a larger overlap with the ground state from a diquark-antidiquark interpolating operator would be interpreted in favor of the associated binding mechanism.
 However, as discussed, proper control over the operator structures and an unbiased basis selection have not yet been shown. Ultimately, the goal should be to infer this information from properly defined and determinable observables, such as the scattering parameters.
The task then returns to the fundamental question of how compositeness can be decided from low-energy scattering properties. The Weinberg criterion briefly highlighted below gives one way to do this for very specific cases, but it is not clear whether a fully general connection can be found.

%\paragraph{Compositeness and the Weinberg criterium.}
First put forward in 1965 in \cite{Weinberg:1965zz}, the Weinberg compositeness relations are based on the idea that if there is a bound state in a system then the wave function $|\Psi\rangle$ is given by the superposition of a compact ($z |\psi_0\rangle$) and a continuum ($\zeta(\vec{p})| \phi_1 \phi_2\rangle_{\vec{p}}$)  part. The probability of finding a compact component is then $|\langle \psi_0| \Psi\rangle|^2=z^2=:Z$, and, assuming the decomposition is complete, $X=1-Z$ the probability of finding a molecular component.
The Weinberg compositeness relations connect this idea with the scattering length and effective range:
\begin{align} 
a_0&=\frac{2(1-Z)}{(2-Z)} R+\mathcal{O}\left(m_\pi^{-1}\right)~,~~
r_0=\frac{-Z}{(1-Z)} R+\mathcal{O}\left(m_\pi^{-1}\right)~,\nonumber\\
\Leftrightarrow a_0 & =-\frac{2 X}{1+X} R+\mathcal{O}\left(m_\pi^{-1}\right)~,~~
r_0 =-\frac{1-X}{X} R+\mathcal{O}\left(m_\pi^{-1}\right)~,
\end{align}
where $R=1/\sqrt{2\mu |E_B|}$ is the classical radius of the state. In the language of a short distance EFT, see also \cite{vanKolck:2022lqz}, they are related to the scattering phase shift:
\begin{equation}
p \cot \delta^{(\ell=0)}(p)=-R\left[1+\frac{Z}{2(1-Z)}\left(1+\frac{p^2}{R^2}\right)\right] + ...~
\end{equation}
where they play the role of a leading term. 
Through determining $E_B$, $a_0$ and $r_0$ also $R$ is fixed and $Z$, likewise $X$, can be computed. Note that since the relations can be understood as terms in an expansion, it was found that the range of $X$ is not necessarily bounded to 1. However, a generalization has been worked out \cite{Li:2021cue}.
Given the derived relations we have a compact state ($X\simeq 0$), in the case where $a_0\sim 0$ (small) and $r_0\ll 0$ (large and negative). This means a tell-tale sign would be a large and negative effective range.
At the same time, for a molecular state ($X\simeq 1$) we should expect $a_0\gg$ (large) and $r_0\gtrsim 0$ (small but usually positive). In both scenarios, an assumption is that the system needs to firstly be a bound state, and secondly that only $S$-wave scattering is considered.
To give a recent example, \cite{Li:2024pfg} uses this reasoning to assess whether the $X(3872)$ is dominantly a $D\bar{D}^*$ molecule. Their study uses two-flavor QCD ensembles to perform a scattering analysis. The scattering parameters they find are $a_0=-4(1)~$fm and $r_0=0.187(38)~$fm with a binding energy of $E_B\simeq-1.3(1)~$MeV. Using the Weinberg criterion, the near-threshold binding energy and small, positive effective range imply a compositeness $X\simeq 1$, indicating a predominantly $D\bar{D}^*$ state. As a side remark, note that the existence of a left-hand cut also complicates this analysis.

\begin{figure}[t!]
\centering
\includegraphics[width=0.48\textwidth]{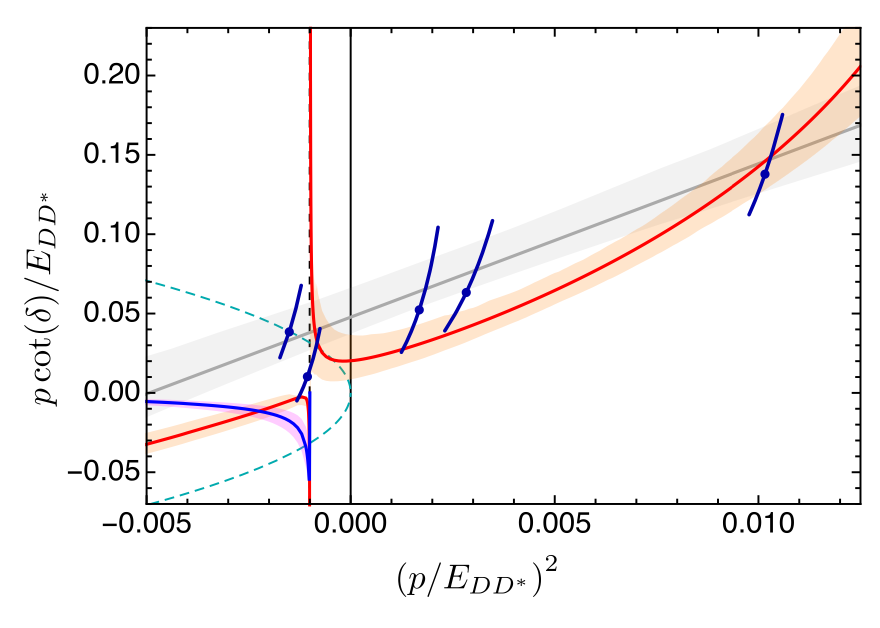}
\caption{\textit{Results from re-analyzing the data of \cite{Padmanath:2022cvl} using an OPE approach in \cite{Du:2023hlu}. The OPE form (red) explicitly incorporates the effects of one-pion exchange and adapts the effective range expansion. This incorporates the information of the left-hand cut into the fit.
}}
\label{fig:lefthandcut-1}
\end{figure}

Overall, this brief discussion illustrates why it is so difficult to make robust statements on the compositeness of a given state from lattice QCD. Indeed, the first sentence of Weinberg's paper \cite{Weinberg:1965zz} reads: "Many physicists believe that low-energy experiments can never decide whether a given particle is composite or elementary." 
Before going on to argue that in the case of the deuteron and very shallowly bound states in $S$-wave scattering, model-independent statements can be made. As shown, lattice QCD can provide the scattering parameters through advanced spectroscopic methods similar to those of low-energy experiments.
An approach in full generality, especially in the presence of non-analyticities in the effective range expansion, as it would be required for robust use in lattice QCD, however, is not (yet) available.
Note that although one is tempted to apply the above reasoning for the compositeness also for the doubly heavy tetraquarks using the tabulated results in Tab.~\ref{tab:summary-table-bb} and Tab.~\ref{tab:summary-table-ot}, it is still too early for such an analysis.

%---------------------------------
\subsection{Left-hand cut and extended finite volume scattering}
\label{sec:lefthandcut}

Despite the highlighted successes in the scattering analysis of $T_{cc}^{ud}$ on the lattice, in many ways, this research is still in the development stage. For example, one should expect a non-analyticity in the effective range expansion due to the presence of a left-hand cut. The left-hand cut appears as the $D^*$ in principle, may exchange a pion with $D$ when it is stable or decays into $D\pi$ when it is not. Both cases are relevant for lattice calculations since the input quark masses control whether the second process is allowed.
The situation causes another problem for lattice calculations as it affects the effective range expansion on the analysis or interpretation side, and the determination of the scattering phase shifts from the lattice finite volume energies. That is because the derivation of the finite volume quantization conditions or L\"uscher formulas \cite{Luscher:1986pf} builds upon assumptions on analyticity properties of the phase shifts, see also \cite{Hansen:2024ffk} for a discussion.
It is then clear, that while throughout the current studies the 2-to-2 finite volume scattering formalism was applied, ultimately one needs to include the three-particle channel $DD\pi$ in some way and that some improved formalism is required.

\begin{figure}[t!]
\centering
\includegraphics[width=0.58\textwidth]{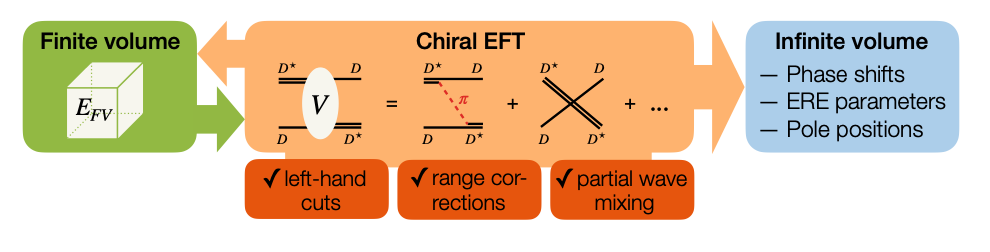}
\caption{\textit{Schematic illustration of the approach developed in \cite{Meng:2023bmz}. From left to right, the pipeline starts with the lattice-determined finite volume energies and evaluates them in an EFT framework to arrive at the scattering phase shifts. These can then be analyzed using the expanded approach from \cite{Du:2023hlu}.  }}
\label{fig:lefthandcut-2}
\end{figure}

\begin{figure}[t!]
\centering
\includegraphics[width=0.88\textwidth]{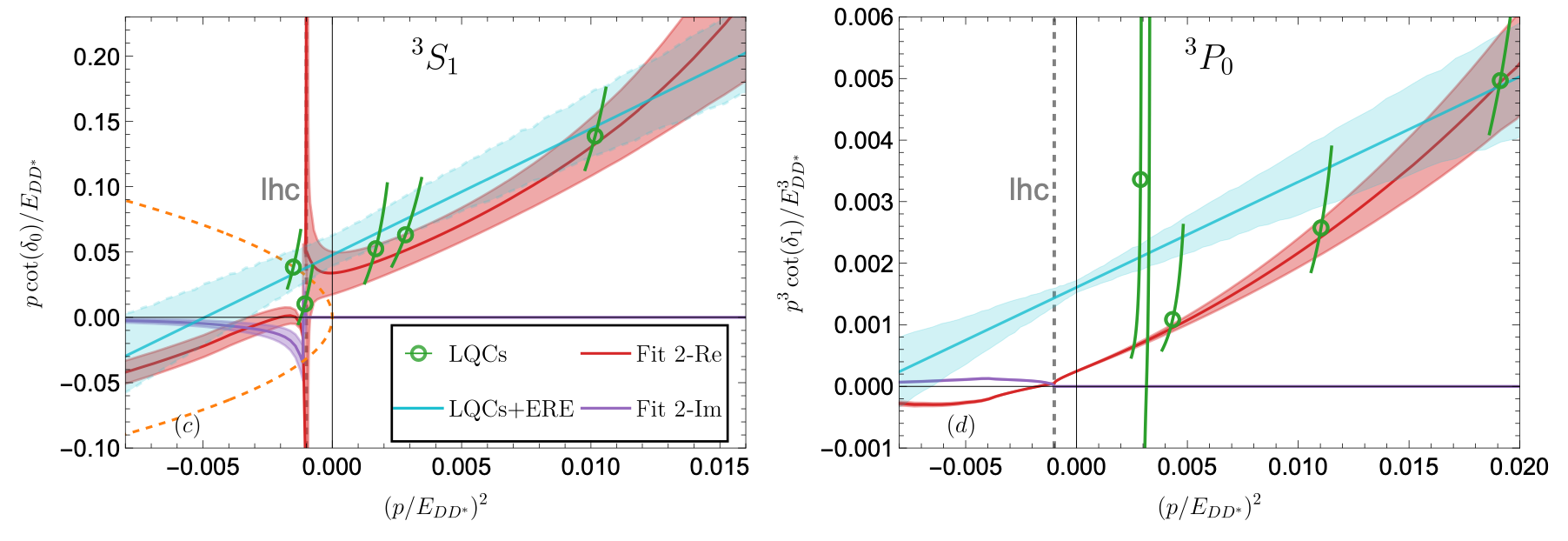}
\caption{\textit{Application of the pipeline presented in Fig.~\ref{fig:lefthandcut-2} in \cite{Meng:2023bmz}. Shown are the results for the $^3S_1$ (left) and $^3P_0$ (right) partial waves calculated from the finite volume energies of \cite{Padmanath:2022cvl}. The panels show the results of fitting this data to the expanded OPE-based approach in red and using the standard effective range ansatz in blue.}}
\label{fig:lefthandcut-3}
\end{figure}

Recently, progress has been made in this direction in \cite{Du:2023hlu,Hansen:2024ffk,Du:2021zzh,Meng:2023bmz,Raposo:2023oru,Bubna:2024izx}. Focusing on just a few, the studies \cite{Du:2023hlu,Meng:2023bmz} use the lattice QCD results of \cite{Padmanath:2022cvl} to study these issues. As such, in \cite{Du:2023hlu}, the authors developed an extended framework to understand the effective range expansion with left-hand cuts present. Their findings are illustrated in Fig.~\ref{fig:lefthandcut-1}, where the lattice determined results in $p\cot \delta$ are analyzed once using the standard effective range expansion, similar to \cite{Padmanath:2022cvl}, in grey and once using their proposed OPE based approach in red. Later in \cite{Meng:2023bmz}, the group around the authors also addressed the problem of converting the lattice finite volume energies to scattering phase shifts. Using the same data as before, they derived an EFT-based approach that can be used as an alternative to the L\"uscher formula-based approach used so far. A schematic representation of the approach is given in Fig.~\ref{fig:lefthandcut-2}. Results applying the new method, as well as using the OPE-based analysis, are given in Fig.~\ref{fig:lefthandcut-3} for $^3S_1$ (left) and $^3P_0$ (right) partial waves.

%---------------------------------
%\subsection{Pentaquarks}

%uudbs not bound?

%---------------------------------
\subsection{The role of diquarks}
\label{sec:diquarks}

\begin{figure}[t!]
\centering
\begin{minipage}{0.45\textwidth}
\begin{tabular}{ccc}
\hline \hline All in $[\mathrm{MeV}]$ & $\delta E_{\text {lat }}\left(m_\pi^{\text {phys }}\right)$ & $\delta E_{\text {pheno }}$ \\
\hline$\delta\left(1^{+}-0^{+}\right)_{u d}$ & $198(4)$ & $206(4)$ \\
$\delta\left(1^{+}-0^{+}\right)_{\ell s}$ & $145(5)$ & $145(3)$ \\
$\delta\left(1^{+}-0^{+}\right)_{s s^{\prime}}$ & $118(2)$ & \\
\hline$\delta\left(Q[u d]_{0^{+}}-\bar{Q} u\right)$ & $319(1)$ & $306(7)$ \\
$\delta\left(Q[\ell s]_{0^{+}}-\bar{Q} s\right)$ & $385(9)$ & $397(1)$ \\
$\delta\left(Q[\ell s]_{0^{+}}-\bar{Q} \ell\right)$ & $450(6)$ & \\
\hline \hline
\end{tabular}
\end{minipage}
\begin{minipage}{0.4\textwidth}
\includegraphics[width=0.99\textwidth]{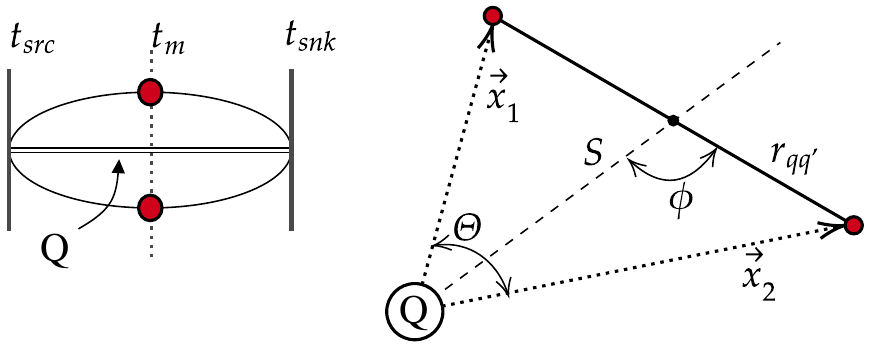}
\end{minipage}
\caption{\textit{Left: Diquark-diquark and Diquark-quark mass differences in the spectator embedding approach \cite{Alexandrou:2005zn}, results shown from \cite{Francis:2021vrr}, see this reference for details. Right: Generic geometry of the density-density correlations was used to study the diquark attractive effect further.}}
\label{figtab:diquarks-1}
\end{figure}

A complementary approach to studying the structure of tetraquarks and shedding light on the effective binding mechanism in a phenomenological sense 
could be to study diquarks and their properties themselves in lattice QCD calculations. Crucial properties would be the diquark splittings between the different possible channels, the confirmation of an attraction in the good diquark channel, and the effective size of the diquark. The latter is useful for quark model building, which needs to introduce this feature if it exists.
Recall that formally, the interpolating operator of a diquark may be written as
$D_\Gamma =q^c C \Gamma q'$
where $q,q'$ denote two different quark flavors, $c,C$ indicate charge conjugation, and $\Gamma$ acts on Dirac space and cycles through the 16 standard combinations of $\gamma_5,\gamma_\mu,...$, as usual.
The interpolating operator highlights a key difficulty for lattice QCD calculations: 
Diquarks are colored objects, i.e., they are not gauge-invariant, and the only direct way to access them is via gauge-fixed approaches \cite{Hess:1998sd,Bi:2015ifa,Babich:2007ah,Teo:1992zu,Negele:2000uk,Alexandrou:2002nn}. However, in this situation, masses and sizes become gauge-dependent quantities. Although it could be that the gauge dependence is small, a gauge-invariant approach would be beneficial.
In \cite{Alexandrou:2005zn,Alexandrou:2006cq}, this issue was addressed by forming a gauge-invariant probe to diquark properties by embedding them in hadrons that contain a single static quark as a spectator. The correlator then formally permits a mass-decomposition \cite{Alexandrou:2005zn,Alexandrou:2006cq,Orginos:2005vr,Green:2010vc}:
\begin{equation}
    C_{\Gamma}(t)=\sum_{\vec x} \Big\langle [D_\Gamma Q](\vec x,t)~[D_\Gamma Q]^\dagger(\vec 0,0)  \Big\rangle~~\rightarrow ~~C_{\Gamma}(t)\sim \exp\left[-t\left(m_{D_{\Gamma}} + m_Q + \mathcal{O}(m_Q^{-1})\right)\right]~~.
\end{equation}
We see that in mass differences the spectator quark is exactly canceled out, and we arrive at a gauge-invariant approach to diquark properties. %Note that absolute values for diquark masses are not accessible in this way, which is also reassuring in some sense, given their gauge-variant nature.
Calculating the possible mass differences between a static-light-light or strange baryon and a static-light or strange meson in \cite{Francis:2021vrr}, the authors arrive at the splittings given on the left of Fig.~\ref{figtab:diquarks-1} (left). The setup uses the PACS-CS ensembles used by a subset of the authors before and includes all pion masses $m_\pi=164,~299,~415,~575$ and 707~MeV as well as a quenched study to compare with previous works.
Overall, the results confirm expectations from other lattice studies and the PDG. 

\begin{figure}
\centering
\begin{minipage}{0.45\columnwidth}
%\vspace{3ex}
\includegraphics[width=0.99\columnwidth]{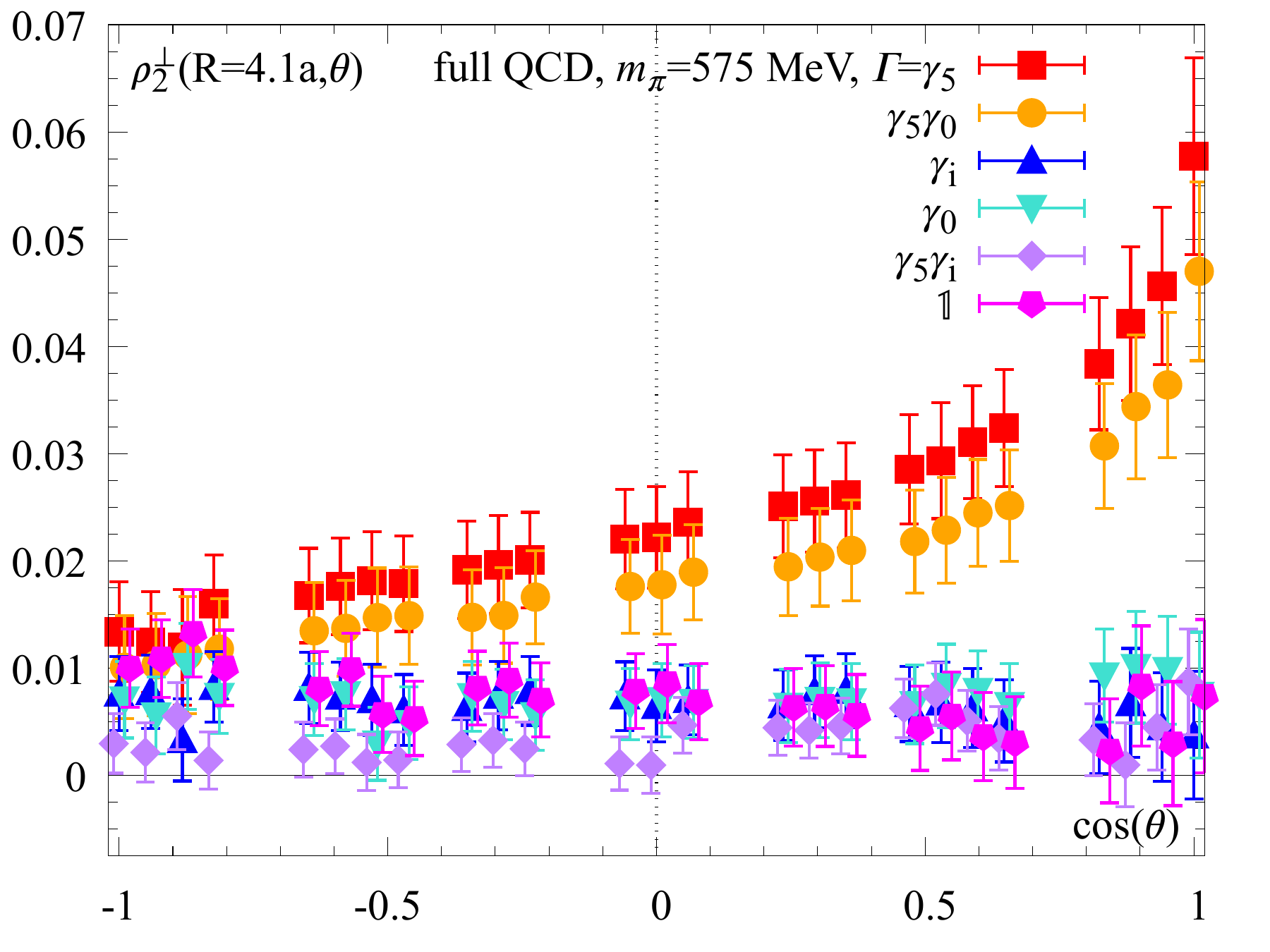}
%\vspace{-8pt}
\end{minipage}
\begin{minipage}{0.45\columnwidth}
\vspace{-0ex}
{\includegraphics[width=0.79\columnwidth]{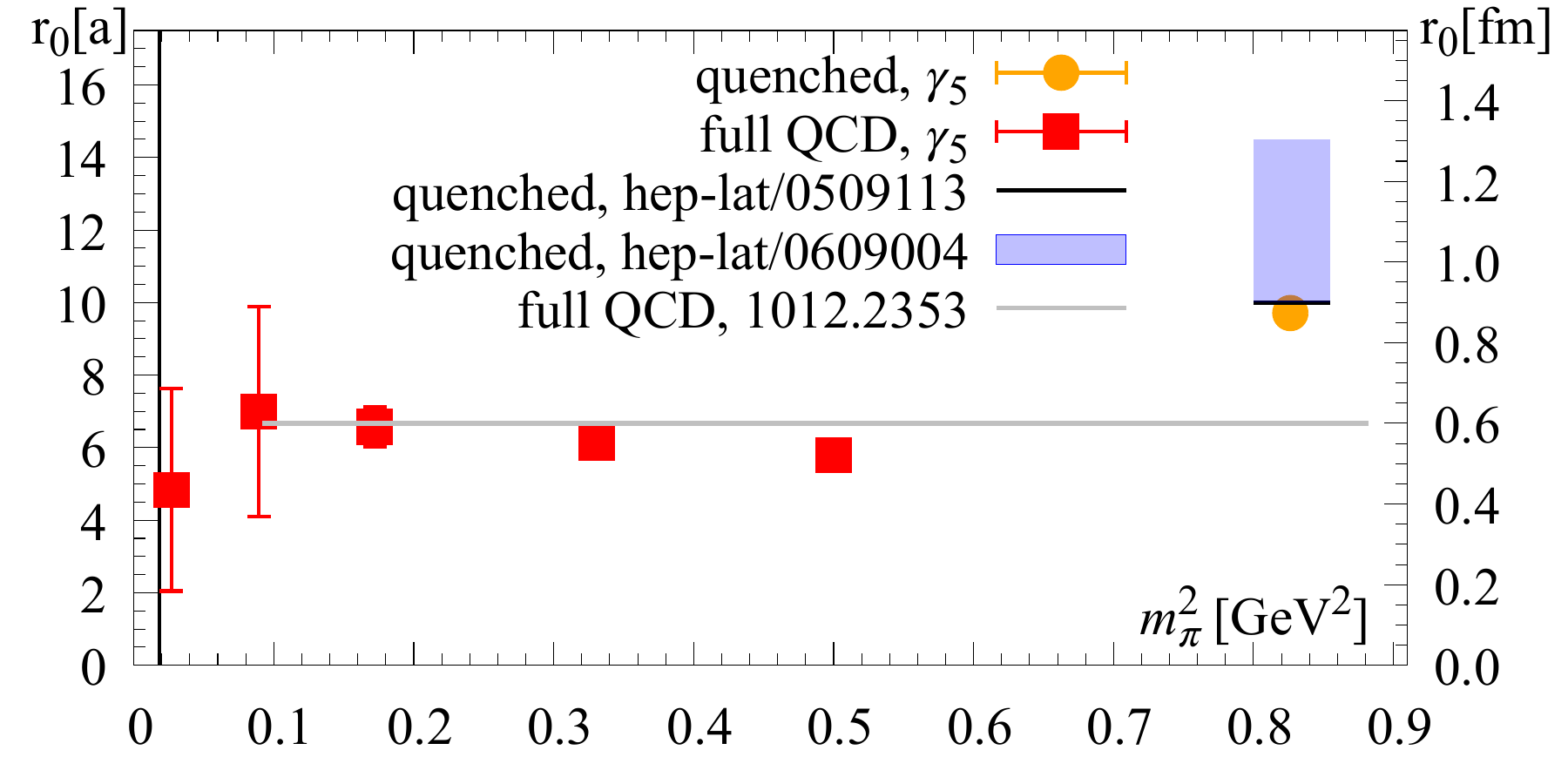}}
{\includegraphics[width=0.79\columnwidth]{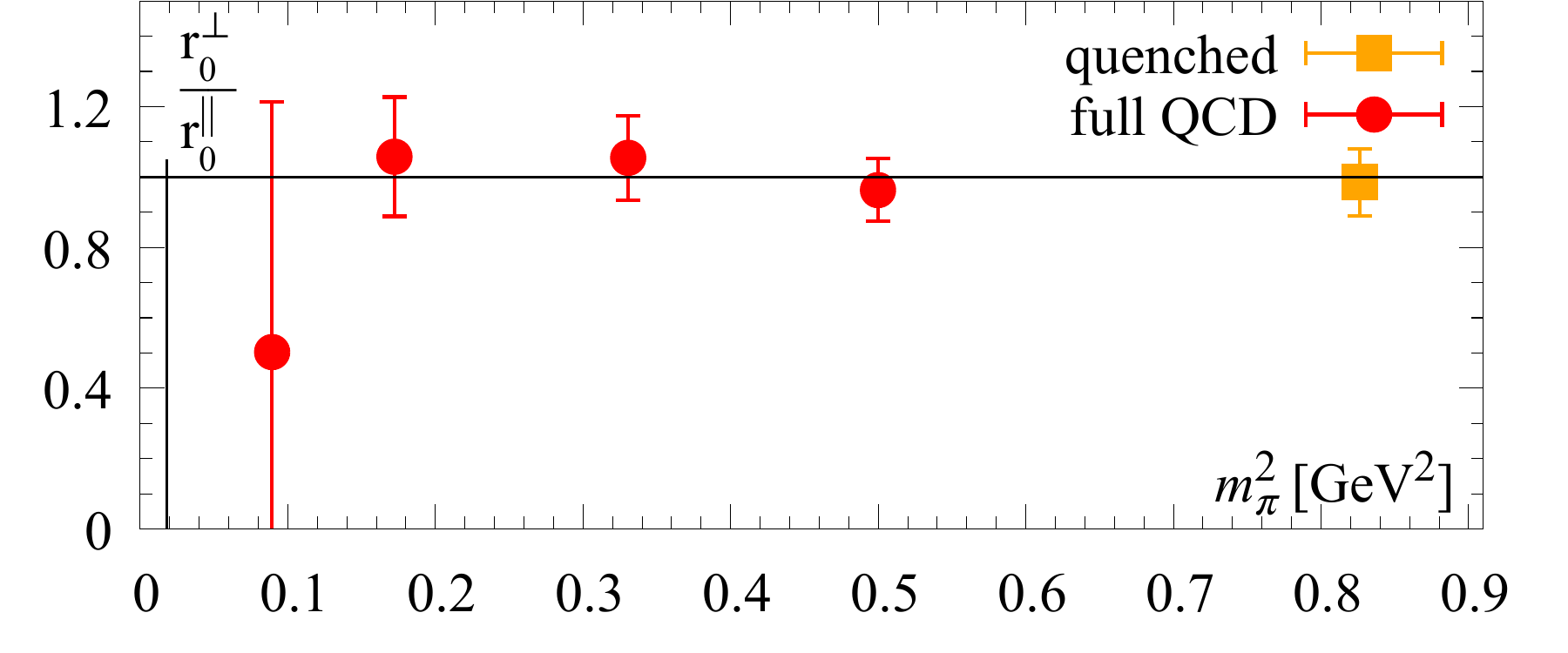}}
\end{minipage}
\caption{\textit{Left: Visualizing the good diquark attractive effect \cite{Francis:2021vrr}.
Top right: Good diquark size $r_0$ versus
$m_\pi^2$. Bottom right: Ratio of radial and tangential sizes over $m_\pi^2$.
}}
\label{fig:diquark-2}
 \end{figure}

Going further, the embedded configuration can be used to define a measure for the diquark structure through density-density correlations. These are lattice correlation functions in which two currents are inserted in the ground state regime:
\begin{align}
    C_{\Gamma}^{dd}(\vec x_1, \vec x_2, t)=
    \Big\langle B_\Gamma(\vec 0,2t)
    \rho(\vec x_1,t)\rho(\vec x_2,t)
    B_\Gamma^\dagger(\vec 0,0)  \Big\rangle
\end{align}
where $\rho(\vec x,t)=\bar{q}(\vec x,t)\gamma_0 q(\vec x,t)$. By studying the spatial correlations of the two current insertions using the spectator as a point of reference, the internal structure of the diquarks can be probed. The general geometry of these correlators and spatial correlations studied is depicted in Fig.~\ref{figtab:diquarks-1} (right); see \cite{Francis:2021vrr} for more details.
A special case of the possible geometries is where spatial correlations are considered a fixed distance $R$ away from the static quark. Relative to one of the current insertions, all spatial correlations with the other current can be quantified by an angle $\theta$ that encodes the distance between the two currents. The results at a fixed pion mass and for all available diquark channels are shown in Fig.~\ref{fig:diquark-2} (left). The case of minimal separation in the figure is $\cos(\theta)=1$. In the short distance regime, an increase in correlation, signaling attraction, is observed exclusively in the channel with $\Gamma=\gamma_5,\gamma_5\gamma_0$, in accordance with the good diquark picture. This is confirmed for all quark masses considered in the study. 
The good diquark can be further studied by considering the distance between the test currents $r_{qq^\prime}=R\sqrt{2(1-\cos(\Theta ))}$ and re-interpreting the data according to an exponential decay with an effective diquark size radii $r_0^d$ as exponent, $\sim \exp(-r_{qq^\prime}/r^d_0)$. The results for all available quark masses are shown in Fig.~\ref{fig:diquark-2} (top right) and indicate an effective hadronic size of the good diquark. Finally, one may further separate the radii of this type in the radial and tangential directions of the sphere around the static quark. The results for all available quark masses are shown in Fig.~\ref{fig:diquark-2} (bottom right) and, in a ratio, show now significant separation, implying an almost spherical shape for the good diquarks.

%\newpage
\section{Brief summary and conclusion}\label{sec:sum}

This review summarizes the research by the lattice QCD community to study doubly heavy tetraquarks. The period covered starts around 1990 and ends with the first few months of 2024. Despite the long period, this field is fresh with many studies ongoing and a steady stream of new work being published or presented. Indeed, since the experimental observation of the $T_{cc}^{ud}$, there has been an accelerating publication rate as the community discovers doubly heavy tetraquarks as a test-bed for new developments in finite volume spectroscopy, both on the technical and formal side. Simultaneously, studying these states poses the highest requirements on the underlying lattice ensembles. With special emphasis on having physical light quark masses and well-controlled heavy quarks alongside volume constraints and the expectation of significant discretization effects, all main lattice systematics must be controlled at the highest levels.
Further, in recent times, a new division of labor has emerged in which lattice data on finite volume energies is being analyzed independently by groups that are experts in scattering phase shift analysis. Without necessarily being lattice practitioners themselves, this is a unique and exciting opportunity for cross-boundary research and an interesting dynamic.
In the future, new studies already underway will further strengthen the gathered catalog of findings. They will, in particular, add more precise and accurate binding energies and hopefully soon also scattering parameters for the $T_{bb}$ tetraquarks, which are still lacking, although these candidates have been studied the most so far.\\
With this review, the author hopes to have gathered a resource that will be useful for practitioners and those wanting an overview. 
It is the nature of these things that some papers might have been missed, and no claim of completeness is being made. Any errors and opinions are the sole responsibility of the author.

%end of the core of the manuscript

%\newpage
\section*{Acknowledgments}
It is a great pleasure to thank Christian Fischer for suggesting to write this review and his endless patience in waiting for its submission.
Since taking up this research in early 2015, the author has had innumerable positive conversations with members of the hadron spectroscopy community, with special mentions going to 
Sinya Aoki, 
Gunnar Bali,
Vanya Belyaev,
Pedro Bicudo,
Raul Brice\~no,
Brian Colquhoun, 
Takumi Doi, 
Philippe de Forcrand,
Jeremy R. Green,
Maxwell T. Hansen, 
Renwick J. Hudspith, 
Parikshit Junnarkar,
Marek Karliner,
Randy Lewis,
Padmanath Madanagopalan,
Kim Maltman,
Nilmani Mathur, 
Mikhail Mikhasenko,
Daniel Mohler,  
%Martin Pflaumer
William Parrott, 
Sasa Prelovsek,
Christopher E. Thomas, 
Marc Wagner and
David Wilson.
This work was supported by the National Science and Technology Council of Taiwan under grant 111-2112-M-A49-018-MY2.

%\section*{Author's contributions \textit{(optional section)}}
%Detailing here the contributions of the authors of the review.

\bibliography{references.bib}
%Please use Bib\TeX\ to generate your bibliography and include DOIs whenever available. Example of bib file: 

%%%%%%%%%%%%%%%%%%%%%%%%%%%%%%%%%%%%%%%%%%%%%%%%%%%%%%%%%%%%%%%%%%%
% Encoding: ISO-8859-1

%@Article{Eichmann:2016yit,
  %author        = {Eichmann, Gernot and Sanchis-Alepuz, Helios and Williams, Richard and Alkofer, Reinhard and Fischer, Christian S.},
  %title         = {{Baryons as relativistic three-quark bound states}},
  %journal       = {Prog. Part. Nucl. Phys.},
  %year          = {2016},
  %volume        = {91},
  %pages         = {1-100},
  %archiveprefix = {arXiv},
  %doi           = {10.1016/j.ppnp.2016.07.001},
  %eprint        = {1606.09602},
  %owner         = {chfi},
  %primaryclass  = {hep-ph},
  %slaccitation  = {%%CITATION = ARXIV:1606.09602;%%},
  %timestamp     = {2018.08.02},
%}

%@Comment{jabref-meta: databaseType:bibtex;}
%%%%%%%%%%%%%%%%%%%%%%%%%%%%%%%%%%%%%%%%%%%%%%%%%%%%%%%%%%%%%%%%%%%

%\newpage
%\appendix
%\renewcommand*{\thesection}{\Alph{section}}

%\section{Appendices, if necessary}\label{appendix}

\end{document}